\documentclass[prc,amsmath,amsfonts,showpacs,eqsecnum,floatfix,twocolumn,nofootinbib]{revtex4}
 \usepackage{graphicx}  
 \usepackage{pstcol}    
 \usepackage{bm}         
 \usepackage{tabularx}

\begin{document}
\hyphenation{Rijken}
\hyphenation{Nijmegen}
 
\title{
    Extended-soft-core Baryon-Baryon ESC08 model \\   
    III. $S=-2$ Hyperon-hyperon/nucleon Interactions }
\author{M.M.\ Nagels}
\affiliation{ Institute of Mathematics, Astrophysics, and Particle Physics \\
 University of Nijmegen, The Netherlands}

\author{Th.A.\ Rijken}
\email[]{t.rijken@science.ru.nl}
\affiliation{ Institute of Mathematics, Astrophysics, and Particle Physics \\
 University of Nijmegen, The Netherlands}
\affiliation{Nishina Center for Accelerator-Based Science, Institute for 
Physical and Chemical Research (RIKEN). Wako, Saitama, 351-0198, Japan}

\author{Y.\ Yamamoto}
\email[]{yamamoto@tsuru.ac.jp}  
\affiliation{Nishina Center for Accelerator-Based Science, Institute for 
Physical and Chemical Research (RIKEN). Wako, Saitama, 351-0198, Japan}


\date{version of: \today}
 
\begin{abstract}
This paper presents the Extended-Soft-Core (ESC) potentials ESC08c           
for baryon-baryon channels with total strangeness $S=-2$.          
The potential models for $S=-2$ are
based on SU(3) extensions of potential models for the $S=0$ and $S=-1$
sectors, which {\it are\/} fitted to experimental data. 
Flavor SU(3)-symmetry is broken only 'kinematically' by the masses   
of the baryons and the mesons. 
For the S=-2 channels no experimental scattering data exist, and 
also the information from hypernuclei is rather limited.
Nevertheless, in the fit to the $S=0$ and $S=-1$ sectors information from the 
NAGARA event and the $\Xi$-well-depth has been used as constraints to determine
the free parameters in the simultaneous fit of the $NN \oplus YN \oplus YY$ data.
Therefore, the potentials for the $S=-2$ sectors are mainly determined
by the NN-, YN-data, and SU(3)-symmetry.
Various properties of the potentials are illustrated by giving results
for scattering lengths, bound states, phase-parameters, 
and total cross sections.\\
Notably is the prediction of a bound state $D^*$ in the 
$\Xi N(^3S_1,I=1)$-channel with a binding energy $B_E= 1.56$ MeV.
This state is "deuteron-like" i.e. a member of the $\{10^*\}$-decuplet. 
As for the normal deuteron $D=pn(^3S_1-^3D_1)$ the strong tensor force is 
responsible for this state.

The features of $\Xi$ hypernuclei predicted by ESC08c are studied on the basis
of the G-matrix approach. The well-depth $U_\Xi = -7.0$ MeV and the $\Xi N-\Lambda\Lambda$
conversion width is $\Gamma^c_\Xi = 4.5$ MeV.\\

\end{abstract}

\maketitle

\section{Introduction}
In this paper, the third in a series of papers following \cite{NRY12a,NRY12b},
henceforth referred to as I and II respectively, on the results and predictions 
of the Extended-soft-core (ESC) model for low energy baryon-baryon interactions.
It presents the next phase in the development of the ESC-models and is the
follow up of the ESC04-models \cite{Rij05,RY05,Rij05b}
and the ESC08a,b-models \cite{YMR10a} for $S=0,-1,-2$.
In \cite{SR99} the Nijmegen soft-core one-boson-exchange (OBE) interactions
NSC97a-f for baryon-baryon (BB) systems for $S=-2,-3,-4$ were presented.

For the S=-2  YY and YN channels hardly any experimental scattering information is available, 
and also the information from hypernuclei is very limited. There are data on
double $\Lambda\Lambda$-hypernuclei, which recently became very much improved
by the observation of the Nagara-event \cite{Tak01}. This event indicates that the 
$\Lambda\Lambda$-interaction is rather weak, in contrast to the estimates
based on the older experimental observations \cite{Dan63,Pro66}. 


In the absense of experimental scattering information, we assume that 
the potentials obey (broken) flavor SU(3) symmetry. As in I and II,
the potentials are parametrized in terms of meson-baryon-baryon, 
and meson-pair-baryon-baryon couplings and gaussian form factors 
as well as diffractive couplings.
This enables us to include in the interaction one-boson-exchange (OBE),
two-pseudoscalar-exchange (TME), and meson-pair-exchange (MPE), 
and diffractive contributions without any new parameters. 
All parameters have been fixed by a simultaneous 
fit to the {\it N\!N} and {\it Y\!N} data, with the constraints 
imposed (i) for $\Lambda\Lambda(^1S_0)$ from the NAGARA event, and
(ii) for the well-depth $U_\Xi$. The latter is assumed to be attractive and is the
main reason for the occurrence of the "deuteron-like" state in the ESC08-model.
For the procedure see the description in I and II. 
This way, each ${\it N\!N} \oplus {\it YN}$-model
leads to a {\it Y\!Y}-model in a well defined way, and the predictions for
the $\Lambda\Lambda$- and $\Xi N$-channels contain no ad hoc free parameters.
We have choosen for ESC08c the options:
SU(3)-symmetry for of coupling constants, and pseudovector coupling for the 
pseudoscalar mesons. 
(In ESC04 also alternative options were investigated, but it appeared that there is no reason
to choose any of these.)
Then, SU(3)-symmetry allows us to define all coupling constants
needed to describe the multi-strange interactions
in the baryon-baryon channels occurring in $\{8\}\otimes\{8\}$.
Quantum-chromodynamics (QCD) is, as is generally accepted now, the physical basis
of the strong interactions. Since in QCD the gluons are flavor blind, SU(3)-symmetry
is a basic symmetry, which is broken by the chiral-symmetry-breaking at 
low energies. This picture supports our assumptions, stated above, on SU(3)-symmetry.
As is shown in \cite{Rij05,RY05} the coupling constants and the $F/(F+D)$-ratio's
used in the ESC04-models follow the predictions of the $^3P_0$-pair creation
model (QPC) \cite{Mic69} rather closely. 
The same is the case for the ESC08-models, see paper I and 
Ref.~\cite{THAR11} for details.
Now, it has been shown that in the 
strong-coupling Hamiltonian lattice formulation of QCD, the flux-tube model, that 
this is indeed the dominant picture in flux-tube breaking \cite{Isg85}.
Therefore, since the ESC-models are very much in line with the Quark-model and QCD, 
the predictions for the $S=-2$-channels should be rather realistic.\\

\noindent The material in this paper is organized by the following considerations:\\
Most of the details of the SU(3) description are well known. In particular
for baryon-baryon scattering the details can be found in papers I, II, and e.g.
\cite{MRS89,RSY99,SR99}. Here we restrict ourselves to a minimal
exposition of these matters that is necessary for the readability of this paper.
Therefore, in Sec.~\ref{sec:2} 
we first review for $S=-2$ the baryon-baryon multi-channel description, and 
present the SU(3)-symmetric interaction Hamiltonian describing
the interaction vertices between mesons and members of the
$J^P={\textstyle     (1/2) }^+$ baryon octet, and define their coupling
constants.  We then identify the various channels which 
occur in the $S=-2$ baryon-baryon systems. 
In appendix~\ref{app:C} the potentials on the isospin basis are given in terms
of the SU(3)-irreps.
In most cases, the
interaction is a multi-channel interaction, characterized by transition
potentials and thresholds. Details were given in \cite{SR99,RSY99}.            
For the details on the pair-interactions, 
we refer to paper I and II \cite{NRY12a,NRY12b}.
In Sec.~\ref{sec:exch} we give a general treatment of the problem
of flavor-exchange forces, which is very helpful to understand 
the proper treatment of exchange forces and the treatment 
of baryon-baryon channels with identical particles.
In Sec.~\ref{sec:thresh} we describe briefly the treatment of the 
multi-channel threshods in the potentials.
In Sec.~\ref{sec:results} we present the results of the ESC08c
potentials for all the sectors with total strangeness $S=-2$.
We give the couplings and $F/(F+D)$-ratio's for OBE-exchanges of ESC08c.   
Similarly, tables with the pair-couplings are shown in appendix~\ref{app:A}.
We give the $S$-wave scattering lengths, discuss the possibility of
bound states in these partial waves. Also, we give the S-matrix information for the
elastic channels in terms of the Bryan-Klarsfeld-Sprung (BKS) phase parameters 
\cite{Bry81,Kla83,Spr85}, or in the Kabir-Kermode (KK) \cite{Kab87} format.
Tables with the BKS-phase parameters are displayed in appendix~\ref{app:B}.                
Such information is very useful for example for the
construction of the $\Lambda$-, $\Sigma$-, and $\Xi$-nucleus potentials.
We also give results for the total cross sections for all leading channels.

Important differences among the different versions of the ESC-models appear in the
$\Xi N$ sectors. Table XXV in Ref.~\cite{RY05} demonstrates that ESC04a,b 
(ESC04c,d) lead to repulsive (attractive) $\Xi$ potentials in nuclear matter.
In ESC08c, and also ESC08a/b/a" \cite{YMR10b}, the 
$\Xi N$ interactions is attractive enough to
produce various $\Xi$ hypernuclei. A notable advantage of ESC08 over ESC04 is the
occurrence of a "deuteron-like" bound state in the I=1 channel, which is 
accessible in a $K^-K^+$-transition $\Xi$-production experiment at JPARC.
Therefore, it is very interesting to study the ESC08 interactions 
in the G-matrix approach to baryonic matter. In Sec.~\ref{sec:gmat-5}, we represent
the $\Xi N$ G-matrix interactions derived from ESC08c as density-dependent
local potentials.
Here, structure calculations for $\Xi$ hypernuclei are performed
with use of $\Xi$-nucleus folding potentials obtained from the G-matrix
interactions. It is discussed how the features of ESC08c appear in the 
level structure of $\Xi$ hypernuclei.
We conclude the paper with a summary and some final remarks in Sec.~\ref{sec:conc}.

\section{Channels, Potentials, and SU(3) Symmetry}
\label{sec:2}

\subsection{Multi-channel Formalism}     
\label{sec:2a}
In this paper we consider the baryon-baryon reactions with $S=-2$
\begin{equation}
 A_1(p_a,s_a)+B_1(p_b,s_b) \rightarrow A_2(p'_a,s'_a)+B_2(p'_b,s'_b)             
\label{eq:2.1}\end{equation}
Like in Ref.'s~\cite{MRS89,RSY99} we will for the {\it YN}-channels 
also refer to $A_1$ and $A_2$
as particles 1 and 3, and to $B_1$ and $B_2$ as particles 2 and 4. For the 
kinematics and the definition of the amplitudes, we refer to paper I 
\cite{Rij05} of this series. Similar material can be found in \cite{MRS89}.
Also, in paper I the derivation of the Lippmann-Schwinger equation 
in the context of the relativistic two-body equation is described.

On the physical particle basis, there are five charge channels:
\begin{eqnarray}
   q=+2:\ \  && \Sigma^+\Sigma^+\rightarrow\Sigma^+\Sigma^+,    \nonumber\\
   q=+1:\ \  && (\Xi^0 p, \Sigma^+\Lambda, \Sigma^0\Sigma^+)\rightarrow
             (\Xi^0 p, \Sigma^+\Lambda, \Sigma^0\Sigma^+),      \nonumber\\
   q=\ \ 0:\ \ && (\Lambda\Lambda , \Xi^0n, \Xi^-p, 
                   \Sigma^0\Lambda, \Sigma^0\Sigma^0, \Sigma^-\Sigma^+)\rightarrow
   \nonumber\\ && (\Lambda\Lambda , \Xi^0n, \Xi^-p, 
                   \Sigma^0\Lambda, \Sigma^0\Sigma^0, \Sigma^-\Sigma^+), \nonumber\\
   q=-1:\ \  && (\Xi^- n, \Sigma^-\Lambda, \Sigma^-\Sigma^0)\rightarrow
             (\Xi^- n, \Sigma^-\Lambda, \Sigma^-\Sigma^0),      \nonumber\\
   q=-2:\ \  && \Sigma^-\Sigma^-\rightarrow\Sigma^-\Sigma^-.
\label{eq:2.2}\end{eqnarray}
Like in \cite{MRS89,RSY99}, the potentials are calculated on the isospin basis.
For $S=-2$ hyperon-nucleon systems there are three isospin channels:
\begin{eqnarray}
I={\textstyle 0}:\ \ && (\Lambda \Lambda, \Xi N, \Sigma \Sigma\rightarrow
          \Lambda \Lambda, \Xi N, \Sigma \Sigma), \nonumber\\      
I={\textstyle 1}:\ \ && (\Xi N, \Sigma\Lambda, \Sigma \Sigma\rightarrow
                           \Xi N, \Sigma\Lambda, \Sigma \Sigma), \nonumber\\
I={\textstyle 2}:\ \ && \Sigma \Sigma\rightarrow\Sigma\Sigma.      
\label{eq:2.3}\end{eqnarray}

  \begin{figure}   
  \resizebox{7.25cm}{8.75cm}
  {\includegraphics[100,500][500,850]{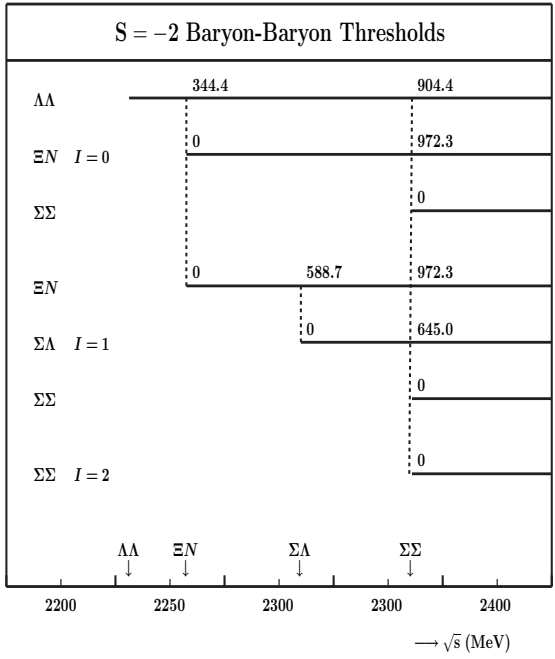}}
 \caption{Thresholds in {\it YN}- and {\it YY}-channels for $S=-2$. }                
 \label{thresh1}
  \end{figure}

For the kinematics of the reactions and the various thresholds, see \cite{RSY99}.
In this work we do not solve the Lippmann-Schwinger equation, but the 
multi-channel Schr\"{o}dinger equation in configuration space, completely 
analogous to \cite{MRS89}.
The multi-channel Schr\"odinger equation for the configuration-space
potential is derived from the Lippmann-Schwinger equation through
the standard Fourier transform, and the equation for the radial
wave function is found to be of the form~\cite{MRS89}
\begin{equation}
   u^{\prime\prime}_{l,j}+(p_i^2\delta_{i,j}-A_{i,j})u_{l,j}
       -B_{i,j}u'_{l,j}=0,           
\label{eq:2.6}\end{equation}
where $A_{i,j}$ contains the potential, nonlocal contributions, and
the centrifugal barrier, while $B_{i,j}$ is only present when non-local
contributions are included. 
The solution in the presence of open and closed channels is given,
for example, in Ref.~\cite{Nag73}.
The inclusion of the Coulomb interaction in the
configuration-space equation is well known and included in the evaluation
of the scattering matrix.

Obviously, the potential on the particle basis for the $q=2$ and
$q=-2$ channels are given by the $I={\textstyle 2}$
$\Sigma\Sigma $ potential on the isospin basis. For $q=0$ and $q=\pm 1$, the
potentials are related to the potentials on the isospin basis by an
isospin rotation. 
Using the indices $a, b, c, d$ for $\Lambda\Lambda, \Xi N, \Sigma\Lambda$, and
$\Sigma\Sigma$ respectively, we have \cite{Mae95}
 \onecolumngrid                                          
 \begin{equation}
 V(q=0) =
  \left(\begin{array}{cccccc}
    V_{aa}
    &  \sqrt{\textstyle\frac{1}{2}}V_{ba}             
    & -\sqrt{\textstyle\frac{1}{2}}V_{ba}             
    & 0
    & -\sqrt{\textstyle\frac{1}{3}}V_{ad}             
    &  \sqrt{\textstyle\frac{1}{3}}V_{ad} \\[2mm]     
      \cdot
    &  {\textstyle\frac{1}{2}}\left[V_{bb}(1)+V_{bb}(0)\right]                
    &  {\textstyle\frac{1}{2}}\left[V_{bb}(1)-V_{bb}(0)\right]                
    &  \sqrt{\textstyle\frac{1}{2}}V_{bc}             
    & -\sqrt{\textstyle\frac{1}{6}}V_{bd}(0)          
    &  \sqrt{\textstyle\frac{1}{6}}V_{bd}(0)                                 
      -{\textstyle\frac{1}{2}}V_{bd}(1) \\[2mm]                         
      \cdot & \cdot
    &  {\textstyle\frac{1}{2}}\left[V_{bb}(1)+V_{bb}(0)\right]                
    &  \sqrt{\textstyle\frac{1}{2}}V_{bc}             
    &  \sqrt{\textstyle\frac{1}{6}}V_{bd}(0)          
    & -\sqrt{\textstyle\frac{1}{6}}V_{bd}(0)                                 
      -{\textstyle\frac{1}{2}}V_{bd}(1) \\[2mm]                         
      \cdot & \cdot & \cdot & V_{cc} & 0  
    &  -\sqrt{\textstyle\frac{1}{2}}V_{cd} \\[2mm]     
      \cdot & \cdot & \cdot & \cdot         
    &  {\textstyle\frac{1}{3}}\left[2V_{dd}(2)+V_{dd}(0)\right]                
    &  {\textstyle\frac{1}{3}}\left[ V_{dd}(2)-V_{dd}(0)\right] \\[2mm]        
      \cdot & \cdot & \cdot & \cdot & \cdot 
    &  {\textstyle\frac{1}{6}}\left[V_{dd}(2)+3V_{dd}(1)+2V_{dd}(0)\right]                
        \end{array}\right)\ ,
\label{eq:2.4} \end{equation}
and for $q=+1$ we have
 \begin{equation}
 V(q=+1) =
  \left(\begin{array}{ccc}
    V_{bb}(1) & V_{bc} & -\sqrt{\textstyle\frac{1}{2}}V_{bd} \\[2mm]     
    V_{bc}    & V_{cc} & -\sqrt{\textstyle\frac{1}{2}}V_{cd} \\[2mm]     
      -\sqrt{\textstyle\frac{1}{2}}V_{bd}(1)          
    & -\sqrt{\textstyle\frac{1}{2}}V_{cd}          
    &  {\textstyle\frac{1}{2}}\left[V_{dd}(1)+V_{dd}(2)\right]                
        \end{array}\right)\ ,
\label{eq:2.5} \end{equation}
 \twocolumngrid                                          
Here, when necessary an isospin label is added in parentheses.

The momentum space and configuration space potentials for the ESC-model
have been described in \cite{Rij05} for baryon-baryon in general.
Therefore, they apply also to hyperon-nucleon and we can refer for that
part of the potential to paper I.  
Also in the ESC-model, the potentials are of such a form that they are exactly 
equivalent in both momentum space and configuration space. 
The treatment of the mass differences among the baryons are handled exactly 
similar as is done in \cite{MRS89,RSY99}. Also, exchange potentials related to
strange meson exchanhes $K, K^*$ etc. , can be found in these references. 

The baryon mass differences in the intermediate states for TME- and MPE-
potentials has been neglected for {\it YN}-scattering. This, although possible
in principle, becomes rather laborious and is not expected to change the 
characteristics of the baryon-baryon potentials.     

\subsection{Potentials and SU(3) Symmetry}
\label{sec:channels}
We consider all possible baryon-baryon interaction channels, where
the baryons are the members of the $J^P={\textstyle\frac{1}{2}}^+$
baryon octet
\begin{equation}
  B = \left( \begin{array}{ccc}
      {\displaystyle\frac{\Sigma^{0}}{\sqrt{2}}+\frac{\Lambda}{\sqrt{6}}}
               &  \Sigma^{+}  &  p  \\[2mm]
      \Sigma^{-} & {\displaystyle-\frac{\Sigma^{0}}{\sqrt{2}}
                   +\frac{\Lambda}{\sqrt{6}}}  &  n \\[2mm]
      -\Xi^{-} & \Xi^{0} &  {\displaystyle-\frac{2\Lambda}{\sqrt{6}}}
             \end{array} \right).
\label{eq:3.1}\end{equation}
The baryon masses, used in this paper, are given in Table~\ref{tabBmass}.
The meson nonets can be written as
\begin{equation}
     P=P_{\rm sin}+P_{\rm oct},
\label{eq:3.2}\end{equation}
where the singlet matrix $P_{\rm sin}$ has elements $\eta_0/\sqrt{3}$
on the diagonal, and the octet matrix $P_{\rm oct}$ is given by
\begin{equation}
   P_{\rm oct} = \left( \begin{array}{ccc}
      {\displaystyle\frac{\pi^{0}}{\sqrt{2}}+\frac{\eta_{8}}{\sqrt{6}}}
             & \pi^{+}  &  K^{+}  \\[2mm]
      \pi^{-} & {\displaystyle-\frac{\pi^{0}}{\sqrt{2}}
         +\frac{\eta_{8}}{\sqrt{6}}}  &   K^{0} \\[2mm]
      K^{-}  &  \overline{{K}^{0}}
             &  {\displaystyle-\frac{2\eta_{8}}{\sqrt{6}}}
             \end{array} \right),
\label{eq:3.3}\end{equation}
and where we took the pseudoscalar mesons with $J^P=0^+$ as a specific
example. 
For the other mesons the octet matrix is obtained by the following 
substitutions: (i) vector mesons
$\pi \rightarrow \rho$, $\eta_8 \rightarrow \phi_8$,
$ K \rightarrow K^*$, 
(ii) scalar $\pi \rightarrow a_0$, $\eta_8 \rightarrow f_{0,8}$,
$ K \rightarrow \kappa$, 
(iii) axial-vector $\pi \rightarrow A_1$, $\eta_8 \rightarrow E_{1,8}$,
$ K \rightarrow K_A$. 
\onecolumngrid
\vspace*{15mm}

Introducing the following notation for the isodoublets,
\begin{eqnarray}
  N &=& \left(\begin{array}{c} p \\ n \end{array} \right), \ \ \
  \Xi=\left(\begin{array}{c} \Xi^{0} \\ \Xi^{-} \end{array} \right), \ \ \
  {\rm and} \ \ \
  K = \left(\begin{array}{c} K^{+} \\ K^{0} \end{array} \right), \ \ \
   K_{c}=\left(\begin{array}{c} \overline{K^{0}} \\
               -K^{-} \end{array} \right),        
\label{eq:3.4} \end{eqnarray}
the most general, SU(3) invariant, interaction Hamiltonian is then
given by~\cite{Swa63}
\begin{eqnarray}
   {\cal H}_{\rm pv}^{\rm oct} &=&
  g_{N\!N\pi}(\overline{N}\bm{\tau}N)\!\cdot\!\bm{\pi}
  -ig_{\Sigma\Sigma\pi}(\overline{\bm{\Sigma}}\!\times\!\bm{\Sigma})
      \!\cdot\!\bm{\pi}
  +g_{\Lambda\Sigma\pi}(\overline{\Lambda}\bm{\Sigma}+
      \overline{\bm{\Sigma}}\Lambda)\!\cdot\!\bm{\pi}
  +g_{\Xi\Xi\pi}(\overline{\Xi}\bm{\tau}\Xi)\!\cdot\!\bm{\pi}
         +  \nonumber\\
 &&g_{\Lambda N\!K}\left[(\overline{N}K)\Lambda
         +\overline{\Lambda}(\overline{K}N)\right]
   +g_{\Xi\Lambda K}\left[(\overline{\Xi}K_{c})\Lambda
         +\overline{\Lambda}(\overline{K_{c}}\Xi)\right] + \nonumber\\
 &&g_{\Sigma N\!K}\left[\overline{\bm{\Sigma}}\!\cdot\!
         (\overline{K}\bm{\tau}N)+(\overline{N}\bm{\tau}K)
         \!\cdot\!\bm{\Sigma}\right]
   +g_{\Xi\Sigma K}\left[\overline{\bm{\Sigma}}\!\cdot\!
       (\overline{K_{c}}\bm{\tau}\Xi)
     +(\overline{\Xi}\bm{\tau}K_{c})\!\cdot\!\bm{\Sigma}\right]
                                +          \nonumber\\
 &&g_{N\!N\eta_{8}}(\overline{N}N)\eta_{8}
   +g_{\Lambda\Lambda\eta_{8}}(\overline{\Lambda}\Lambda)\eta_{8}
   +g_{\Sigma\Sigma\eta_{8}}(\overline{\bm{\Sigma}}\!\cdot\!
       \bm{\Sigma})\eta_{8}
   +g_{\Xi\Xi\eta_{8}}(\overline{\Xi}\Xi)\eta_{8}  +    \nonumber\\
 &&g_{N\!N\eta_{0}}(\overline{N}N)\eta_{0}
   +g_{\Lambda\Lambda\eta_{0}}(\overline{\Lambda}\Lambda)\eta_{0}
   +g_{\Sigma\Sigma\eta_{0}}(\overline{\bm{\Sigma}}\!\cdot\!
       \bm{\Sigma})\eta_{0}
   +g_{\Xi\Xi\eta_{0}}(\overline{\Xi}\Xi)\eta_{0},  
\label{eq:3.5}\end{eqnarray}
\twocolumngrid
where we again took the pseudoscalar mesons as an example, dropped
the Lorentz character of the interaction vertices,
and introduced the charged-pion mass to make the pseudovector coupling
constant $f$ dimensionless.
All coupling constants can be expressed in terms of only four parameters.
The explicit expressions can be found in Ref.~\cite{RSY99}.
The $\Sigma$-hyperon is an isovector with phase chosen such~\cite{Swa63}
that
\begin{equation}
  \bm{\Sigma}\!\cdot\!\bm{\pi} = \Sigma^{+}\pi^{-}
       +\Sigma^{0}\pi^{0}+\Sigma^{-}\pi^{+}.
\label{eq:3.6}\end{equation}
This definition for $\Sigma^+$ differs from the standard Condon and
Shortley phase convention~\cite{Con35} by a minus sign. This means that,
in working out the isospin multiplet for each coupling constant in
Eq.~(\ref{eq:3.5}), each $\Sigma^+$ entering or leaving an interaction
vertex has to be assigned an extra minus sign. However, if the potential
is first evaluated on the isospin basis and then, via an isospin
rotation, transformed to the potential on the physical particle basis
(see below), this extra minus sign will be automatically accounted for.

In appendix~\ref{app:C}, Table~\ref{tab.irrep1} and Table~\ref{tab.irrep2}
we give the relation between the potentials on the isospin-basis, see
(\ref{eq:2.4})-(\ref{eq:2.5}), and the SU(3)-irreps.

Given the interaction Hamiltonian (\ref{eq:3.5}) and a theoretical
scheme for deriving the potential representing a particular Feynman
diagram, it is now straightforward to derive the one-meson-exchange
baryon-baryon potentials.
We follow the Thompson approach~\cite{Tho70,Rij91,Rij92,Rij96} and
expressions for the potential in momentum space can be found in
Ref.~\cite{MRS89}. 
Since the nucleons have strangeness $S=0$, the hyperons $S=-1$,
and the cascades $S=-2$, the possible baryon-baryon interaction
channels can be classified according to their total strangeness,
ranging from $S=0$ for $N\!N$ to $S=-4$ for $\Xi\Xi$.
Apart from the wealth of accurate $N\!N$ scattering data for the total
strangeness $S=0$ sector, there are only a few, and not very accurate,
$Y\!N$ scattering data for the $S=-1$ sector, while there are no data
at all for the $S<-1$ sectors. We therefore believe that at this stage
it is not yet worthwhile to explicitly account for the small mass
differences between the specific charge states of the baryons and
mesons; i.e., we use average masses, isospin is a good quantum number,
and the potentials are calculated on the isospin basis. The possible
channels on the isospin basis are given in 
(\ref{eq:2.3}).

However, the Lippmann-Schwinger or Schr\"odinger equation is solved for
the physical particle channels, and so scattering observables are
calculated using the proper physical baryon masses.
The possible channels on the physical particle basis can be classified
according to the total charge $Q$; these are given in
(\ref{eq:2.2}).
The corresponding potentials are obtained
from the potential on the isospin basis by making the appropriate
isospin rotation. The matrix elements of the isospin rotation matrices
are nothing else but the Clebsch-Gordan coefficients for the two baryon
isospins making up the total isospin. (Note that this is the reason why
the potential on the particle basis, obtained from applying an isospin
rotation to the potential on the isospin basis, will have the correct
sign for any coupling constant on a vertex which involves a $\Sigma^+$.)

In order to construct the potentials on the isospin basis, we need
first the matrix elements of the various OBE exchanges between particular
isospin states. 
Using the iso-multiplets (\ref{eq:3.3}) and the Hamiltonian (\ref{eq:3.5})
the isospin factors can be calculated.
The results are given in Table~\ref{tab.iso},
where we use the pseudoscalar mesons as a specific example.
The entries contain the flavor-exchange operator $P_f$, which is
$+1$ for a flavor symmetric and $-1$ for a flavor anti-symmetric two-baryon
state. Since two-baryons states are totally anti-symmetric, one has
$P_f=-P_{x} P_{\sigma}$. Therefore, the exchange operator $P_f$ has the value
$P_f=+1$ for even-$L$ singlet and odd-$L$ triplet partial waves, and
$P_f=-1$ for odd-$L$ singlet and even-$L$ triplet partial waves.
In order to understand Table~\ref{tab.iso} fully, we have given in the 
following section Sec.~\ref{sec:exch} a general treatment of exchange forces.
This treatment shows also how to deal with the case where the initial/final
state involves identical particles and the final/inition state does not.

Second, we need to evaluate the TME and the MPE exchanges. The method
we used for these is the same as for hyperon-nucleon, and is described
in \cite{RY05}, Sec. IID.

\section{Exchange Forces}       
\label{sec:exch} 
The proper treatment of the flavor-exchange forces is for the $S \leq -2$-channels
more difficult than for the $S=0,-1$-channels. The extra complication is the occurrence
of couplings between channels with identical and channels with non-identical particles.
In order to understand the several $\sqrt{2}$-factors, see \cite{SR99}, we give 
here a systematic treatment of the flavor-exchange potentials.
The method followed is using a multi-channel framework, which starts starts by
ordering the two-particle states by assigning $A_i$ and $B_i$ for the channel
labeled with the index $i$, like in eq.~(\ref{eq:2.1}). The particles $A_i$ and
$B_i$ have CM-momenta $p_i$ and $-p_i$, spin components $s_{A,i}$ and $s_{B,i}$.
The two-baryon states $|A_i B_i\rangle$ and $|B_i A_i\rangle$ are considered to
be distinct, leading to distinct two-baryon channels. The 'direct' and
the 'exchange' T-amplitudes are given by the T-matrix elements
\begin{equation}
 \langle A_j B_j| T_d|A_i B_i\rangle,\ \ \langle B_j A_j|T_e|A_iB_i\rangle\ ,
\label{exch.1}\end{equation}
and similarly for the direct and flavor-exchange potentials $V_d$ and $V_e$.
It is obvious from rotation invariance that 
\begin{eqnarray}
 \langle A_j B_j| T_d|A_i B_i\rangle &=& \langle B_j A_j|T_d|B_iA_i\rangle\ ,
 \nonumber\\ 
 \langle B_j A_j| T_e|A_i B_i\rangle &=& \langle A_j B_j|T_e|B_iA_i\rangle\ .    
\label{exch.2}\end{eqnarray}
A similar definition (\ref{exch.1}) and relation (\ref{exch.2}) apply for 
the direct and flavor-exchange potentials $V_d$ and $V_e$.
 \begin{figure}[ht]
  \resizebox{4.5cm}{!}
 {\includegraphics[200, 325][400,600]{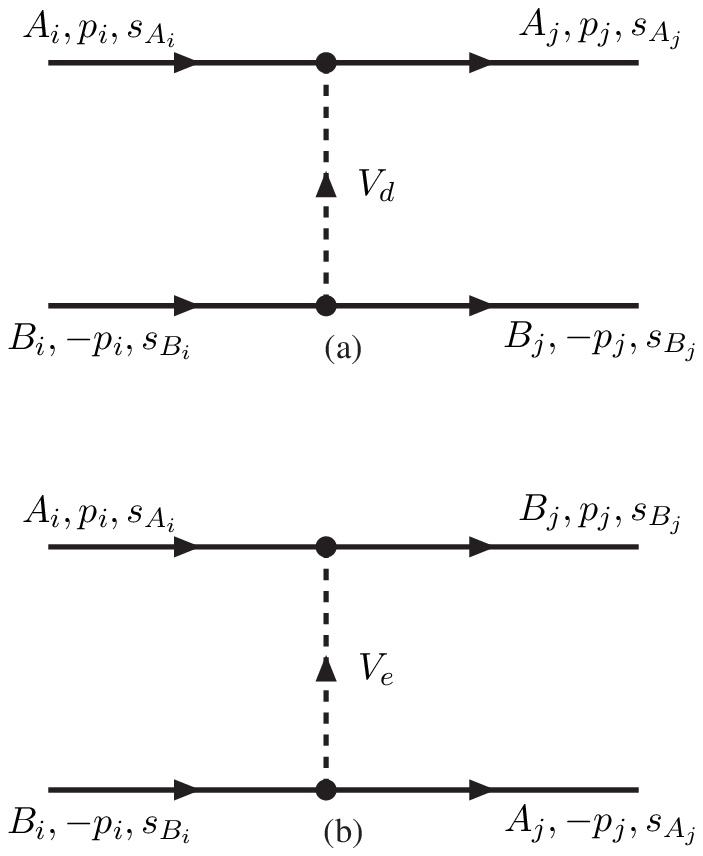}}
 \caption{$V_d$ (a) and $V_e$ (b) in CM-system.}         
 \label{fig:vevd}
 \end{figure}
We notice that in interchanging A and B there is no exchange of momenta or spin-components, 
see Fig.~\ref{fig:vevd}. This is necessary for the application of Lippmann-Schwinger type
of integral equations, which can produce only one type of spectral function e.g.
$\rho(s,t)$. So, the momentum transfer for $V_d$ and for $V_e$ is the same. 
Viewed from the coupled-channel scheme this is the standard situation. 

\noindent The integral equations with two-baryon unitarity, e.g. 
the Thompson-, Lippmann-Schwinger-equation etc., read for the $T_d$- and $T_e$-operator
\onecolumngrid
\begin{subequations}
\label{exch.3}
\begin{eqnarray}
\langle A_jB_j|T_d|A_iB_i\rangle &=& \langle A_jB_j|V_d|A_iB_i\rangle + \sum_k \left[
\langle A_jB_j|V_d|A_kB_k\rangle\ G_k\ \langle A_kB_k|T_d|A_iB_i\rangle 
 \right.\nonumber\\ && \left. + 
\langle A_jB_j|V_e|B_kA_k\rangle\ G_k\ \langle B_kA_k|T_e|A_iB_i\rangle\ \right]\ , 
\\ 
\langle B_jA_j|T_e|A_iB_i\rangle &=& \langle B_jA_j|V_e|A_iB_i\rangle + \sum_k \left[
\langle B_jA_j|V_d|B_kA_k\rangle\ G_k\ \langle B_kA_k|T_e|A_iB_i\rangle 
 \right.\nonumber\\ && \left. + 
\langle B_jA_j|V_e|A_kB_k\rangle\ G_k\ \langle A_kB_k|T_d|A_iB_i\rangle\ \right]\ . 
\end{eqnarray}
\end{subequations}
\twocolumngrid
These coupled equations can be diagonalized by introducing the T$^\pm$- and V$^\pm$-operators
\begin{equation}
 T^{\pm} = T_d \pm T_e\ ,\ \ V^{\pm} = V_d \pm V_e\ .    
\label{exch.4}\end{equation}
which, as follows from (\ref{exch.3}), satisfy separate integral equations
\begin{eqnarray}
&& \hspace*{-7mm}
 \langle A_jB_j|T^\pm|A_iB_i\rangle = \langle A_jB_j|V^\pm|A_iB_i\rangle 
 + \nonumber\\ && \hspace*{-7mm}
 \sum_k
\langle A_jB_j|V^\pm|A_kB_k\rangle\ G_k\ \langle A_kB_k|T^\pm|A_iB_i\rangle.
\label{exch.5}\end{eqnarray}
Notice that on the basis of states with definite flavor symmetry
\begin{equation}
 |A_i B_i\rangle_{\pm} = \frac{1}{\sqrt{2}}\left[ |A_i B_i \rangle 
 \pm |B_i A_i\rangle\right]\ , 
\label{exch.6}\end{equation}
the $T^\pm$ and $V^\pm$ matrix elements are also given by
\begin{equation}
  T^\pm_{ij} = _\pm\!\!\langle A_i B_i | T | A_j B_j\rangle_\pm\ ,\ \  
  V^\pm_{ij} = _\pm\!\!\langle A_i B_i | V | A_j B_j\rangle_\pm\ .     
\label{exch.7}\end{equation}
\twocolumngrid

\subsection{Identical Particles}
Sofar, we considered the general case where $A_i \neq B_i$ for all channels.
In the case that $A_i=B_i$ for some $i$, one has
$\langle B_iA_i|V_e  |A_iB_i\rangle = 0$, because there is no distinct 
physical state corresponding to the 'flavor exchange-state'. For example for
a flavor single channel like $pp$ one deduces from (\ref{exch.3}) that then also
$T_e=0$, and one has in this case the integral equation
\begin{eqnarray}
&& \langle A_jB_j|T_d|A_iB_i\rangle = \langle A_jB_j|V_d|A_iB_i\rangle +
 \nonumber\\ && \sum_k 
\langle A_jB_j|V_d|A_kB_k\rangle\ G_k\ \langle A_kB_k|T_d|A_iB_i\rangle\ , 
\label{exch.8}\end{eqnarray}
where the labels $i$ and $j$ now denote e.g. the spin-components.

\subsection{Coupled $\Lambda\Lambda$ and $\Xi N$ system}
This multi-channel system represents the case where there is mixture of channels with
identical and with non-identical particles. The three states we distinguish are
$|\Lambda\Lambda\rangle, |\Xi N\rangle$, and $|N\Xi\rangle$. Choosing the same 
ordering, the potential written as a $3\times 3$-matrix reads
\begin{equation}
V = \left(\begin{array}{ccc} 
 (\Lambda\Lambda|V|\Lambda\Lambda) & (\Lambda\Lambda|V| \Xi N) & 
 (\Lambda\Lambda|V| N \Xi \\
 (\Xi N |V|\Lambda\Lambda) & (\Xi N |V| \Xi N) & (\Xi N |V| N \Xi \\
 (N \Xi |V|\Lambda\Lambda) & (N \Xi |V| \Xi N) & (N \Xi |V| N \Xi   
 \end{array}\right)\ .
\label{exch.9}\end{equation}
With a similar notation for the T-matrix, 
the Lippmann-Schwinger equation can be written compactly as a $3\times 3$-matrix
equation:
\begin{equation}
   T = V + V\ G\ T\ ,\ {\rm with}\ G_{ij}= G_i\ \delta_{ij}\ .
\label{exch.10}\end{equation}
Next, we make a transformation to states, which are either symmetric or anti-symmetric
for particle interchange. Then, according to the discussion above, we can separate them
in the Lippmann-Schwinger equation. This is achieved by the transformation
\onecolumngrid
\begin{equation}
 \left(\begin{array}{c} 
 \Lambda\Lambda \\ \Xi N \\ N \Xi \end{array}\right) \Rightarrow
 \left(\begin{array}{c} 
 \Lambda\Lambda \\ (\Xi N + N \Xi)/\sqrt{2} \\ 
 (\Xi N - N \Xi)/\sqrt{2} \end{array}\right) =             
 \left(\begin{array}{ccc} 1 & 0 & 0 \\ 0 & 
 1/\sqrt{2} & 1/\sqrt{2} \\
 0 & 1/\sqrt{2} &-1/\sqrt{2}    
 \end{array}\right)
 \left(\begin{array}{c} 
 \Lambda\Lambda \\ \Xi N \\ N \Xi \end{array}\right)\ .             
\label{exch.11}\end{equation}
one gets in the transformed basis for the potential
\begin{equation}
 UVU^{-1} = \left(\begin{array}{cll} 
 V_{\Lambda\Lambda;\Lambda\Lambda} & 
 (V_{\Lambda\Lambda;\Xi N} + V_{\Lambda\Lambda;N \Xi})/\sqrt{2} & 
 (V_{\Lambda\Lambda;\Xi N} - V_{\Lambda\Lambda;N \Xi})/\sqrt{2} \\
 (V_{\Xi N;\Lambda\Lambda} + V_{N \Xi;\Lambda\Lambda})/\sqrt{2} & 
 (V_{\Xi N;\Xi N} + V_{\Xi N; N \Xi}) & \hspace{1.43cm} 0 \\
 (V_{\Xi N;\Lambda\Lambda} - V_{N \Xi;\Lambda\Lambda})/\sqrt{2} &\hspace{1.43cm}  0 &
 (V_{\Xi N;\Xi N} - V_{\Xi N; N \Xi}) \\
\end{array}\right)\ ,
\label{exch.12}\end{equation}
and of course, a similar form is obtained for the T-matrix on the transformed basis.
\twocolumngrid
Now, obviously we have that $V_{\Lambda\Lambda;\Xi N} = V_{\Lambda\Lambda;N \Xi}$ and          
$V_{\Xi N;\Lambda\Lambda} = V_{N \Xi;\Lambda\Lambda}$. Therefore, one sees that
the even and odd states under particle exchange are decoupled in (\ref{exch.12}).
Also $(V_{\Xi N;\Lambda\Lambda} + V_{N \Xi;\Lambda\Lambda})/\sqrt{2}= \sqrt{2}
V_{\Xi N;\Lambda\Lambda}$, etc. showing the           
appearance of the $\sqrt{2}$-factors, mentioned before. Indeed, they appear
in a systematic way using the multi-channel framework.

\subsection{The K-exchange Potentials}                         
Consider for example the $(\Lambda\Lambda, \Xi N)$-system, having $I=0$.
Mesons with strangeness, i
$K(495)$, $K^*(892)$, $\kappa(841)$, $K_A(1273)$, $K_B(1400)$,
are obviously the only ones that can give transition potentials, i.e. 
$ V_{\Lambda\Lambda;\Xi N} \neq 0$ and $ V_{\Xi N;\Lambda\Lambda} \neq 0$.  
The $\Xi N(I=0)$-states anti-symmetric and
symmetric in flavor are respectively:
\begin{subequations}
\label{exch.13}
\begin{eqnarray}
&& P_f=-1:\ \frac{1}{\sqrt{2}}\left[|\Xi N(I=0)\rangle - |N \Xi(I=0)\rangle\right]\ ,
\\ 
&& P_f=+1:\ \frac{1}{\sqrt{2}}\left[|\Xi N(I=0)\rangle + |N \Xi(I=0)\rangle\right]\ .
\end{eqnarray}
\end{subequations}
Analyzing the $^1S_0$-state one has because of the anti-symmetry of the two-fermion state
w.r.t. the exchange of all quantum labels, $P_f = -P_\sigma P_x = +1$, where
$P_f$ denotes the flavor-symmetry.  
Taking here the $K(495)$ as a generic example, and using (\ref{eq:3.4}) and (\ref{eq:3.5}),
one finds that
\begin{eqnarray}
\langle \Xi^0 n|V(K)|\Lambda\Lambda\rangle &=&
 + g_{K\Lambda N} g_{K\Xi \Lambda}, \\  
\langle \Xi^- p|V(K)|\Lambda\Lambda\rangle &=&
 - g_{K\Lambda N} g_{K\Xi \Lambda}\ .  
\label{exch.14}\end{eqnarray}
Then, since
$|\Xi N(I=0)\rangle = \left[|\Xi^0 n\rangle - |\Xi^- p\rangle\right]/\sqrt{2}$, 
one obtains for the 'direct' potential the coupling
\begin{equation}
V_{\Xi N;\Lambda\Lambda} = \langle \Xi N(I=0)\rangle | V_d(K)|
\Lambda\Lambda\rangle \Leftarrow \sqrt{2} g_{K\Lambda N} g_{K\Xi \Lambda}\ .
\label{exch.15}\end{equation}
The same result is found for the 'exchange' potential $V_{N\Xi;\Lambda\Lambda}$. 
Therefore
\begin{eqnarray}
\frac{1}{\sqrt{2}}\left(V_{\Xi N;\Lambda\Lambda} + V_{N \Xi;\Lambda\Lambda} 
\right) &=& \langle \Xi N(I=0) | V(K)| \Lambda\Lambda\rangle 
 \nonumber\\ &&
\Leftarrow 2 g_{K\Lambda N} g_{K\Xi \Lambda}\ ,
\label{exch.16}\end{eqnarray}
which has indeed the $(1+P_f)$-factor given in Table~\ref{tab.iso},  
and is identical to Table IV in \cite{SR99}, for
$(\Lambda\Lambda|K|\Xi N)$.                  

For 
$\langle\Xi N|K|\Sigma\Lambda\rangle$ the entry for $I=1$ consists of two parts. These 
correspond to $V_d \propto g_{\Lambda N K} g_{\Xi\Sigma K}$ and  
$V_e \propto g_{\Sigma N K} g_{\Xi\Lambda K}$ respectively, i.e. the direct and      
exchange contributions involve different couplings. Therefore, they are not
added together.

\subsection{The $\eta$- and $\pi$-exchange Potentials}                         
Next, we discuss briefly the calculation of the 
entries for $\eta$- and $\pi$-exchange in Table~\ref{tab.iso}. 
First, the entries with --- indicate that the corresponding physical state does
not exist. Next we give further specific remarks and calculations:
\begin{enumerate}
\item[a.] For $\eta,\eta'$-exchange one has that $V_e=0$. The matrix elements for the
$\Lambda\Lambda$- and $\Xi N$-state are easily seen to be correct. For the 
$\Sigma\Sigma$-states one has $P_f=1$ for $I_{\Sigma\Sigma}=0,2$, and 
$P_f=-1$ for $I_{\Sigma\Sigma}=1$. This explains the $\Sigma\Sigma$ matrix 
element.
\item[b.] For $\langle\Xi N|\pi|\Xi N\rangle$ the calculation is identical to that 
for NN, in particular $p n$.
\item[c.] For $\langle\Sigma\Sigma|\pi|\Sigma\Sigma\rangle$ consider the $I=0,I_3=0$ and
$I=1,I_3=0$ matrix elements. In these cases one has $V_e=0$ as one can easily check.
Then, using the cartesian base, we have for 
$\langle\Sigma_i\Sigma_m|\pi|\Sigma_j\Sigma_n\rangle \Rightarrow
-g_{\Sigma\Sigma\pi}^2 \sum_{p=1}^3 \epsilon_{jip}\epsilon_{nmp}=
 -g_{\Sigma\Sigma\pi}^2 (\delta_{jn}\delta_{im}-\delta_{jm}\delta_{in})$. 
Employing the states $|I=0,I_3=0\rangle \sim -\sum_{i,m=1}^3\delta_{im}|\Sigma_i\Sigma_m\rangle
/\sqrt{3}$ and 
$|I=1,I_3=0\rangle \sim -i\sum_{i,m=1}^3\epsilon_{im3}|\Sigma_i\Sigma_m\rangle/\sqrt{2}$,
one obtains the results in Table~\ref{tab.iso}.
\end{enumerate}
With the ingredients given above one can easily check the other entries in Table~\ref{tab.iso}.
 
\vspace{3mm}

\twocolumngrid
 
\begin{table}
\caption{Isospin factors for the various meson exchanges in the
         different total strangeness and isospin channels.
         $P_f$ is the flavor-exchange operator.
         The $I=2$ case only contributes to $S=-2$ $\Sigma\Sigma$
         scattering, where the isospin factors can collectively be
         given by $(\Sigma\Sigma|\eta,\eta',\pi|\Sigma\Sigma)=
         {\protect\textstyle\frac{1}{2}}(1+P_f)$, and so they are
         not separately displayed in the table.
         Non-exixisting channels are marked by a long-dash.
         }
\begin{ruledtabular}
\begin{tabular}{ccc} 
  $S=-2$ & $I=0$ & $I=1$ \\
\colrule 
  $(\Lambda\Lambda|\eta,\eta'|\Lambda\Lambda)$
                       & ${\textstyle\frac{1}{2}}(1+P_f)$ & --- \\[.5mm]
  $(\Xi N|\eta,\eta'|\Xi N)$ & ${\textstyle\frac{1}{2}}(1+P_f)$ & 1 \\[.5mm]
  $(\Sigma\Sigma|\eta,\eta'|\Sigma\Sigma)$
                       & ${\textstyle\frac{1}{2}}(1+P_f)$
                       & ${\textstyle\frac{1}{2}}(1-P_f)$ \\
  $(\Sigma\Lambda|\eta,\eta'|\Sigma\Lambda)$ & --- & 1 \\
  $(\Xi N|\pi|\Xi N)$  & $-3$ & 1 \\
  $(\Sigma\Sigma|\pi|\Sigma\Sigma)$ & $-(1+P_f)$
                       & $-{\textstyle\frac{1}{2}}(1-P_f)$ \\[.5mm]
  $(\Lambda\Lambda|\pi|\Sigma\Sigma)$
                       & $-{\textstyle\frac{1}{2}}\sqrt{3}(1+P_f)$ & --- \\
  $(\Sigma\Lambda|\pi|\Lambda\Sigma)$ & --- & $P_f$ \\
  $(\Sigma\Sigma|\pi|\Sigma\Lambda)$ & --- & $(1-P_f)$ \\
  $(\Lambda\Lambda|K|\Xi N)$ & $1+P_f$ & --- \\
  $(\Sigma\Sigma|K|\Xi N)$   & $\sqrt{3}(1+P_f)$ & $\sqrt{2}(1-P_f)$ \\
  $(\Xi N|K|\Sigma\Lambda)$  & --- & $\sqrt{2};-P_f\sqrt{2}$ \\
\end{tabular}
\end{ruledtabular}
\label{tab.iso}
\end{table}
\section{ Short-range Phenomenology}                               
\label{sec:5} 
For a detailed discussion and description of the short-range region 
we refer to paper II \cite{NRY12b}. Here, the meson- and diffractive-exchange and
the quark-core in the ESC08c-modeling has been described. In this section 
we give the quark-core phenomenology for the S=-2 baryon-baryon channels.
\subsection{Relation S=-2 YN,YY-states and SU$_{fs}$(6)-irreps}  
\label{sec:5a} 
The relation between the SU$_f$(3)-irreps and SU$_{fs}$(6)-irreps has 
been derived in paper II \cite{NRY12b}
In Appendix~\ref{app:C} the S=-2 BB-potentials are given in terms of the 
SU(3)$_f$-irreps. Combining these two things gives the representation 
of the S=-2 potentials in terms of the SU$_{fs}$(6)-irreps as displayed 
in Tables~\ref{tab.su6.a} and \ref{tab.su6.b}.
 \begin{table}[ht]
\renewcommand{\arraystretch}{2.0}
\begin{center}
\caption{$SU(6)_{fs}$-contents spin-space odd $^1S_0, ^3P, ^1D_2, ...$ 
 potentials on the spin-isospin basis.}  
\begin{ruledtabular}
\begin{tabular}{ccl} 
   & $(S,I)$ & $V = a V_{[51]} + b V_{[33]}$ \\ \hline
$ \Lambda\Lambda \rightarrow \Lambda\Lambda $  & $ (0,0) $
   & $ V_{\Lambda\Lambda,\Lambda\Lambda} = 
       \frac{1}{2}V_{[51]}+\frac{1}{2}V_{[33]}  $ \\
$ \Xi N \rightarrow \Xi N $  & $ (0,0) $
   & $ V_{\Xi N,\Xi N} = \frac{1}{3}V_{[51]}+\frac{2}{3}V_{[33]}  $ \\
$ \Sigma\Sigma \rightarrow \Sigma\Sigma $  & $ (0,0) $
   & $ V_{\Sigma\Sigma,\Sigma\Sigma} = 
       \frac{11}{18}V_{[51]}+\frac{7}{18}V_{[33]}  $ \\
$ \Xi N \rightarrow \Xi N $  & $ (0,1) $
   & $ V_{\Xi N,\Xi N} = \frac{7}{9}V_{[51]}+\frac{2}{9}V_{[33]}  $ \\
$ \Sigma\Lambda\rightarrow \Sigma\Lambda$  & $ (0,1) $
   & $ V_{\Sigma\Lambda,\Sigma\Lambda} = 
       \frac{2}{3}V_{[51]}+\frac{1}{3}V_{[33]}  $ \\
$ \Sigma\Sigma \rightarrow \Sigma\Sigma $  & $ (0,2) $
   & $ V_{\Sigma\Sigma,\Sigma\Sigma} = 
       \frac{4}{9}V_{[51]}+\frac{5}{9}V_{[33]}  $ \\
\end{tabular}
\end{ruledtabular}
\end{center}
\renewcommand{\arraystretch}{1.0}
\label{tab.su6.a} \end{table}
 \begin{table}[ht]
\renewcommand{\arraystretch}{2.0}
\begin{center}
\caption{
$SU(6)_{fs}$-contents of the spin-space even $^3S_1, ^1P_1, ^3D, ... $ 
 potentials on the spin-isospin basis.}   
\begin{ruledtabular}
\begin{tabular}{ccl} 
   & $(S,I)$ & $V = a V_{[51]} + b V_{[33]}$ \\ \hline
$ \Xi N \rightarrow \Xi N $  & $ (1,0) $
   & $ V_{\Xi N,\Xi N} = \frac{5}{9}V_{[51]}+\frac{4}{9}V_{[33]}  $ \\
$ \Xi N \rightarrow \Xi N $  & $ (1,1) $
   & $ V_{\Xi N,\Xi N} = \frac{17}{27}V_{[51]}+\frac{10}{27}V_{[33]}  $ \\
$ \Sigma\Lambda\rightarrow \Sigma\Lambda$  & $ (1,1) $
   & $ V_{\Sigma\Lambda,\Sigma\Lambda} = 
       \frac{2}{3}V_{[51]}+\frac{1}{3}V_{[33]}  $ \\
$ \Sigma\Sigma \rightarrow \Sigma\Sigma $  & $ (1,1) $
   & $ V_{\Sigma\Sigma,\Sigma\Sigma} = 
       \frac{16}{27}V_{[51]}+\frac{11}{27}V_{[33]}  $ \\
\end{tabular}
\end{ruledtabular}
\end{center}
\renewcommand{\arraystretch}{1.0}
\label{tab.su6.b} \end{table}
\subsection{ Parametrization Quark-core effects}                      
\label{sec:5b} 
As introduced in II, 
the repulsive short-range Pomeron-like YN,YY potential is splitted 
linearly in a diffractive (Pomeron) and a quark-core component by writing
\begin{eqnarray}
 V_{PBB} &=& V_{BB}(POM)+V_{BB}(PB)
\label{eq:51.1}\end{eqnarray}
where $V_{BB}(POM)$ represents the genuine Pomeron and $V_{BB}(PB)$ 
the structural effects of the quark-core forbidden [51]-configuration, 
i.e. a Pauli-blocking (PB) effect. 
Since the Pomeron is a unitary-singlet its contribution is the same
for all BB-channels (apart from some small baryon mass breaking effects), 
i.e. $V_{BB}(POM)=V_{NN}(POM)$.
Furthermore the PB-effect for the BB-channels is assumed
to be proportional to the relative weight of the forbidden
 [51]-configuration compared to its weight in NN 
\begin{equation}
 V_{BB}(PB) = a_{PB}\ \left(w_{BB}[51]/w_{NN}[51]\right)\cdot V_{NN}(PB)
\label{eq:51.3}\end{equation}
where $a_{PB}$ denotes the quark-core fraction w.r.t. the pomeron 
potential for the NN-channel, i.e. $V_{NN}(PB)= a_{PB}\ V_{PNN}$.
Then we have
\begin{equation}
 V_{PBB} = (1-a_{PB}) V_{PNN} + a_{PB} 
 \left(\frac{w_{BB}[51]}{w_{NN}[51]}\right)\cdot V_{PNN}
\label{eq:51.4}\end{equation}
 \begin{table}
\renewcommand{\arraystretch}{2.0}
\begin{center}
\caption {Effective Pomeron+PB contribution\\
  \hspace*{2.5cm} on the spin,isospin basis. }
\begin{tabular}{cccc} \hline \hline
   & $(S,I)$ & $V_{PBB}/V_{PNN}$ & $ESC08c$ \\ \hline
$ NN \rightarrow NN $  & $ (0,1) $
                & $ 1 $ & $ 1.000 $\\
$ NN \rightarrow NN $  & $ (1,0) $
                & $ 1 $ & $ 1.000 $ \\
\hline 
$ \Lambda \Lambda \rightarrow \Lambda \Lambda $  & $(0,0)$ &
                   $ 1+\frac{1}{8} a_{PB}$ & $ 1.034 $\\
$ \Xi N \rightarrow \Xi N $  & $(0,0)$ &
                   $ 1-\frac{1}{4} a_{PB}$ & $ 0.931 $ \\
$ \Sigma \Sigma \rightarrow \Sigma \Sigma $  & $(0,0)$ &
                   $ 1+\frac{3}{8} a_{PB}$ & $ 1.103 $\\
$ \Xi N \rightarrow \Xi N $  & $(0,1)$ &
                   $ 1+\frac{3}{4} a_{PB}$ & $ 1.206$  \\
$ \Sigma \Lambda \rightarrow \Sigma \Lambda $  & $(0,1)$ &
                   $ 1+\frac{1}{2} a_{PB}$ & $ 1.138 $\\
$ \Sigma \Sigma \rightarrow \Sigma \Sigma $  & $(0,2)$ &
                   $ 1 $ & $ 1.000 $\\
\hline 
$ \Xi N \rightarrow \Xi N $  & $(1,0)$ &
                   $ 1+\frac{1}{4} a_{PB}$ & $ 1.069 $ \\
$ \Xi N \rightarrow \Xi N $  & $(1,1)$ &
                   $ 1+\frac{5}{12} a_{PB}$ & $ 1.115 $ \\
$ \Sigma \Lambda \rightarrow \Sigma \Lambda $  & $(1,1)$ &
                   $ 1+\frac{1}{2} a_{PB}$ & $ 1.138 $\\
$ \Sigma \Sigma \rightarrow \Sigma \Sigma $  & $(1,1)$ &
                   $ 1+\frac{1}{3} a_{PB}$ & $ 1.092 $\\
\hline \hline
\end{tabular}
\end{center}
\renewcommand{\arraystretch}{1.0}
\label{tab.su6.c} \end{table}

\noindent A subtle treatment of $all$ BB channels according to this linear 
scheme is characteristic for the ESC08c-model. 
The value of the PB factor $a_{PB}$ is searched in the fit 
to the NN- and YN-data.  The parameter $a_{PB}$ turns out 
to be about 27.5\%. This means that the Quark-core repulsion is
roughly 34\% of the genuine Pomeron repulsion. 
Then, the PB effects in the S=-2 channels are entirely determined.
From Eqn.~(\ref{eq:51.4}) the ratio $V_{PBB}/V_{PNN}$ is given by the 
weights of the [51]-irrep and $a_{PB}$. 
In Table \ref{tab.su6.c} we give this ratio for the various S=-2 BB channels
in the ESC08c model, With only one exception, the effective pomeron 
repulsion is stronger than in the NN-channels.


\section{Multi-channel Thresholds and Potentials}
\label{sec:thresh}
\subsection{Thresholds}
Clearly, the $S=-2$ two-baryon channels represent a number of separate 
coupled-channel systems, separated by the charge, see (\ref{eq:2.2}).
A further subdivision is according to the total isospin. 
The different thresholds have been discussed in detail in \cite{SR99}, and 
we show these thresholds here in Fig.~\ref{thresh1} for  the purpose of 
general orientation. Their presence turns the
Lippmann-Schwinger and Schr\"odinger equation into a coupled-channel
matrix equation, where the different channels open up at different energies.
In general one has a combination of 'open' and 'closed' channels. For a 
discussion of the solution of such a mixed system, we refer to \cite{Swa71}.

\subsection{Threshold- and Meson-mass corrections in Potentials}
As discussed in \cite{SR99}, the one-meson-exchange Feynman-graph consists
actually of two three-dimensional time-ordered graphs. 
The energy denominator from these two diagrams reads
\begin{equation}
   D(\omega)=\frac{1}{2\omega}\left[\frac{1}{E_2+E_3-W+\omega}
             +\frac{1}{E_1+E_4-W+\omega}\right],
\label{4.1} \end{equation}
where, $W=\sqrt{s}$ is the total energy and $\omega^2={\bf k}^2+m^2$,
with $m$ the meson mass and ${\bf k}={\bf p}'-{\bf p}$ the momentum
transfer. 
From (\ref{4.1}) it is clear that the potential is energy dependent.
We use the static approximation $E_i\rightarrow M_i$ and
$W\rightarrow M^0_1+M^0_2$, where the superscript 0 refers to the masses of the 
lowest threshold of the particular coupled-channel system q, see (\ref{eq:2.2}). 
They are in general not 
equal to the masses $M_1$ and $M_2$ occurring in the time-ordered
diagrams. For example, the potential for the $\Sigma\Sigma$
contribution in the coupled-channel $\Lambda\Lambda$ system has
$M_1=M_2=M_{\Sigma}$, but $M^0_1=M^0_2=M_{\Lambda}$.
Denoting $a \equiv E_2+E_3-W \approx M_2+M_3-M^0_1-M^0_2>0$, and similarly
for $E_1+E_4-W$, we have for the 'propagators' \cite{Rij92} for $a>0$
\begin{equation}
  \frac{1}{\omega(\omega+a)} =\frac{2}{\pi}\int_0^{\infty}
          \frac{ad\lambda}{(a^{2}+\lambda^2)(\omega^2+\lambda^2)}\ .
\label{4.2} \end{equation}
For $a<0$ there is the extra term 
$+2\theta(-a)/(\omega^2-a^2)$ on the r.h.s. in (\ref{4.2}). 
This integral representation makes it possible to deal with it numerically
rather exactly. However, we think that such a sophistication is unnecessary 
at present nor for a description of the $S=-1$ scattering data, nor for $S=-2$.
where there are virtually no data at all.
Therefore, we handle with this energy dependence approximately as follows:\\

\noindent 1.\ \underline{Elastic potentials}: In this case we use (\ref{4.2}), and 
in (\ref{4.1}) one has $E_1=E_3 \approx M_i$ and $E_2=E_4 \approx M'_i$, for
the elastic channel, label $i$. Here $a \approx M_i+M'_i -M_1^0-M_2^0 \geq 0$. 
Then,  
\begin{eqnarray}
   D_i(\omega)&=&\frac{1}{\omega^2} + \Delta_i(\omega,a),\ \  
   \Delta_i(\omega,a) = \frac{2}{\pi}\int_0^{\infty}
   \frac{a d\lambda}{a^2+\lambda^2}\cdot \nonumber \\ && \times
 \left[ \frac{1}{\omega^2}-\frac{1}{\omega^{2}+\lambda^2}\right]\ ,
\label{4.3} \end{eqnarray}
for $0<a<m$. 
Because of this condition we can not apply this to the pseudoscalars,
but is possible for the vector-, scalar-, and axial-mesons. 
The largest effect is for $\Lambda\Lambda$-scattering, 
where the $\Sigma\Sigma$-channel potential is somewhat reduced by this effect. 
This because the $\Sigma\Sigma$-channel is rather far away from the others. 
In this paper we neglect the effects of a finite $a$ 
in all elastic channels and for all mesons.\\                  

\noindent 2.\ \underline{Inelastic potentials}: In this case, like in \cite{SR99} and all
other papers on the Nijmegen potentials, we use the approximation of \cite{Swa62},
using the fact that $M_1^0+M_2^0$ is mostly rather close to the average of the
initial and final-sate baryon masses. Then, the propagator can be written as
\begin{equation}
   D(\omega)\rightarrow \frac{1}{\omega^2-{\textstyle\frac{1}{4}}
            (M_3-M_4+M_2-M_1)^2},      
\label{4.4} \end{equation}
which amounts to introducing an effective meson mass $\overline{m}$
\begin{equation}
    m^2 \rightarrow \overline{m}^2=m^2-{\textstyle\frac{1}{4}}
                 (M_3-M_4+M_2-M_1)^2.
\label{4.5} \end{equation}
For more details of this effect on the exchanged meson masses, we refer to 
\cite{SR99}.

\begin{table}
\caption{Baryon masses in MeV/$c^2$.}
\begin{ruledtabular}
\begin{tabular}{lcr}
  Baryon & & Mass \\
\hline      
  Nucleon & $p$            &  938.2796  \\
          & $n$            &  939.5731  \\
  Hyperon & $\Lambda$      & 1115.60    \\
          & $\Sigma^+$     & 1189.37    \\
          & $\Sigma^0$     & 1192.46    \\
          & $\Sigma^-$     & 1197.436   \\
  Cascade & $\Xi^0$        & 1314.90    \\
          & $\Xi^-$        & 1321.32    \\
          &                &            \\
\end{tabular}
\end{ruledtabular}
\label{tabBmass}
\end{table}
The used baryon masses are about the same as in \cite{SR99}, and are given in
Table~\ref{tabBmass}.
The used meson masses are the same as in paper II \cite{NRY12b}, as well as the 
cut-off mases.

\section{Results}
\label{sec:results}
The main purpose of this paper is to present the properties of the
ESC08c potentials for the $S=-2$ sector. 
As described above, the free parameters in each model are fitted mainly to the $N\!N$ and $Y\!N$
scattering data for the $S=0$ and $S=-1$ sectors, respectively.
Given the expressions for the coupling constants in terms of the octet
and singlet parameters and their values for the six different models
as presented in Ref.~\cite{RSY99}, it is straightforward to evaluate
all possible baryon-baryon-meson coupling constants needed for the
$S\leq-2$ potentials. A complete set of coupling constants for models
ESC08c is given in Table~\ref{tab.cop08a}.

\begin{figure}   
  \resizebox{3.5cm}{!}
 {\includegraphics[200, 50][400,750]{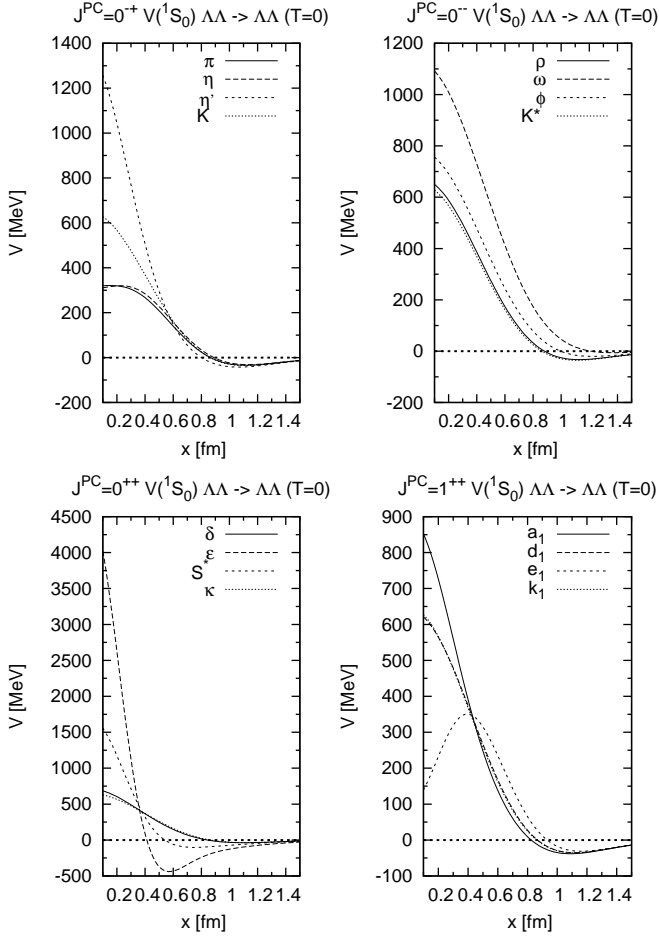}}
\caption{ESC08c: OBE contributions to the $\Lambda\Lambda(^1S_0, I=0)$ potentials                 
 for the PS, V, S, and A meson nonets.}
\label{obefig1}
 \end{figure}

\begin{figure}   
  \resizebox{3.5cm}{!}
 {\includegraphics[200, 50][400,750]{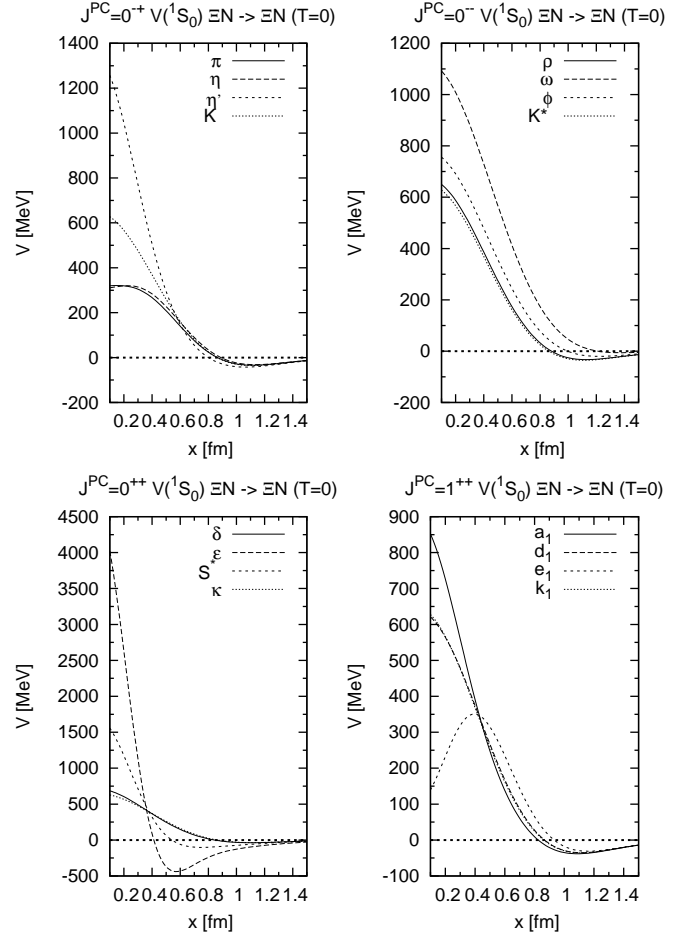}}
\caption{ESC08c: OBE contributions to the $\Xi N(^1S_0, I=0)$ potentials    
 for the PS, V, S, and A meson nonets.}
 \label{obefig2}
 \end{figure}

In Fig's~\ref{obefig1} and Fig.~\ref{obefig2} we display the OBE potentials
for the individual pseudoscalar, vector, scalar, and axial mesons in the
case of model ESC08c.

\onecolumngrid
\begin{center}
\begin{table}[hbt]
\caption{Coupling constants for model ESC08c, divided by
         $\protect\sqrt{4\pi}$. $M$ refers to the meson.
         The coupling constants are listed in the order pseudoscalar,
         vector ($g$ and $f$), axial vector A ($g$ and $f$), scalar, axial vector B, and diffractive.}
\label{tab.cop08a}
\begin{tabular}{ccrrrrcrrrr} \hline\hline
      & $M$ & $N\!N\!M$ & $\Sigma\Sigma M$ & $\Sigma\Lambda M$ & $\Xi\Xi M$
 & $M$ & $\Lambda N\!M$ & $\Lambda\Xi M$ & $\Sigma N\!M$ & $\Sigma\Xi M$ \\
\hline 
 $f$ & $\pi$    &   0.2687  &   0.1961  &   0.1970  & --0.0725 
     & $K$      & --0.2683  &   0.0714   &   0.0725  & --0.2687  \\
 $g$ & $\rho$   &   0.6446  &   1.2892  &   0.0000  &   0.6446  
     & $K^*$    & --1.1165  &   1.1165   & --0.6446  & --0.6446   \\
 $f$ &          &   3.7743  &   3.5639  &   2.3006  & --0.2104  
     &          & --4.2367  &   1.9362   &   0.2104  & --3.7743   \\
 $g$ & $a_1$    & --0.7895  & --0.4929  & --0.6271  &   0.2967  
     & $K_{1A}$ &   0.7404  & --0.1133   & --0.2967  &   0.7895   \\
 $f$ &          & --0.8192  & --0.5114  & --0.6507  &   0.3078  
     &          &   0.7683  & --0.1175   & --0.3078  &   0.8192   \\
 $g$ & $a_0$    &   0.5852  &   1.1705  &   0.0000  &   0.5852  
     & $\kappa$ & --1.0137  &   1.0137   & --0.5852  & --0.5852   \\
 $f$ & $b_1$    & --1.3743  & --1.0991  & --0.9523  &   0.2746  
     & $K_{1B}$ &   1.4280  & --0.4758   & --0.2746  &   1.3743   \\
 $g$ & $a_2$    &   0.00000 &   0.00000 &   0.00000 &   0.00000
     & $K^{**}$ &   0.00000 &   0.00000 &   0.00000 &   0.00000 \\[2mm]
      & $M$ & $N\!N\!M$ & $\Lambda\Lambda M$ & $\Sigma\Sigma M$ & $\Xi\Xi M$
 & $M$ & $N\!N\!M$ & $\Lambda\Lambda M$ & $\Sigma\Sigma M$ & $\Xi\Xi M$ \\
 $f$ & $\eta$        &   0.1265  & --0.1349  &   0.2490  & --0.2045  
     & $\eta'$       &   0.2309  &   0.2912  &   0.2026  &   0.3073   \\
 $g$ & $\omega$      &   3.4570  &   2.7589  &   2.7589  &   2.0608  
     & $\phi$        & --1.3390  & --2.2103  & --2.2103  & --3.0816   \\
 $f$ &               & --0.8574  & --3.5064  & --0.6296  & --4.7170  
     &               &   3.1678  & --0.1386  &   3.4522  & --1.6497   \\
 $g$ & $f_1$         & --0.7613  & --0.1942  & --1.1549  & --0.1074  
     & $f'_1$        &   0.7311  &   1.2070  &   0.4008  &   1.2798   \\
 $f$ &               & --0.4467  &   0.1418  & --0.8551  &   0.2319  
     &               &   0.3495  &   0.8433  &   0.4008  &   1.2798   \\
 $g$ & $\varepsilon$ &   4.1461  &   3.5609  &   3.5609  &   2.9758  
     & $f_0$         & --1.6898  & --2.5176  & --2.5176  & --3.3453   \\
 $f$ & $h_1$         & --0.2277  &   0.5968  & --0.5030  &   0.8714  
     & $h'_1$        & --0.4221  &   0.7443  & --0.8109  &   1.1331   \\
 $g$ & $P$           &   3.5815  &   3.5815  &   3.5815  &   3.5815 
     & $f_2$         &   0.0000  &   0.0000  &   0.0000  &   0.0000  \\
 $g$ & $O$           &   4.6362  &   4.6362  &   4.6362  &   4.6362    
     &               &           &           &           &           \\
 $f$ &               & --4.7602  & --4.7602  & --4.7602  & --4.7602  
     &               &           &           &           &             
 \\ \hline
\end{tabular}
\end{table}
\end{center}
\twocolumngrid

In the following we will present the model predictions for scattering
lengths, bound states, and cross sections. 

 \begin{figure}[ht]
 \includegraphics*[width= 8cm,height=10cm]{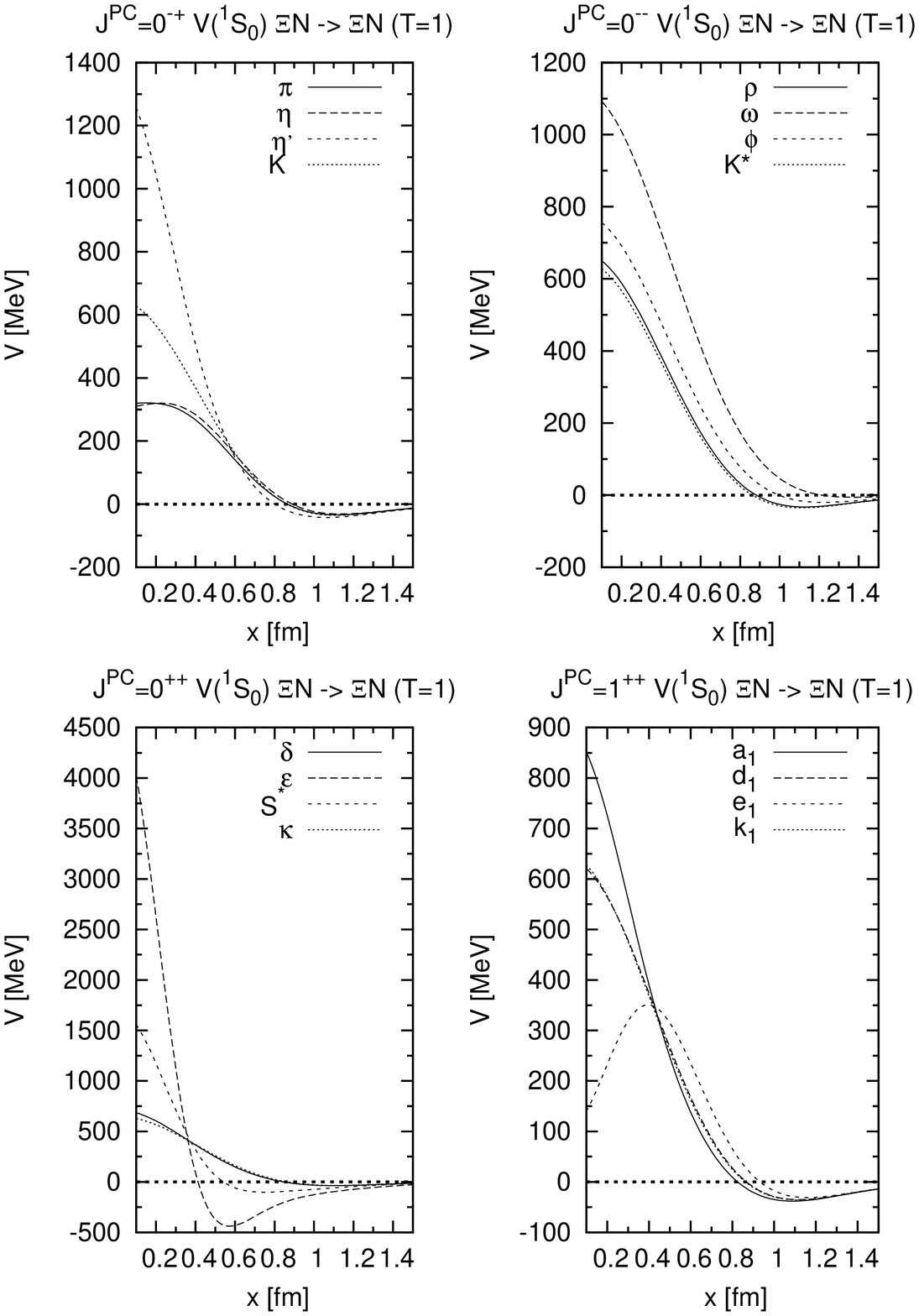} 
\caption{ $\Xi N(^1S_0,I=1)$ potentials}                          
 \label{obefig3}
 \end{figure}
 \begin{figure}[ht]
 \includegraphics*[width= 8cm,height=10cm]{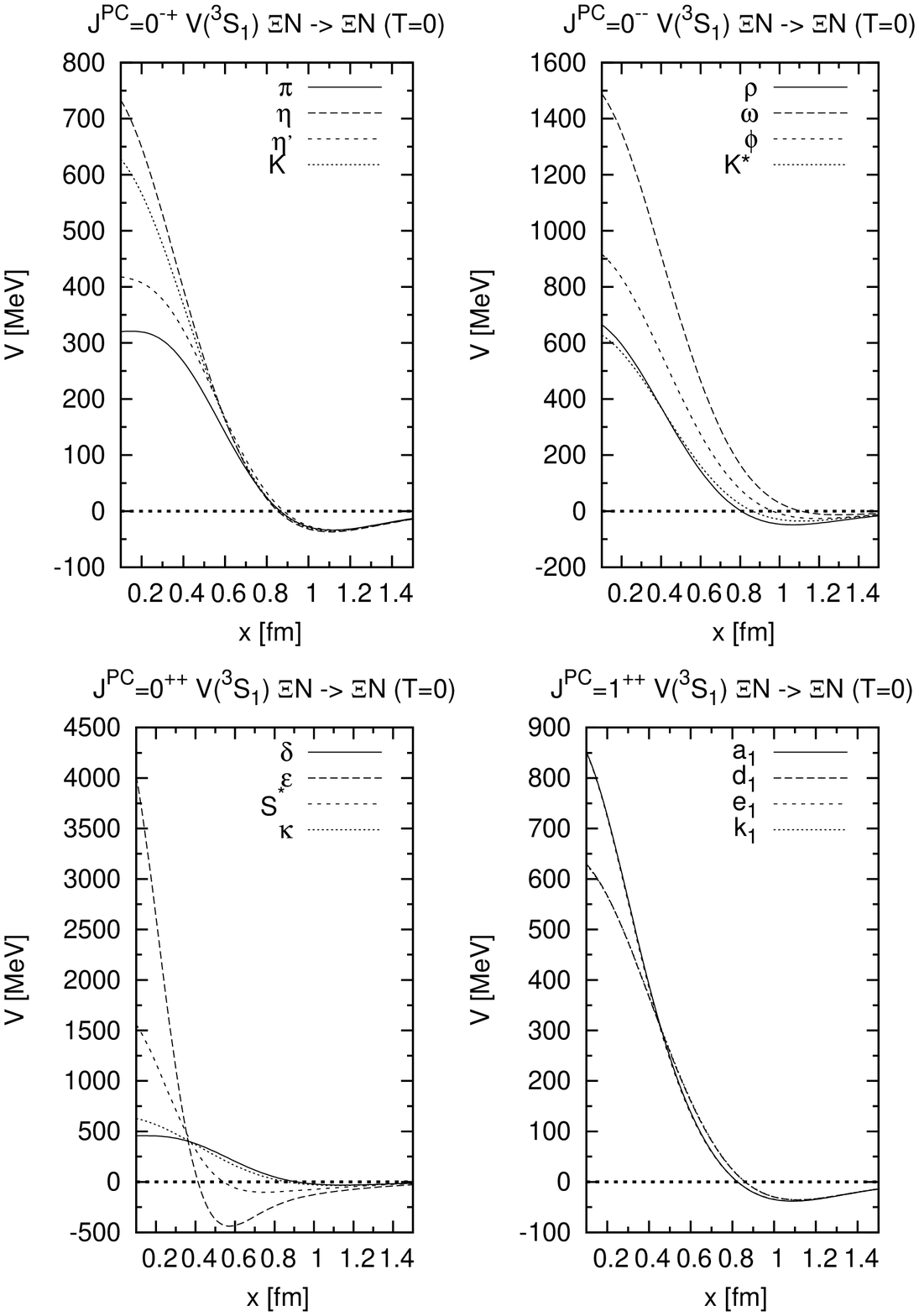} 
\caption{ $\Xi N(^3S_1,I=0)$ potentials}                          
 \label{obefig4}
 \end{figure}
 \begin{figure}[ht]
 \includegraphics*[width= 8cm,height=10cm]{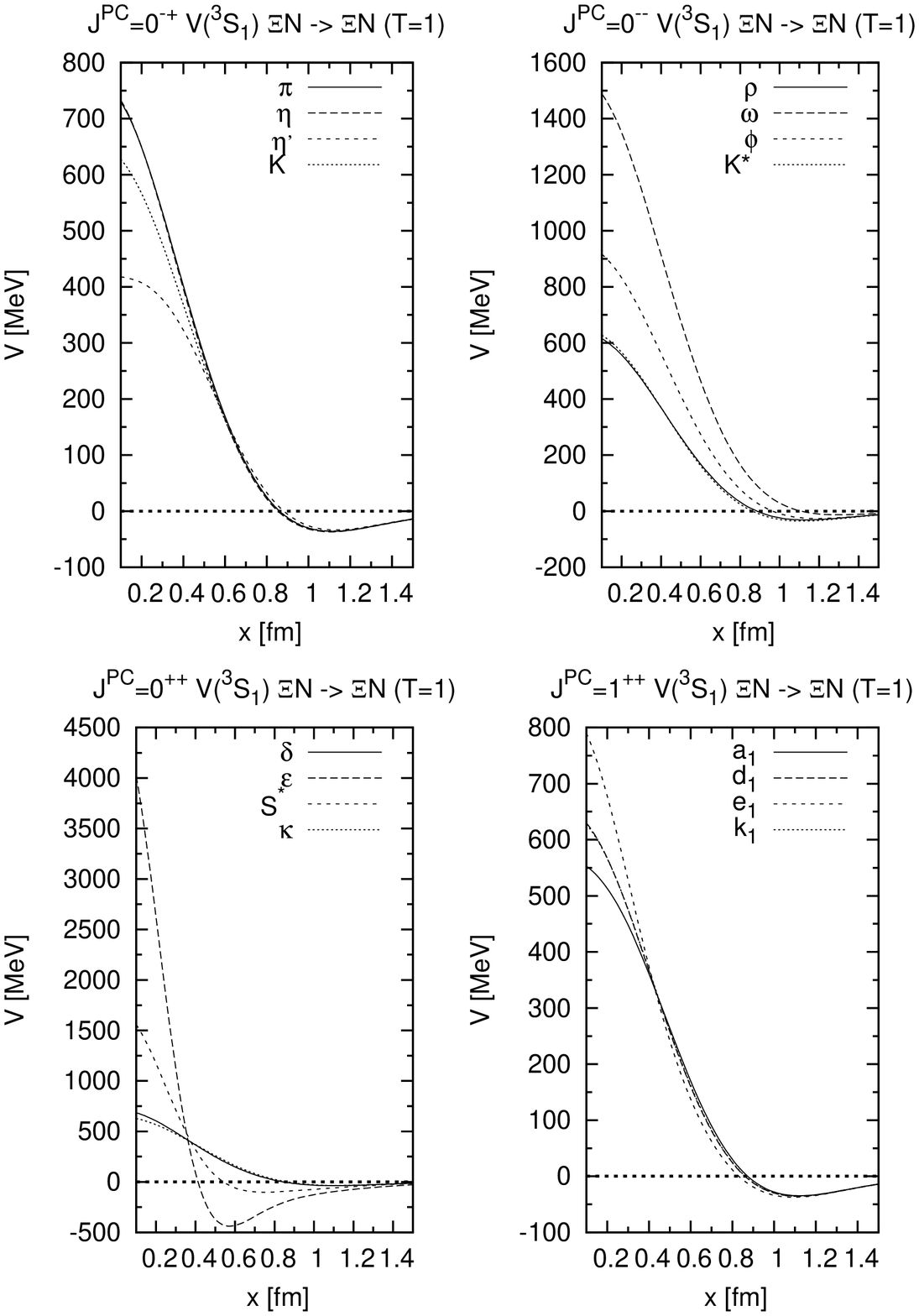} 
\caption{ $\Xi N(^3S_1,I=1)$ potentials}                          
 \label{obefig5}
 \end{figure}
 \begin{figure}[ht]
 \includegraphics*[width= 8cm,height=10cm]{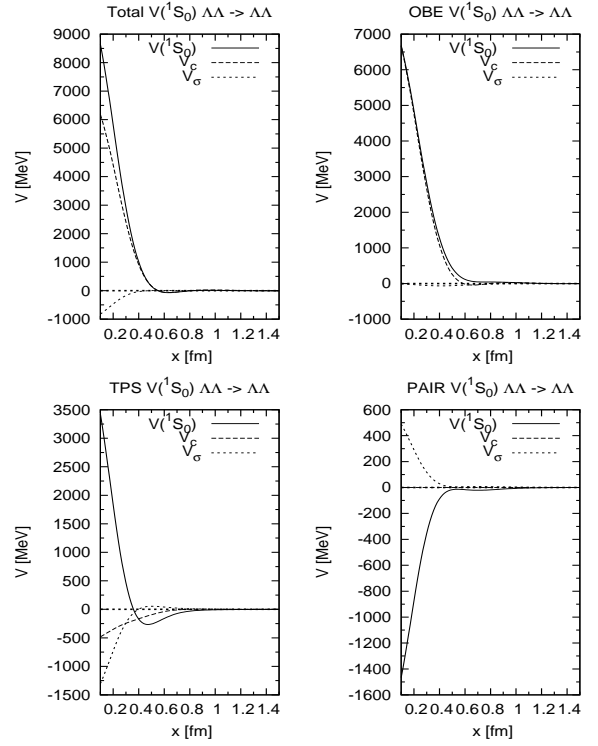} 
\caption{ $\Lambda\Lambda(^1S_0,I=0)$ potentials}                          
\label{pots10.a}
 \end{figure}
 \begin{figure}[ht]
 \includegraphics*[width= 8cm,height=10cm]{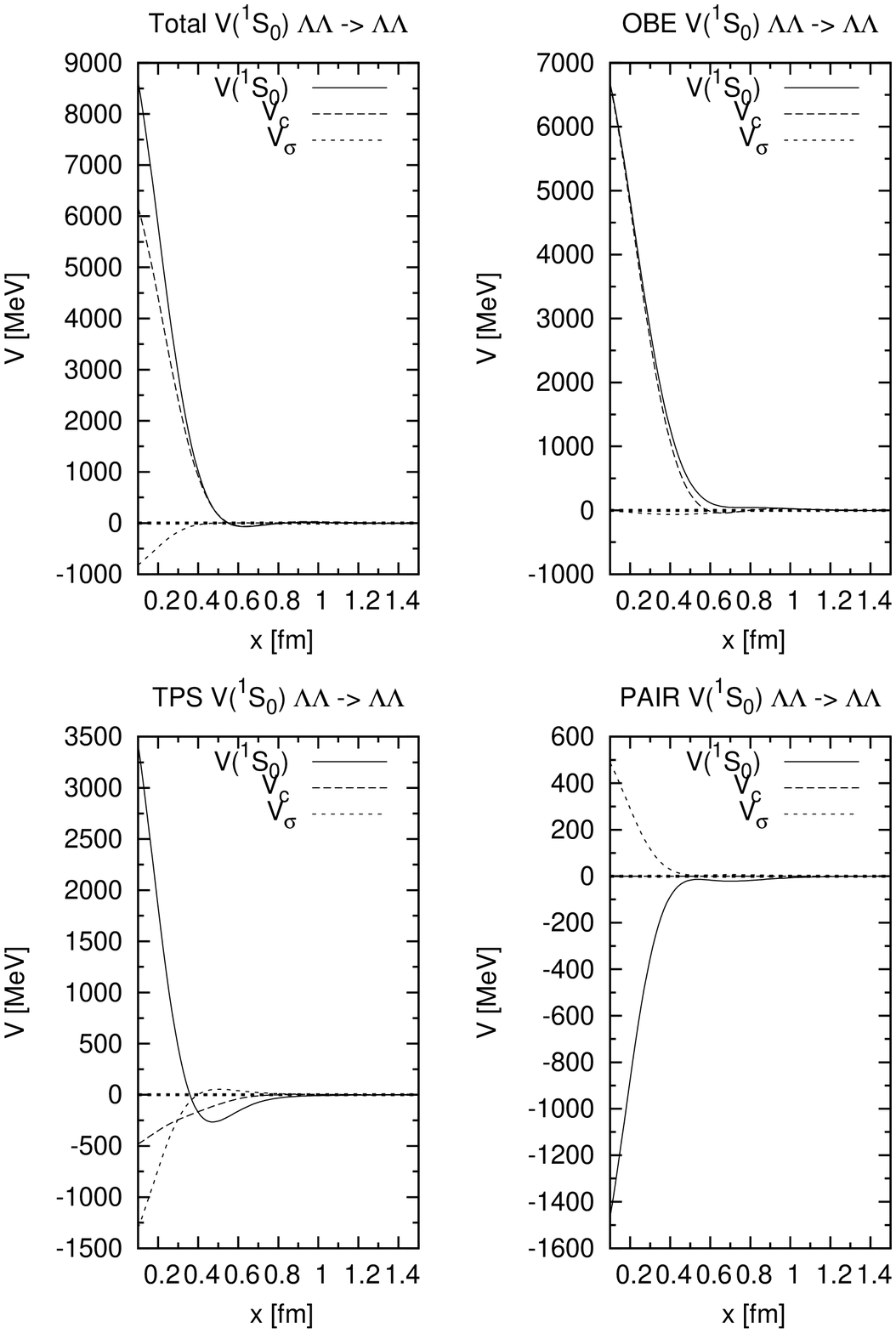} 
\caption{ $\Xi N(^1S_0,I=0)$ potentials}                          
\label{pots10.b}
 \end{figure}
 \begin{figure}[ht]
 \includegraphics*[width= 8cm,height=10cm]{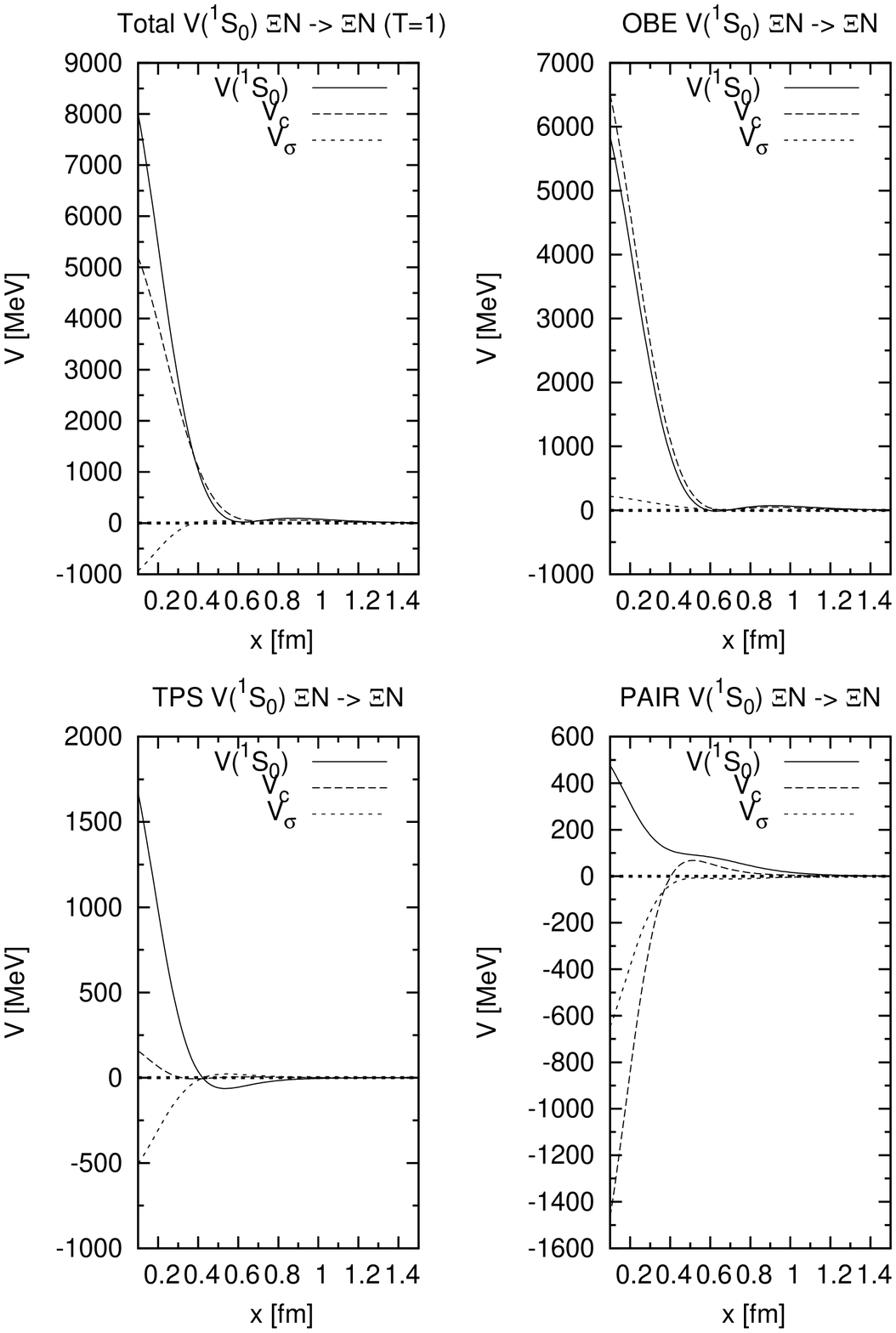} 
\caption{ $\Xi N(^1S_0,I=1)$ potentials}                          
\label{pots10.c}
 \end{figure}
 \begin{figure}[ht]
 \includegraphics*[width= 8cm,height=10cm]{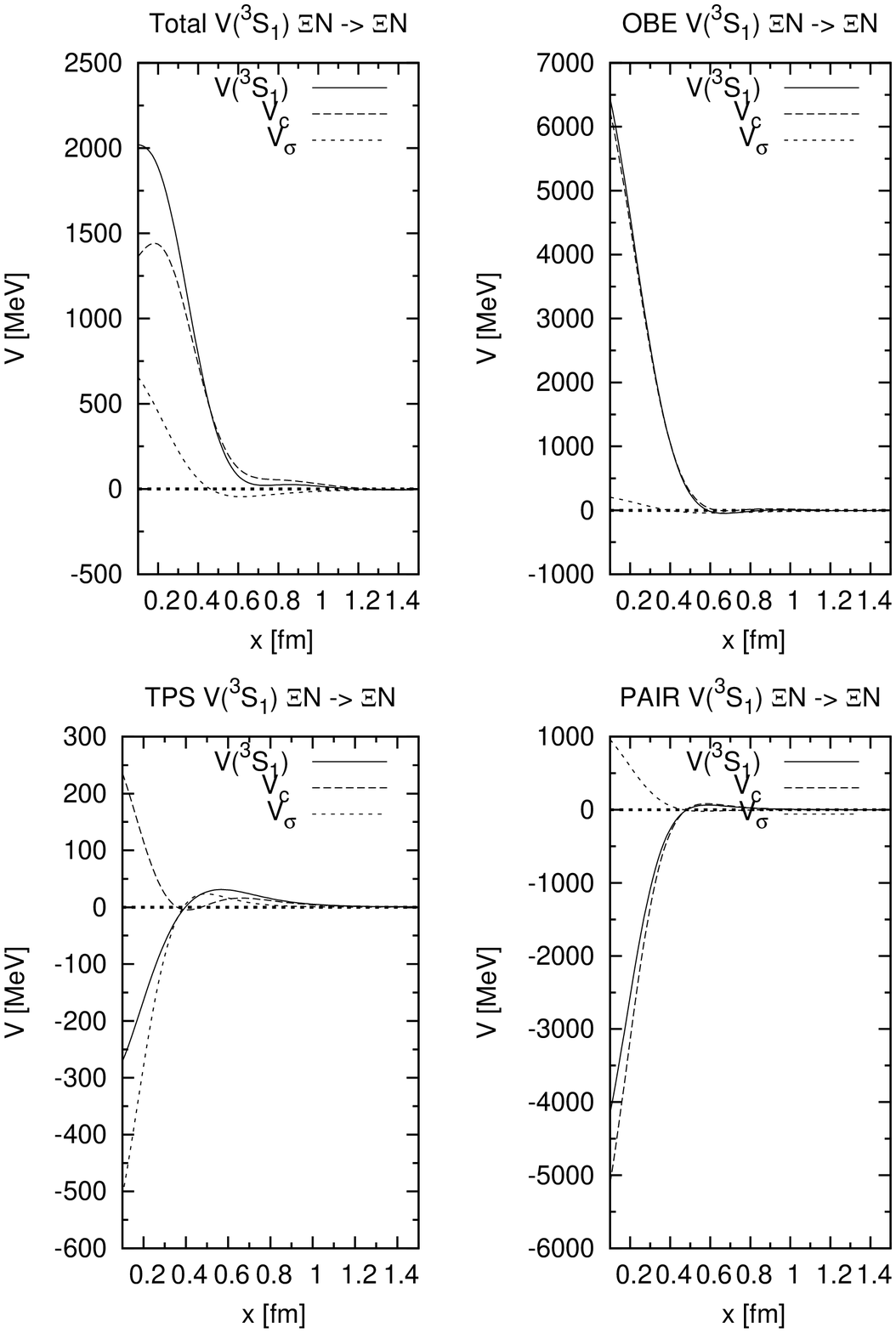} 
\caption{ $\Xi N(^3S_1,I=0)$ potentials}                          
\label{pots10.d}
 \end{figure}
 \begin{figure}[ht]
 \includegraphics*[width= 8cm,height=10cm]{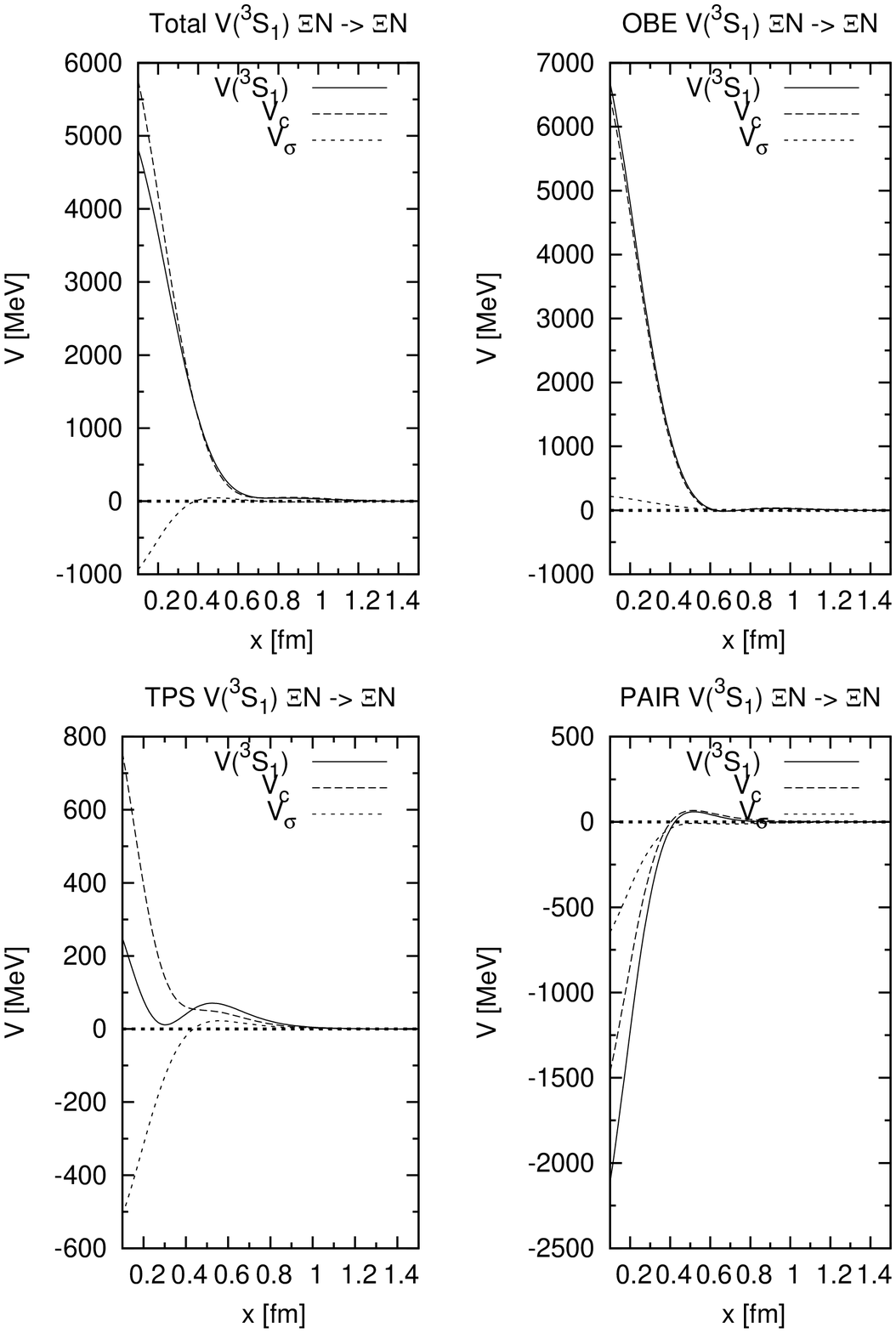} 
\caption{ $\Xi N(^3S_1,I=1)$ potentials}                          
\label{pots10.e}
 \end{figure}

 \begin{figure}   
 \begin{center}
 \resizebox{9.5cm}{11.25cm} 
 {\includegraphics[000,175][400,625]{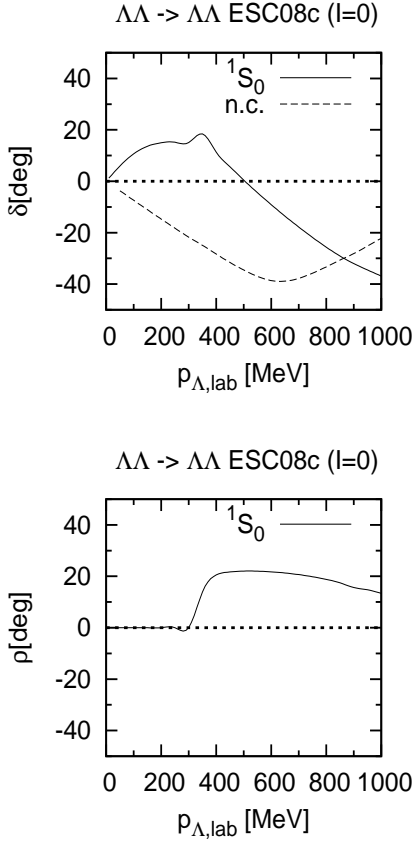}}
\caption{ESC08c $(^1S_0, I=0)$ $\Lambda\Lambda$-phases.
 The dashed curve n.c. is the case with no coupling to 
 the $\Xi N, \Sigma\Sigma$ channels.  }  
 \label{phasll0} 
 \end{center}
 \end{figure}

 \begin{figure}   
 \resizebox{7.0cm}{11.25cm} 
 {\includegraphics[100,175][500,625]{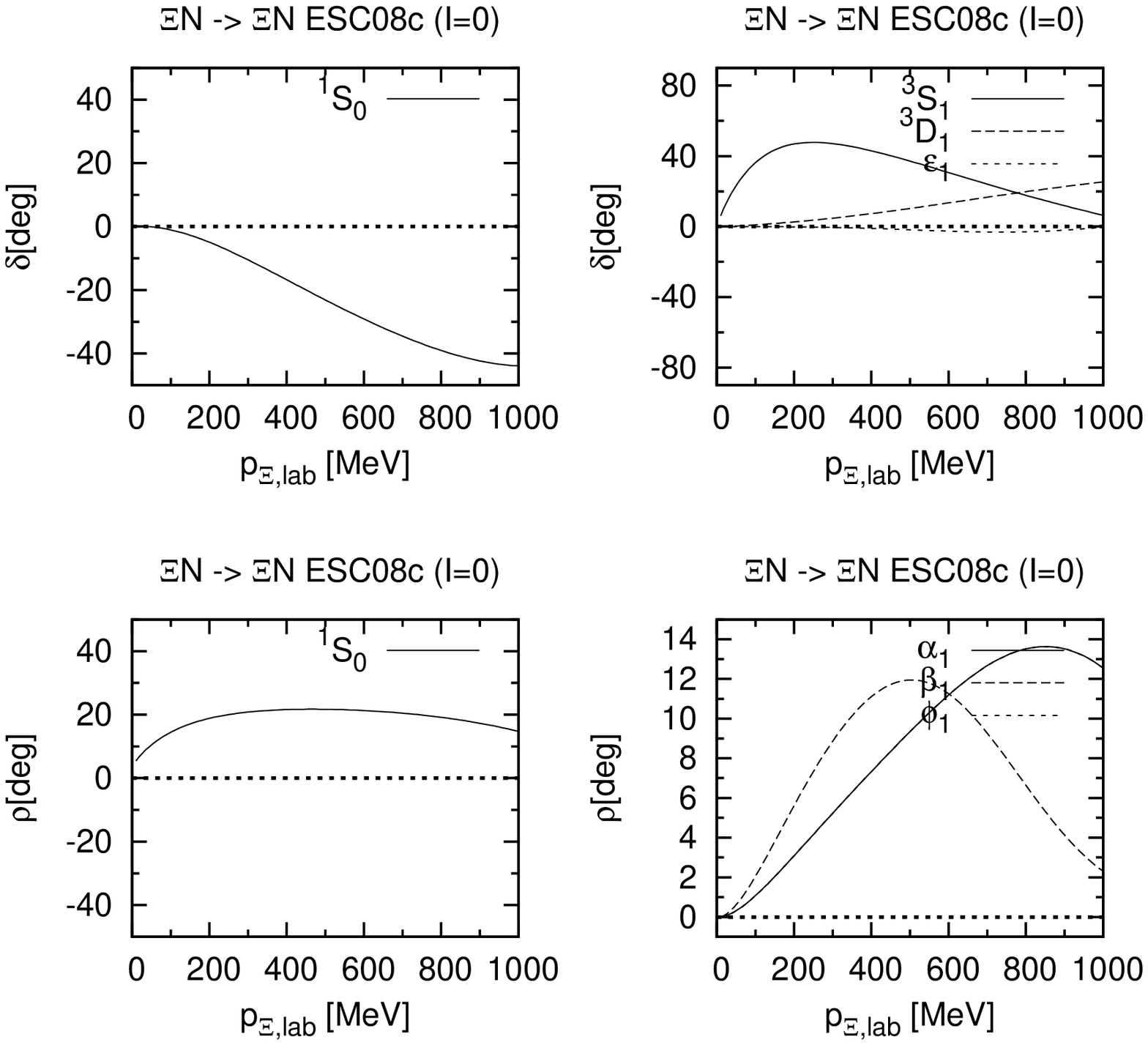}}
\caption{ESC08c $I=0$ $\Xi N$-phases.}  
 \label{phasxn0} 
 \end{figure}

 \begin{figure}   
 \resizebox{7.0cm}{11.25cm} 
 {\includegraphics[100,175][500,625]{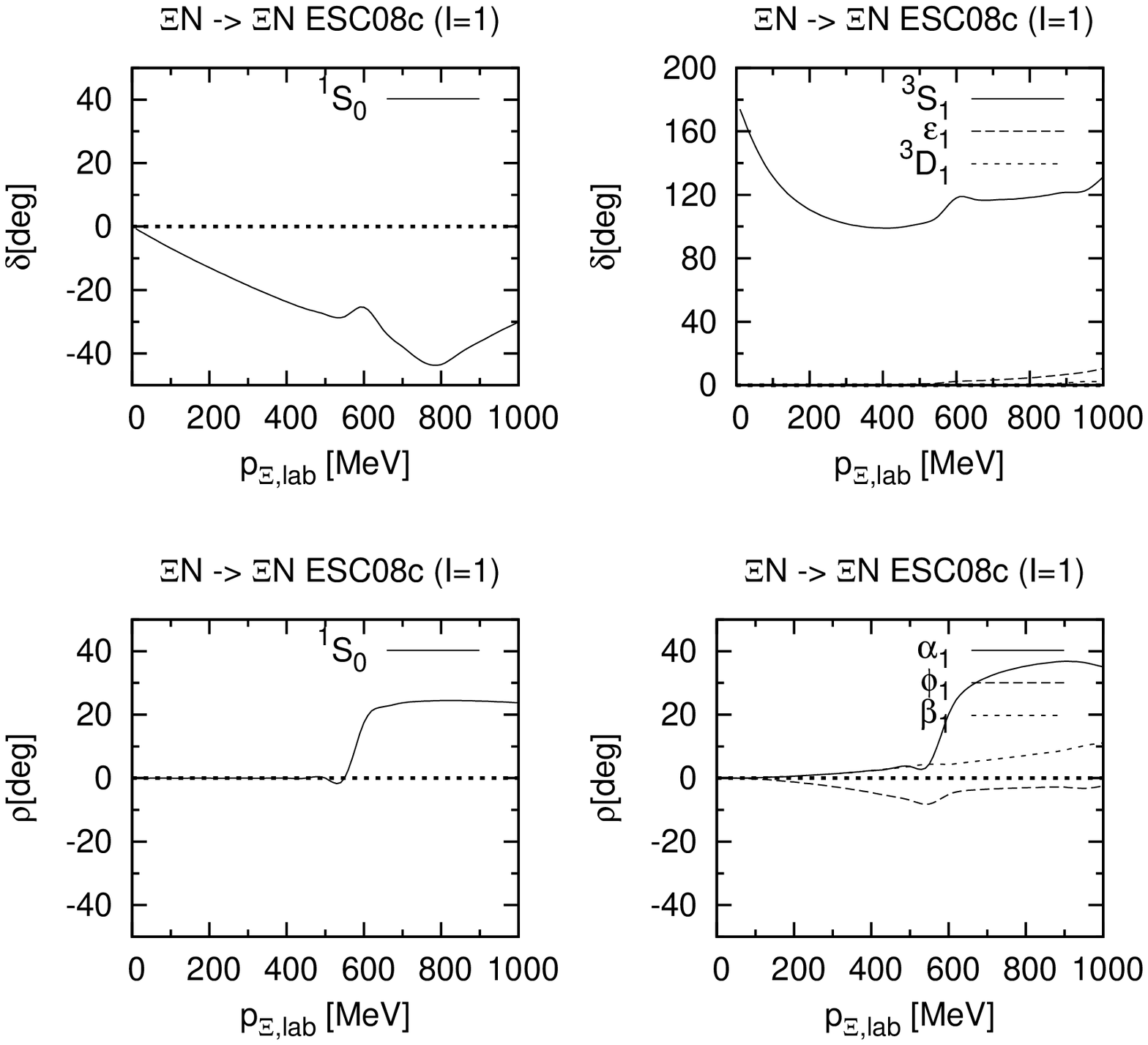}}
\caption{ESC08c $I=1$ $\Xi N$-phases.}
 \label{phasxn1} 
 \end{figure}

 \begin{figure}   
 \resizebox{7.0cm}{11.25cm} 
 {\includegraphics[100,175][500,625]{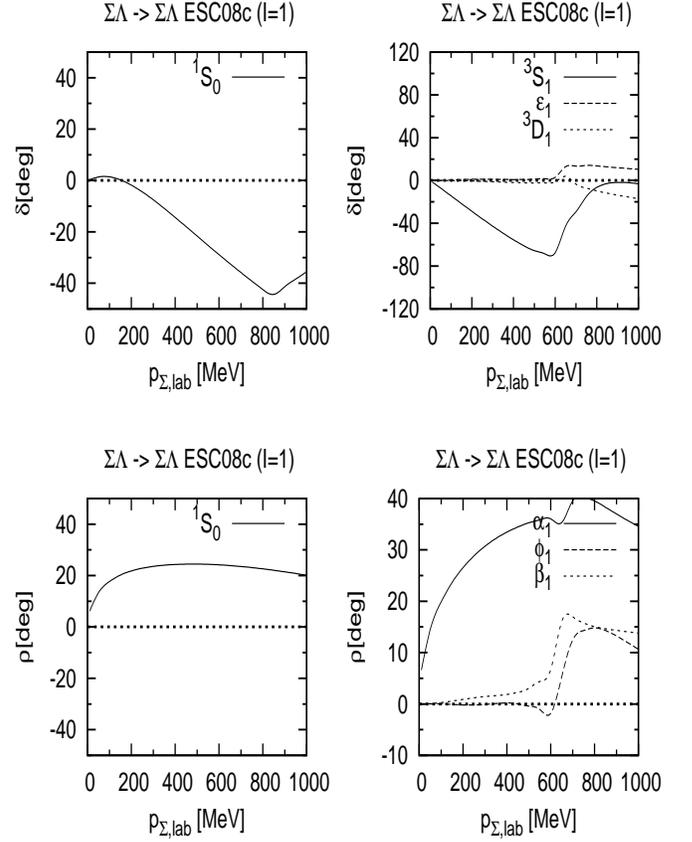}}
\caption{ESC08c $I=1$ $\Sigma\Lambda$-phases.}  
 \label{phassl1} 
 \end{figure}


\subsection{Effective-range parameters}
For ESC08c the $I=0$ low-energy parameters are
\begin{eqnarray*}
 a_{\Lambda\Lambda}(^1S_0) &=& -0.853\ [fm]\ ,\ \ 
 r_{\Lambda\Lambda}(^1S_0)  =   5.126\ [fm]\ .    
\label{le.1} \end{eqnarray*}
\begin{eqnarray*}
 a_{\Xi N}(^3S_1) &=& -5.357\ [fm]\ ,\ \ 
 r_{\Xi N}(^3S_1)  =   1.434\ [fm]\ .    
\label{le.2} \end{eqnarray*}

\begin{table}
\caption{ ESC08c: Inverse-scattering-length and effective-range matrices at (i) the $\Xi N$  
         threshold for I=0, and (ii) the $\Lambda\Sigma$ threshold for I=1.
         The order of the states (1-2) reads $\Lambda\Lambda(^1S_0),
         \Xi N(^1S_0)$, and $\Xi N(^1S_0), \Lambda \Sigma(^1S_0)$ for respectively I=0 and I=1.
         The dimension of the matrix elements are in [fm]$^{-1}(A^{-1})$
         and [fm]$(R)$.}
\begin{ruledtabular}
\begin{tabular}{r|rr|rr}
  & \multicolumn{2}{c|}{ $\Xi N$-threshold } & \multicolumn{2}{c}{ $\Lambda\Sigma$-threshold }\\ 
  & \multicolumn{1}{c}{$A^{-1}$}&\multicolumn{1}{c|}{$R$} 
  & \multicolumn{1}{c}{$A^{-1}$}&\multicolumn{1}{c}{$R$}\\
\colrule \colrule
 11  &  0.472  &13.001  &  0.062  & 11.774 \\
 12  &  1.591  & 2.088  &--1.436  &  9.744 \\
 22  &  0.870  & 3.276  &--0.736  &  9.659 \\
\end{tabular}
\end{ruledtabular}
\label{tabscat-2a} 
\end{table}
\begin{table}
\caption{ $I=1$: Inverse-scattering-length and effective-range matrices at the $\Lambda\Sigma$  
         threshold. The order of the states (1-2) reads $\Xi N(^3S_1),
         \Xi N(^3D_1),\Lambda\Sigma(^3S_1)$. 
         The dimension of the matrix elements are in [fm]$^{-1-l-l'}(A^{-1})$
         and [fm]$^{1-l-l'}(R)$.}
\begin{ruledtabular}
\begin{tabular}{r|rr}
  & \multicolumn{2}{c}{ ESC08c }\\ 
  & \multicolumn{1}{c}{$A^{-1}$}&\multicolumn{1}{c}{$R$}\\
\colrule \colrule
 11  &   1.302 & 1.454   \\    
 12  & --9.122 &  18.305   \\    
 13  &   0.504 &  1.709    \\    
 22  & 239.128 &--590.173 \\    
 23  &   4.252 &--16.637  \\    
 33  &   1.030 &  1.540  \\    
\end{tabular}
\end{ruledtabular}
\label{tabscat-2c} 
\end{table}
For $I=1$ we have for ESC08c:
\begin{eqnarray*}
 a_{\Xi N}(^1S_0) &=&  0.579\ [fm]\ ,\ \ 
 r_{\Xi N}(^1S_0)  =  -2.521\ [fm]\ ,    
\label{le.5} \end{eqnarray*}
\begin{eqnarray*}
 a_{\Xi N}(^3S_1) &=&  4.911\ [fm]\ ,\ \ 
 r_{\Xi N}(^3S_1)  =   0.527\ [fm]\ ,    
\label{le.6} \end{eqnarray*}
and for $I=2$ we have for ESC08c:
\begin{eqnarray*}
 a_{\Sigma^\pm\Sigma^\pm}(^1S_0) &=& +8.10\ (-0.65)\ [fm]\ ,\\  
 r_{\Sigma^\pm\Sigma^\pm}(^1S_0) &=& -65.36\ (19.97)\ [fm]\ .    
\end{eqnarray*}
The values in parentheses indicate the values without Coulomb.
The results at the $\Xi N$ threshold and at the $\Lambda\Sigma$ threshold
are given in Table~\ref{tabscat-2a}-\ref{tabscat-2c}.
The $\Lambda\Lambda(^1S_0)$ scattering lengths are found to be larger 
in absolute value than in the NSC97 models \cite{SR99}, 
indicating a more attractive $\Lambda\Lambda$ interaction.

The old experimental information seemed to indicate a
separation energy of $\Delta B_{\Lambda\Lambda}=4-5$ MeV,
corresponding to a rather strong attractive $\Lambda\Lambda$
interaction. As a matter of fact, an estimate for the $\Lambda\Lambda$
$^1S_0$ scattering length, based on such a value for $\Delta
B_{\Lambda\Lambda}$, gives $a_{\Lambda\Lambda}(^1S_0)\approx-2.0$
fm~\cite{Tan65,Bod65}.
However, in recent years the experimental information and interpretation 
of the ground state levels of 
$_{\Lambda\Lambda}^{\,\ 6}$He, $_{\Lambda\Lambda}^{\;10}$Be, and
$_{\Lambda\Lambda}^{\;13}$B~\cite{Dal89}, has been changed drastically.
This because of the Nagara-event \cite{Tak01}, 
identified uniquely as $_{\Lambda\Lambda}^{\,\ 6}$He
\cite{Tak01}, which established that the $\Lambda\Lambda$-interaction is weaker
($\Delta B_{\Lambda\Lambda} \approx 0.7$ MeV).

In NSC97 \cite{RSY99} it was only possible to increase the attraction in
the $\Lambda\Lambda$ channel by modifying the scalar-exchange
potential. If the scalar mesons are viewed as being mainly 
$q\bar{q}$ states, one finds that the (attractive) scalar-exchange
part of the interaction in the various channels satisfies
\begin{equation}
   |V_{\Lambda\Lambda}| < |V_{\Lambda N}| < |V_{N\!N}|,
\label{le.7} \end{equation}
suggesting indeed a rather weak $\Lambda\Lambda$-potential.
The NSC97 fits to the $Y\!N$ scattering data~\cite{RSY99} give values
for the scalar-meson mixing angle which seem to point to almost ideal
mixing for the scalars as $q\bar{q}$ states. We found that an increased
attraction in the $\Lambda\Lambda$ channel would give rise to
(experimentally unobserved) deeply bound states in the $\Lambda N$ channel.
On the other hand, in the ESC-models there are in principle more possibilities 
because of the presence of meson-pair potentials.
As one sees from the values of the $a_{\Lambda\Lambda}(^1S_0)$ 
in the ESC08c model of this paper, we can produce 
the apparently required attraction in the $\Lambda\Lambda$
interaction without giving rise to $\Lambda N$ bound states.
Notice that also in ESC08 we have ideal scalar mixings, akin to NSC97.

\begin{table}
\caption{SU(3) content of the different interaction channels.
         $S$ is the total strangeness and $I$ is the isospin.
         The upper half refers to the space-spin symmetric
         states $^3S_1$, $^1P_1$, $^3D$, \ldots, while the
         lower half refers to the space-spin antisymmetric
         states $^1S_0$, $^3P$, $^1D_2$, \ldots}
\begin{tabular}{cccc}
  \multicolumn{4}{c}{Space-spin symmetric}                              \\
 $S$ & $I$ & Channels                  & SU(3)-irreps                   \\
\colrule
   0 &   0 & $N\!N$                    & $\{10^*\}$                     \\
 --1 & 1/2 & $\Lambda N$, $\Sigma N$   & $\{10^*\}$, $\{8\}_a$          \\
     & 3/2 & $\Sigma N$                & $\{10\}$                       \\
 --2 &   0 & $\Xi N$                   & $\{8\}_a$                      \\
     &   1 & $\Xi N$, $\Sigma\Sigma$   & $\{10\}$, $\{10^*\}$, $\{8\}_a$\\
     &     & $\Sigma\Lambda$           & $\{10\}$, $\{10^*\}$           \\
  \multicolumn{4}{c}{Space-spin antisymmetric}                          \\
 $S$ & $I$ & Channels                  & SU(3)-irreps                   \\
\colrule
   0 &   1 & $N\!N$                    & $\{27\}$                       \\
 --1 & 1/2 & $\Lambda N$, $\Sigma N$   & $\{27\}$, $\{8\}_s$            \\
     & 3/2 & $\Sigma N$                & $\{27\}$                       \\
 --2 &   0 & $\Lambda\Lambda$, $\Xi N$, $\Sigma\Sigma$
                                       & $\{27\}$, $\{8\}_s$, $\{1\}$   \\
     &   1 & $\Xi N$, $\Sigma\Lambda$  & $\{27\}$, $\{8\}_s$            \\
     &   2 & $\Sigma\Sigma$            & $\{27\}$                       \\

\end{tabular}
\label{tabirrep}
\end{table}

 \begin{figure}[ht]
 \includegraphics*[width= 8cm,height=10cm]{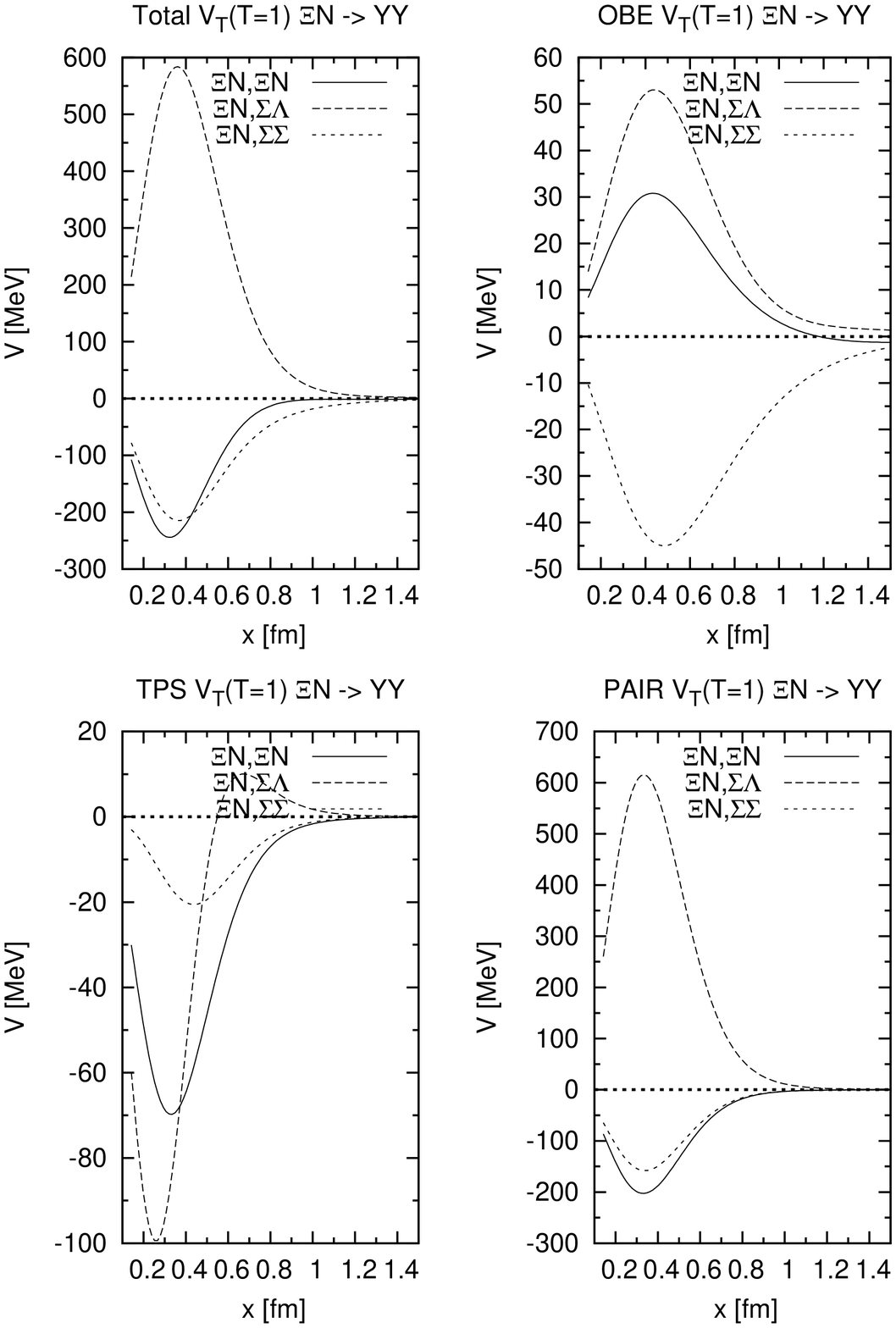} 
\caption{ $\Xi N(^3S_1,T=1)$ tensor potentials}                          
 \label{allten1}
 \end{figure}

\subsection{Deuteron state in $\Xi N(^3S_1-^3D_1,I=1)$}
A discussion of the possible bound-states, using the SU(3) content 
of the different $S=0, -1, -2$ channels is given in \cite{SR99}.
As in \cite{SR99}, for a general orientation, 
we list in Table~\ref{tabirrep} all
the irreps to which the various baryon-baryon channels belong.    
In ESC08c we find a deuteron with isospin I=1 and strangeness S=-2, 
belonging to the $\{10^*\}$ SU(3)-irrep, which is
a $\Xi N$ bound state in the $^3S_1$-$^3D_1$ coupled partial wave. 
In model ESC04d \cite{Rij05b}, however, there occurs a
$\Xi N$ bound state in the $\Xi N(^3S_1$-$^3D_1r,(I=0)$ partial wave. 
From Table~\ref{tabirrep} one sees that this is a $\{8_a\}$-state, which
was a little bit surprising, because the OBE-potential one expects
to be rather repulsive in the irrep $\{8_a\}$, see \cite{MRS89}.
In the ESC04 models this occurrence was ascribed to the inclusion of the 
potentials of the axial-vector-mesons, and the meson pairs.    
Since ESC04a-c did not show such a bound state it is considered 
to be accidental.
However, the situation in ESC08c is completely different. Here the bound state is in the 
deuteron-like states where strong tensor forces are present, which causes the binding
similarly to the np-deuteron. In Fig.~\ref{allten1} the tensor potentials are shown, where
it appears that also the $\Sigma\Lambda$ tensor potential is important. 
This is similar to the situation in $\Lambda N$ below the 
$\Sigma N$-threshold where a large cusp occurs. 
The calculated binding energy $B_E(D^*)= 1.56$ MeV. 
 \begin{figure}[ht]
 \includegraphics*[width= 8cm,height=10cm]{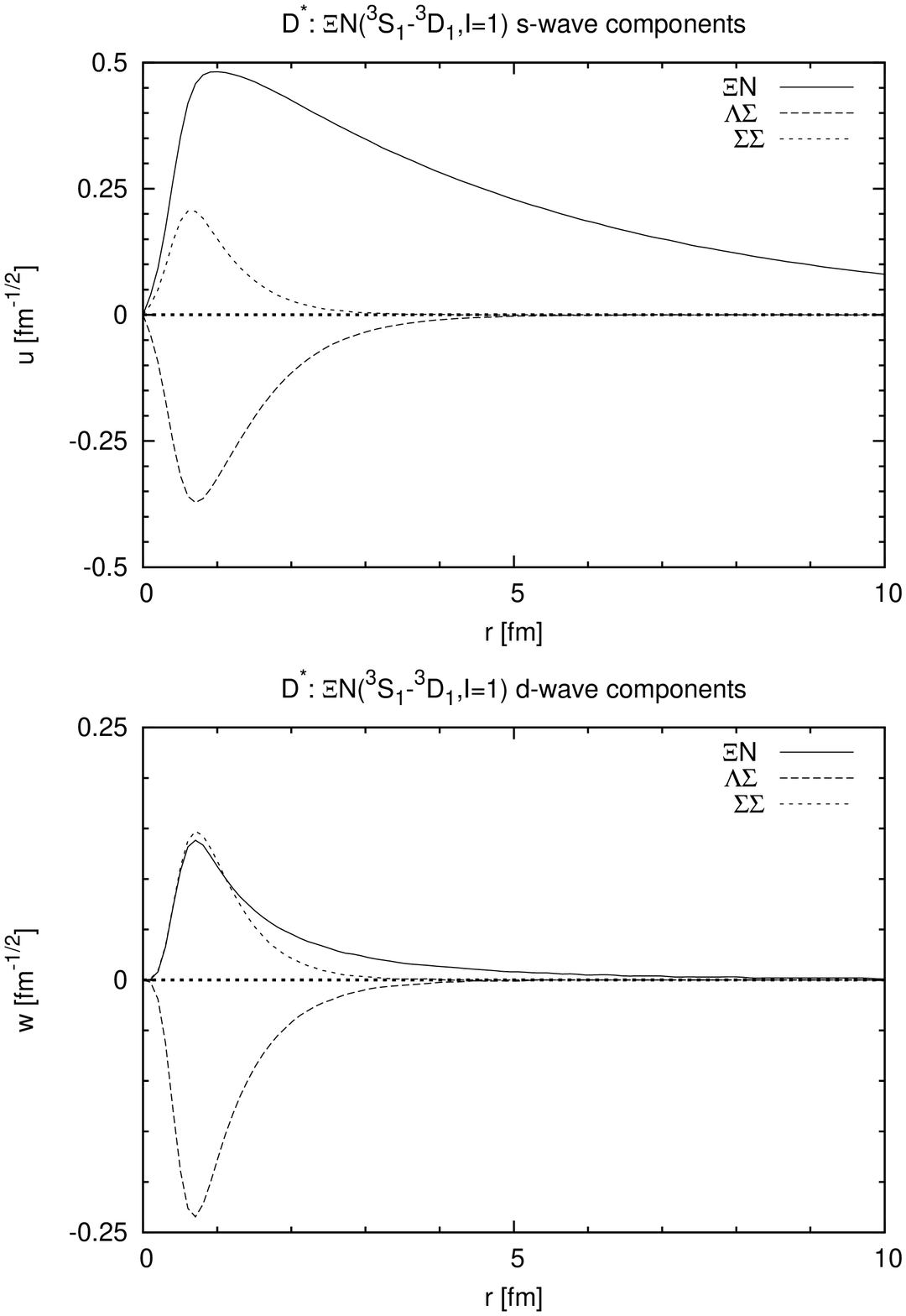} 
\caption{ $\Xi N(^3S_1-^3D_1,I=1)$ deuteron wave functions}             
\label{fig.deut1}
 \end{figure}

 \begin{figure}[ht]
 \includegraphics*[width= 8cm,height=10cm]{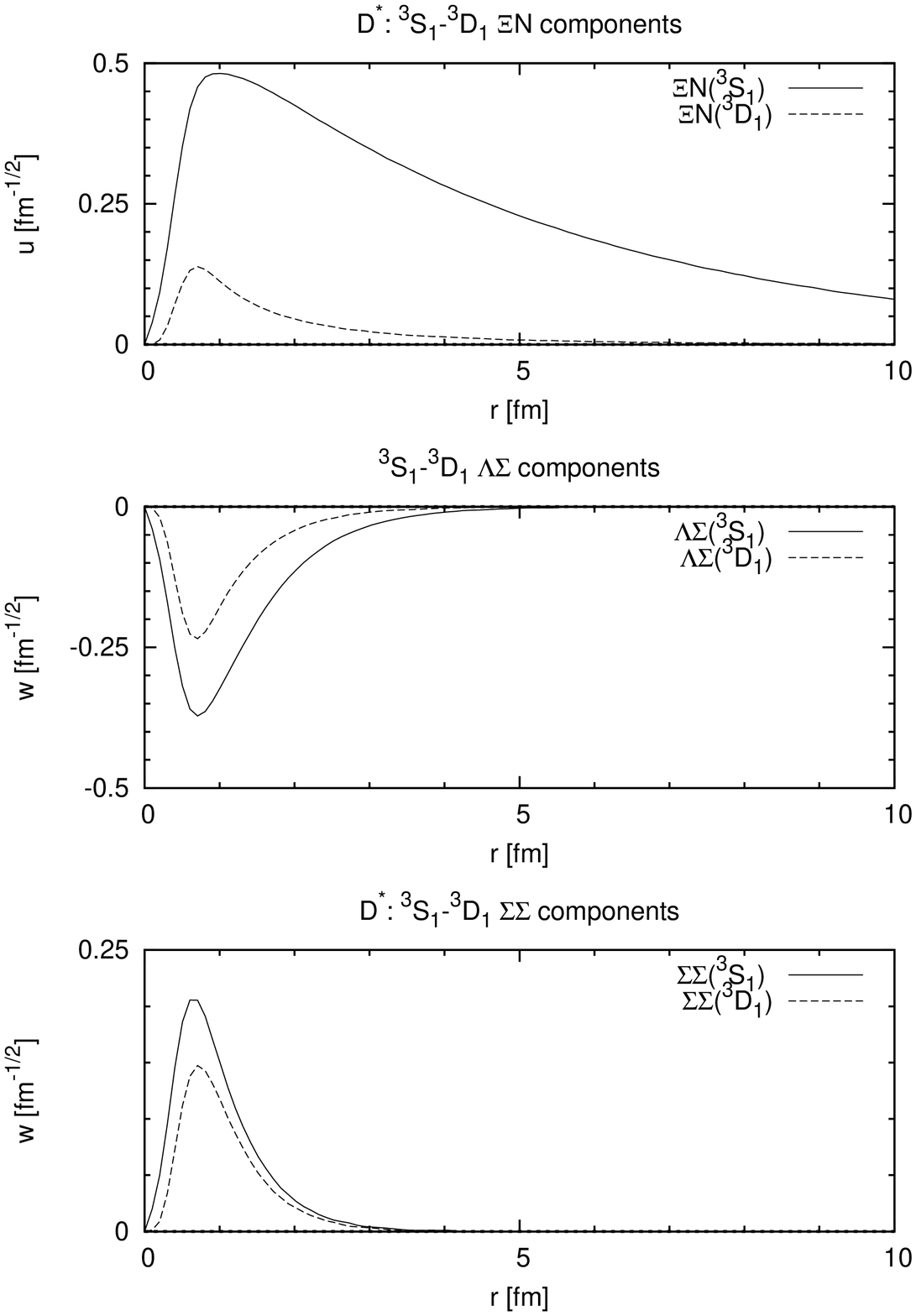} 
\caption{ $\Xi N(^3S_1-^3D_1(I=1)$ deuteron wave functions}             
\label{fig.deut2}
 \end{figure}

\subsection{Partial Wave Phase Parameters}    

For the {\it BB}-channels below the inelastic threshold we use for the parametrization
of the amplitudes the standard nuclear-bar phase shifts \cite{SYM57}.
The information on the
elastic amplitudes above thresholds is most conveniently given using the BKS-phases
\cite{Bry81,Kla83,Spr85}. For uncoupled partial waves, the elastic {\it BB}
$S$-matrix element is parametrized as
\begin{equation}
 S = \eta e^{2i \delta}\ ,\ \ \eta = \cos(2\rho)\ .
\label{pw.1} \end{equation}
For coupled partial waves the elastic {\it BB}-amplitudes are $2\times 2$-matrices.
The BKS $S$-matrix parametrization, which is of the type-S variety, is given by
\begin{equation}
 S = e^{i\delta} e^{i\epsilon} N\ e^{i\epsilon} e^{i\delta}\ ,
\label{pw.2} \end{equation}
where 
\begin{equation}
 \delta = \left(\begin{array}{cc} \delta_\alpha & 0 \\ 
          0 & \delta_\beta \end{array}\right)\ ,\ \
 \epsilon= \left(\begin{array}{cc} 0 & \epsilon \\ 
           \epsilon & 0 \end{array}\right)\ ,
\label{pw.3} \end{equation}
and $N$ is a real, symmetric matrix parametrize as
\begin{equation}
  N = \left(\begin{array}{cc} \eta_{11} & \eta_{12} \\ \eta_{12} & \eta_{22}
  \end{array}\right)\ .
\label{pw.4} \end{equation}
From the various parametrizations of the $N$-matrix, we choose the Kabir-Kermode
parametrization \cite{Kab87} to represent the $N$-matrix in the figures. 
Then, the $N$-matrix is given by the inelasticity 
parameters $(\alpha,\beta,\varphi)$, called $\rho$-parameters, as follows
\begin{equation}
  N = \left(\begin{array}{cc} \cos(2\alpha) & \sin(\varphi+\xi) \\              
                              \sin(\varphi+\xi) & \cos(2\beta)               
  \end{array}\right)\ , 
\label{pw.5} \end{equation}
where  
\begin{eqnarray}
&& \alpha = \pm \frac{1}{2}\cos^{-1}(\eta_{11})\ \ ,\ \ 
 \beta = \pm\frac{1}{2}\cos^{-1}(\eta_{22})\ , \nonumber\\
&& \varphi = \sin^{-1}(\eta_{12})-sgn(\eta_{12})\sin^{-1} Q\ \nonumber\\ 
&&   \xi     = sgn(\eta_{12})\sin^{-1} Q\ .        
\label{pw.6} \end{eqnarray}
Here 
\begin{equation}
 Q^2 = 1 - |\eta_{11}+\eta_{22}|+ \eta_{11} \eta_{22}\ .
\label{pw.7} \end{equation}

In Fig's~\ref{phasll0}-\ref{phassl1} the BKS-phases and coupling parameters
$(\alpha,\beta,\varphi)$ for ESC08c are shown.
In Fig~\ref{phasll0} and Fig.~\ref{phasxn1} we also show the $^1S_0$-phases (n.c.) 
for the case with no coupling to the other two-particle channels. 
For $\Lambda\Lambda$ the n.c.-curve 
shows that the potential is repulsive, which is mainly due to the $\{1\}$-irrep.
The attraction comes in particular from the coupling to the $\Xi N$-channel.

In the Tables~\ref{tab.bks.a}-\ref{tab.bks.g}, 
we give for ESC08c the phases and inelasticity parameters $\rho$ and 
$\eta_{11}, \eta_{12}, \eta_{22}$, which
enable the reader to construct the $N$-matrix most directly.




\subsection{Total cross sections}
We next present the predictions for the total cross section for several
channels. 
We suppose always that the beam as well as the target are unpolarized.
Therefore, we incuded the statistical factors, which are $1/4$           
for the spin-singlet and $3/4$ for the spin-triplet case.



For those cases where both baryons are charged, we do not
include the purely Coulomb contribution to the total cross section,
nor do we include the Coulomb interference to the nuclear amplitude.
The cross section is calculated by summing the contributions from
partial waves with orbital angular momentum up to and including $L=2$.
We find this to be sufficient for all the $S\neq0$ sectors; inclusion
of any higher partial waves has no significant effect. 
Inclusion of higher partial waves will shift the total cross section
to slightly higher values without changing the overall shape. Of course,
their inclusion would be necessary if a detailed comparison with real
experimental data were to be made.




In Table~\ref{tab.sigt0a} we show the $\Lambda\Lambda \rightarrow \Lambda\Lambda, \Xi N$
total X-sections as a function of the laboratory momentum $p_\Lambda$.
Being dominantly S-wave, 
there is in principle has a (sharp) cusp at the $\Xi N$-threshold, 
i.e. $p_\Lambda=344.4$ MeV/$c^2$, which indeed is visible in the table.
In Table~\ref{tab.sigt0a} we also show the $\Xi N \rightarrow \Xi N, \Lambda\Lambda$
total X-sections as a function of the laboratory momentum $p_\Xi$.
In Table~\ref{tab.sigt1a} we show the total X-sections for the $\Xi N \rightarrow \Xi N, \Sigma\Lambda$
and the $I=1,L=0$ $\Sigma \Lambda \rightarrow \Sigma\Lambda, \Xi N, \Sigma\Sigma$ reactions 
as a function of the laboratory momentum $p_\Xi$.


\begin{table}[ht]
\caption{ ESC08c $(I=0,L=0)$ total X-sections 
in [mb] as a function of the laboratory momentum $p_\Lambda$
         in [MeV]}
\begin{ruledtabular}
\begin{tabular}{r|rr|rr}
  & \multicolumn{2}{c|}{ $\Lambda\Lambda\rightarrow \Lambda\Lambda,\Xi N$} & 
    \multicolumn{2}{c}{ $\Xi N \rightarrow \Xi N, \Lambda\Lambda$ }\\ 
$p_{\Lambda}$  & \multicolumn{1}{c}{$\Lambda\Lambda$}&\multicolumn{1}{c|}{$ \Xi N $} 
               & \multicolumn{1}{c}{$\Xi N$}&\multicolumn{1}{c}{$ \Lambda\Lambda $} \\
\colrule   
 10  & 22.65 & --- & 630.95 & 2468.24 \\
 50  & 20.98 & --- & 114.35 & 1817.61 \\
 100 & 16.83 & --- &  49.78 &  990.75 \\
 200 &  8.58 & --- & 100.01 &  607.52 \\
 300 &  5.16 & --- &  58.81 &  405.65 \\
 350 &  6.42 & 2.28&  28.35 &  161.84 \\
 400 &  6.96 & 6.68&  16.15 &   85.65 \\
 500 & 11.03 &23.39&  10.96 &   55.60 \\
 600 &  6.26 &17.50&   8.86 &   41.69 \\
 700 &  4.92 &12.78&   8.13 &   33.69 \\
 800 &  4.87 &10.40&   7.68 &   29.07 \\
 900 &  5.41 & 9.05&   6.93 &   27.48 \\
1000 &  5.69 & 7.38&   6.51 &   26.94 \\
\end{tabular}
\end{ruledtabular}
\label{tab.sigt0a}
\end{table}

\begin{table}[ht]
\caption{ESC08c $(I=1,L=0)$ total X-sections 
 $\Xi N \rightarrow \Xi N, \Sigma\Lambda$
         in [mb] as a function of the laboratory momentum $p_\Xi$
         in [MeV]}
\begin{ruledtabular}
\begin{tabular}{r|rr|rrr}
  & \multicolumn{2}{c|}{ $\Xi N \rightarrow \Xi N, \Sigma \Lambda$} & 
    \multicolumn{3}{c}{  $\Sigma\Lambda \rightarrow \Sigma\Lambda, \Xi N, \Sigma\Sigma$}\\ 
$p_{\Xi}$ & \multicolumn{1}{c}{$\Xi N$}&\multicolumn{1}{c|}{$\Sigma\Lambda$} 
          & \multicolumn{1}{c}{$\Sigma\Lambda$}
          & \multicolumn{1}{c}{$\Xi N$}&\multicolumn{1}{c}{$\Sigma\Sigma$}\\
\colrule   
 10  & 66.52 & --- & 690.58 & 10.61 & --- \\    
 50  & 66.55 & --- & 124.73 &  8.51 & --- \\    
 100 & 66.75 & --- &  58.40 &  6.83 & --- \\    
 200 & 67.88 & --- &  20.46 & 10.28 & --- \\    
 300 & 69.36 & --- &  13.78 & 17.53 & --- \\    
 400 & 70.27 & --- &  12.96 & 21.97 & --- \\    
 500 & 69.80 & --- &  11.86 & 36.62 & --- \\    
 600 & 86.79 & 2.21&  11.70 & 42.12 & --- \\
 700 & 50.02 & 3.84&  10.19 & 19.67 & 0.99\\
 800 & 46.79 & 5.59&   9.84 & 15.49 & 1.94\\
 900 & 42.87 & 6.58&   8.67 & 26.72 & 6.32\\
 950 & 40.71 & 7.18&   7.58 & 23.74 & 5.60\\
1000 & 45.79 & 6.37&   7.56 & 49.80 & 7.83\\
\end{tabular}
\end{ruledtabular}
\label{tab.sigt1a}
\end{table}

\subsection{Flavor SU(3)-irrep potentials}
\begin{figure}[hbt]
 \includegraphics*[width= 8cm,height=10cm]{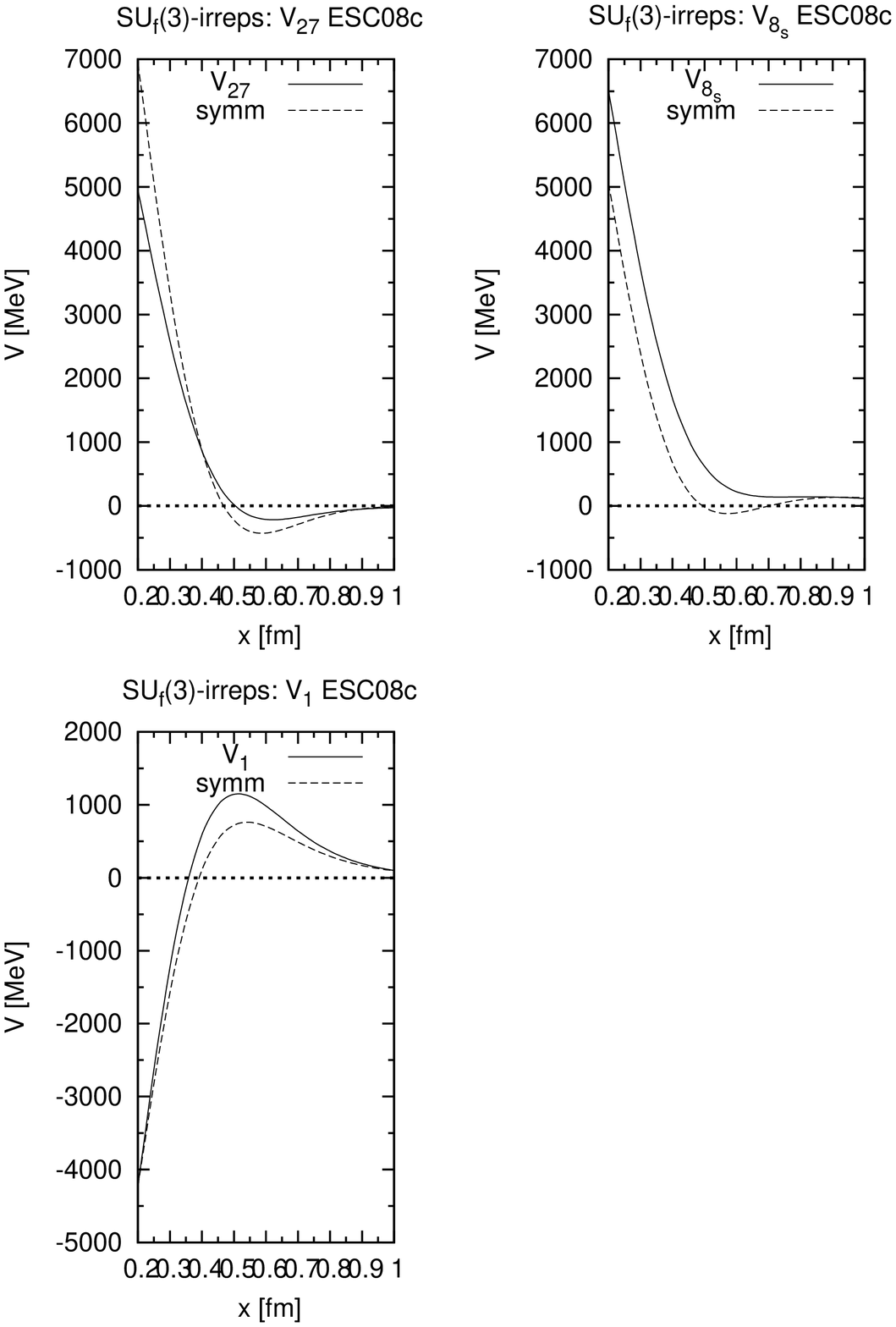}
\caption{ Solid line average SU(3)-irrep potentials in particle basis. Dashed
lines potentials with exact flavor SU(3)-symmetry}
\label{irrep0.fig}
 \end{figure}

\begin{figure}[hbt]
 \includegraphics*[width= 8cm,height=10cm]{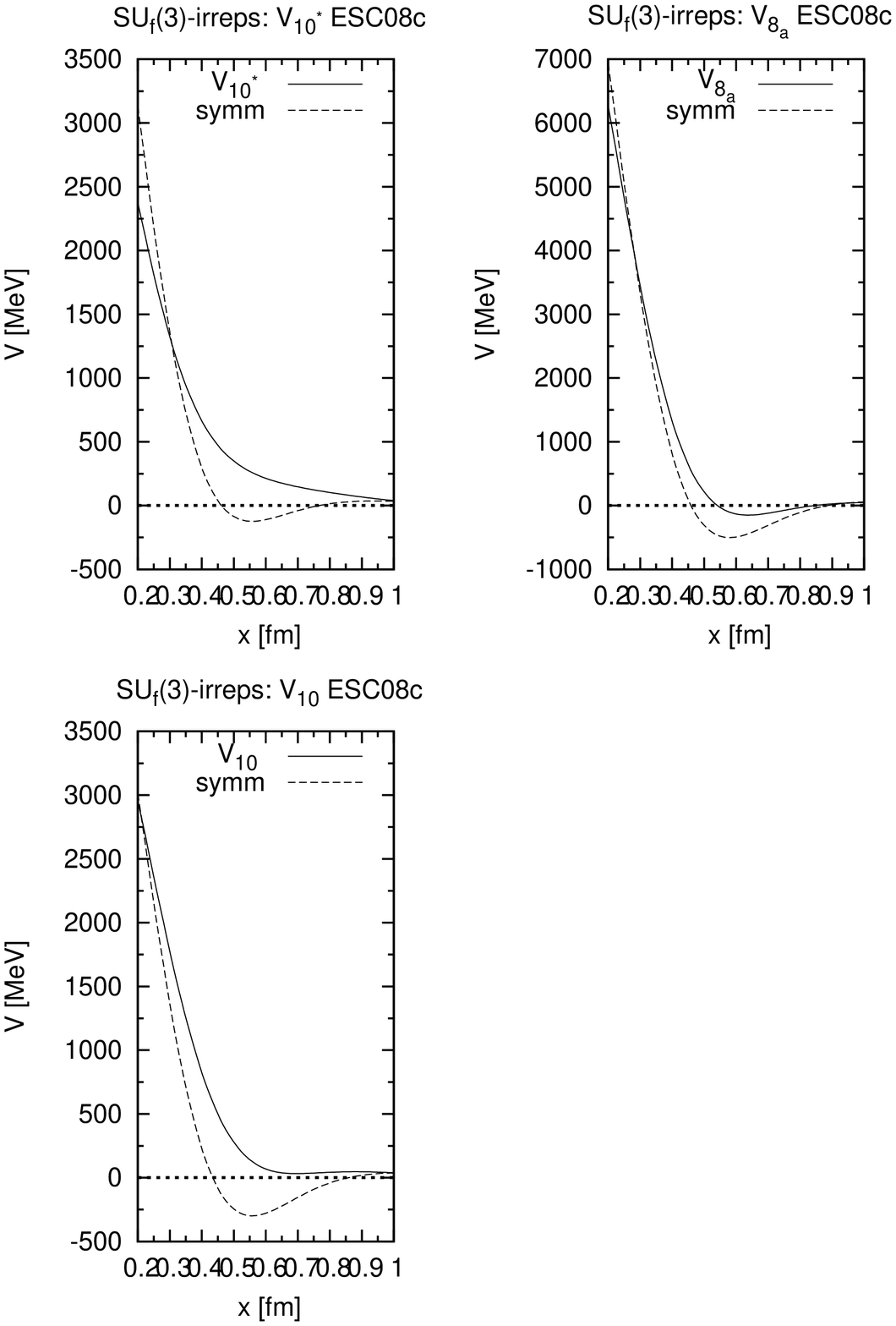}
\caption{ Solid line average SU(3)-irrep potentials in particle basis. Dashed
lines potentials with exact flavor SU(3)-symmetry}
\label{irrep1.fig}
 \end{figure}

In Fig.~\ref{irrep0.fig} and \ref{irrep1.fig} the SU(3)-irreps are
displayed. 
The solid lines show averages of the SU(3)-irrep potentials
using the potentials on the particle basis. The dashed lines are the
irrep potentials in an SU(3) limit, where  
 {\green $M_N=M_\Lambda= M_\Sigma=M_\Xi = 1115.6$ MeV},  
 {\red   $m_\pi=m_K=m_\eta=m_{\eta'}= 410$ MeV},
 {\green $m_\rho=m_{K^*}=m_\omega=m_\phi= 880$ MeV}, and
 {\green $m_{a0}=m_{\kappa}=m_\sigma=m_{f'_0}= 880$ MeV}.
Comparison with the results from LQCD \cite{Inou11,Inou12} shows qualitatively 
very similar results. The exception is the SU(3)-singlet $\{1\}$-irrep.
Here LQCD potential is atractive for $0 < r <\infty$, whereas in ESC08c
there is an attractive pocket for $r \leq 0.5$ fm and is repulsive 
for $r > 0.5$ fm. This shape is due to the behavior of the spin-spin 
potentials from pseudoscalar and vector exchange, which have zero
volume integrals.
In the $\{1\}$-irrep for the SU(3)-broken potential (solid line) there is
no bound state, i.e. no H-particle \cite{Jaffe77}. This is in agreement
with the recent experimental result studying $\Upsilon(1S,2S)$-decay \cite{Kim13}.

\section{$\Xi N$ G-matrix interaction and $\Xi$-nucleus states
\protect\footnote{ In this section we denote isospin by T, 
the nuclear physics notation.}
}
\label{sec:gmat-5}
We calculate $\Xi$ potential energies $U_\Xi$ and derive
$\Xi N$ G-matrix interactions in nuclear matter with the use of ESC08c.
G-matrix calculations are performed with the continuous (CON) choice,
where off-shell potentials are taken into account continuously from 
on-shell ones in intermediate propagations of correlated pairs.
Then, a two-body state is specified by
spin $S$, isospin $T$, orbital and total angular momenta 
$L$ and $J$, respectively.
The imaginary parts of G-matrices appear due to energy-conserving 
transitions from $\Xi\!N$ to $\Lambda\!\Lambda$ channels
in the $T=0$ $^{1}S_1$ and $^{3}P_J$ states.
The conversion width $\Gamma_\Xi^c$ is obtained from
the imaginary part of $U_\Xi$ multiplying by $-2$.

\begin{table}[ht]
\caption{$U_\Xi(\rho_0)$ and partial wave contributions
for ESC08c calculated with the CON choice. 
$\Gamma_\Xi^c$ denotes $\Xi\!N$-$\Lambda\!\Lambda$ conversion width.
All entries are in MeV.}
\label{Gmat-X1}
\begin{center}
 \begin{tabular}{|crrrrrr|r|r|}
 \hline\hline
  $T$ & $^1S_0$ & $^3S_1$ & $^1P_1$ & $^3P_0$ & $^3P_1$ & $^3P_2$ 
& $U_\Xi$ & $\Gamma_\Xi^c$ \\
 \hline
 $0$&   1.4 & $-$8.0  & $-$0.3 & 1.8 &   1.4 & $-$2.1 &  & \\
 $1$&  10.7 & $-$11.1 &  1.1 &  0.7 & $-$2.6 & $-$0.0 & $-$7.0 & 4.5 \\
 \hline
\end{tabular}
\end{center}
\end{table}

Table~\ref{Gmat-X1} shows the potential energy $U_\Xi$
and its partial-wave contributions at normal density $\rho_0$.
The $U_\Xi$ values turn out to be given by the strong cancellation between 
attractive contributions in $^3S_1$ ($T=0,1$) states and repulsive contributions 
in $^1S_0$ ($T=0,1$) states. Eventually, values of $U_\Xi$ become far less 
attractive than those of $U_\Lambda$. The calculated value of 
$\Gamma_\Xi^c(\rho_0)$ is also given in the Table~\ref{Gmat-X1}, 
the dominant contribution of which comes from the 
$\Lambda\!\Lambda$-$\Xi\!N$-$\Sigma\!\Sigma$ coupling interaction
in the $T=0$ $^1S_0$ state. 

\begin{figure}[ht]
\begin{center}
\includegraphics*[width=10cm,height=10cm]{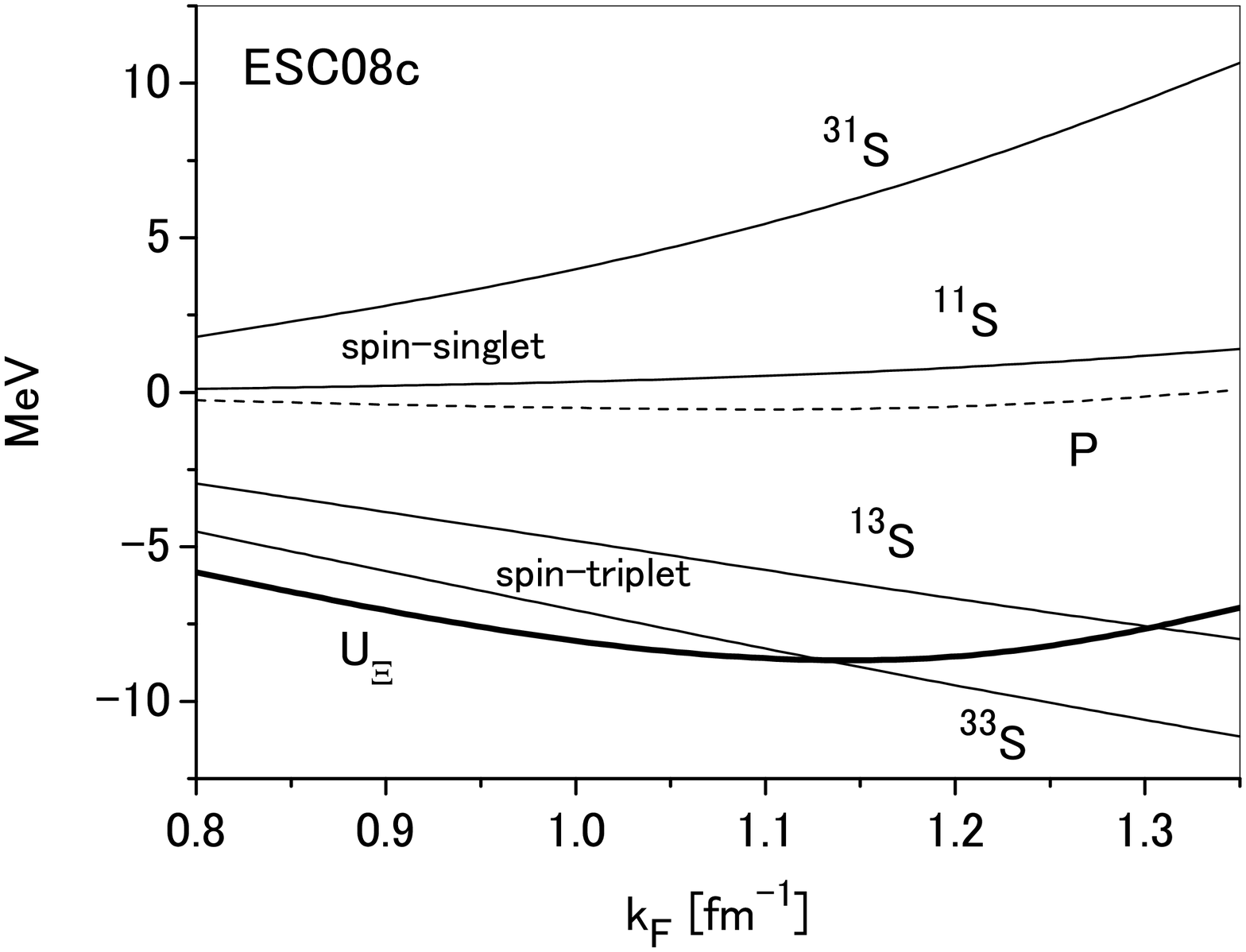}
\caption{$U_\Xi$ and partial-wave contributions in
$^{(2T+1)(2S+1)}L_J$ states are drawn
as a function of $k_F$. $U_\Xi(k_F)$ is shown by
a bold curve. Attractive contributions in spin-triplet 
states ($^{33}S_1$ and $^{13}S_1$) and repulsive ones
in spin-singlet states ($^{31}S_1$ and $^{11}S_1$)
are shown by thin curves. The $P$-state contribution,
summed for $(T,S,J)$, is shown by a dashed curve.
}
\label{UXi-kF}
\end{center}
\end{figure}

In Fig.~\ref{UXi-kF}, 
$U_\Xi$ values and partial-wave contributions
are drawn as a function of $k_F$.
Here, $U_\Xi(k_F)$ is shown by a bold curve, and 
contributions in $^{33}S_1$, $^{31}S_1$, $^{13}S_1$ and 
$^{31}S_1$ states are shown by thin curves. 
$P$-state contribution summed for $(S,T,J)$ states is 
shown by a dashed curve.

As well as in Table~\ref{Gmat-X1}, we see here the cancellation between 
attractive contributions in spin-triplet $S$ states and repulsive ones 
in spin-singlet $S$ states.  
Especially, the attraction in the $^3S_1$ $T=1$ state is due to the 
$\Xi N$-$\Lambda \Sigma$-$\Sigma \Sigma$ tensor-coupling interactions 
in this state. If these tensor parts in this channel are switched off, 
the value of $U_\Xi$ becomes strongly repulsive.
On the other hand, the $P$-state contributions are small.

It should be noted that the $U_{\Xi}$ curve becomes substantially
attractive in the low density region due to the strong density dependence.
This feature works favorably for $\Xi$ binding energies in light systems.

For applications to finite $\Xi$ systems,
$\Xi\!N$-$\Xi\!N$ central parts of the complex G-matrix interactions 
for ESC08c are represented in Gaussian forms, 
whose coefficients are given as a function of $k_F$.
The determined parameters are given in Table~\ref{YNG}.
\begin{table}[ht]
\begin{center}
\caption{${\cal G}(r;k_F)= \sum_{i} 
(a_i+b_i k_F+c_i k_F^2) \exp -(r/\beta_i)^2 $ 
in $^{(2S+1)(2T+1)}E$ and $^{(2S+1)(2T+1)}O$ states.
}
\label{YNG}
\vskip 0.2cm
\begin{tabular}{|c|cccc|}\hline
&  $\beta_i$ (fm) & 0.50 & 0.90 & 2.00  \\
\hline
         & a  & $-$540.0 &   $210.4-66.76i$ & $-$5.59  \\
$^{11}E$ & b  &    4975. &  $-988.7+19.32i$ &  0.0  \\
         & c  & $-$2500. &   $490.4-2.675i$ &  0.0  \\
\hline
         & a  & $-$4353. &    562.8    &   0.215  \\
$^{13}E$ & b  &    6760. & $-$1065.    &   0.0    \\
         & c  & $-$2456. &    408.9    &   0.0    \\
\hline
         & a  & 0.0 &    70.88  &  $-$5.59   \\
$^{11}O$ & b  & 0.0 &    3.319  &   0.0  \\
         & c  & 0.0 &    8.351  &   0.0  \\
\hline
         & a  & 0.0 &   $15.98-77.30i$  & 0.215 \\
$^{13}O$ & b  & 0.0 &   $-214.4+151.5i$  &  0.0  \\
         & c  & 0.0 &   $185.2-95.77i$  &  0.0  \\
\hline
\hline
         & a  & $-$343.9 &    111.2  &   0.357  \\
$^{31}E$ & b  &    133.5 & $-$18.08  &     0.0  \\
         & c  & $-$10.00 &    23.38  &     0.0  \\
\hline
         & a  & $-$809.8 &    50.29  &  $-$1.76  \\
$^{33}E$ & b  &    1619. & $-$213.5  &     0.0   \\
         & c  & $-$633.3 &    95.44  &     0.0   \\
\hline
         & a  & 0.0 &    42.44  &    0.357  \\
$^{31}O$ & b  & 0.0 &    1.531  &    0.0    \\
         & c  & 0.0 &    8.844  &    0.0    \\
\hline
         & a  & 0.0 & $-$17.26  &  $-$1.76   \\
$^{33}O$ & b  & 0.0 &    8.867  &     0.0  \\
         & c  & 0.0 &    6.355  &     0.0  \\
\hline
\end{tabular}
\end{center}
\end{table}

As demonstrated in Ref.~\cite{Yam10},
the observed spectra of $\Lambda$ hypernuclei are described 
successfully with the $\Lambda$-nucleus folding potentials
derived from the $\Lambda N$ G-matrix interactions.
Here, the same method is applied to $\Xi^-$-nucleus systems.
A $\Xi$-nucleus folding potential in a finite system is 
obtained from ${\cal G}^{TS}_{(\pm)}(r;k_F)$ as follows:
\begin{eqnarray}
U_\Xi({\bf r},{\bf r'}  ) &=& U_{dr} +U_{ex} \ ,
\nonumber \\
U_{dr} &=& \delta({\bf r}-{\bf r'})  
\int d{\bf r''} \rho({\bf r''}) \,
V_{dr}(|{\bf r}-{\bf r''}|;k_F  )
\nonumber \\
U_{ex} &=& \rho({\bf r},{\bf r'}) 
V_{ex}(|{\bf r}-{\bf r'}|;k_F  ) \ ,
             \label{eq:fold1}
\end{eqnarray}
\begin{eqnarray}
\left(\begin{array}{c}
V_{dr} \\
V_{ex}
\end{array}\right)
&=&\frac{1}{2(2t_Y+1)(2s_Y+1)}\,
\sum_{TS} (2T+1)(2S+1) \cdot\nonumber\\ && \times
 [{\cal G}^{TS}_{(\pm)} \pm {\cal G}^{TS}_{(\mp)}] \ ,
             \label{eq:fold2}
\end{eqnarray}
where $(\pm)$ denote parity quantum numbers. Here, core nuclei are assumed 
to be spherical, and densities $\rho(r)$ and mixed densities $\rho(r,r')$ are 
obtained from Skyrme-HF wave functions.
The isospin-dependence of ${\cal G}_{(\pm)}^{TS}(r;k_F)$ 
leads to the Lane term. 
In this work, only the diagonal parts of the 
${\bf t_\Xi}\cdot {\bf T_c}$ term are taken into account.

For $k_F$ included in ${\cal G}(r;k_F)$, we use
the averaged-density approximation (ADA):
An averaged value of $\rho(r)$ is defined by
$\bar \rho=\langle \phi_\Xi(r)|\rho(r)|\phi_\Xi(r) \rangle$
by using a $\Xi$-state function $\phi_\Xi(r)$.
Then, an averaged value of $k_F$ is given by
$\bar k_F=(1+\alpha) \left(1.5\pi^2 \bar \rho \right)^{1/3}$.
This value $\bar k_F$ is put into ${\cal G}(r;k_F)$ and 
determined self-consistently for each $\Xi$ state, and
$\alpha$ is a parameter fixed by a fine tuning  
to the experimental data. Hereafter, we investigate the two
cases of $\alpha=0.0$ and $0.1$.

\begin{table}[ht]
\centering 
\setlength{\textwidth}{50mm} 
\caption{Calculated quantities in $\Xi^- + ^{12}$C and
$\Xi^- + ^{14}$N systems for $\alpha$=0.0 and 0.1.
Binding energies $B_{\Xi^-}$ and conversion width $\Gamma^c_{\Xi^-}$
are in MeV. R.m.s. radius $\sqrt{\langle r^2 \rangle}$ is in fm.
}
\label{result}
\vskip 0.2cm
 \begin{tabular}{lccc}
 \hline\hline
$\Xi^- + ^{12}$C  &   &   $\alpha=0.0$ & $\alpha=0.1$   \\
 \hline
$1S$  & $B_{\Xi^-}$                  &  5.18 &  3.83    \\
      & $\Gamma^c_{\Xi^-}$           &  1.77 &  1.49    \\
      & $\sqrt{\langle r^2 \rangle}$ &  2.93 &  3.34    \\ 
\hline
$2P$  & $B_{\Xi^-}$                  &  1.10 &  0.68    \\
      & $\Gamma^c_{\Xi^-}$           &  0.75 &  0.44    \\
      & $\sqrt{\langle r^2 \rangle}$ &  5.37 &  7.54    \\
 \hline\hline
 $\Xi^- + ^{14}$N  &  & $\alpha=0.0$  &  $\alpha=0.1$   \\
 \hline
$1S$  & $B_{\Xi^-}$                 &   6.30 &  4.82    \\
      & $\Gamma^c_{\Xi^-}$           &  2.15 &  1.87    \\
      & $\sqrt{\langle r^2 \rangle}$ &  2.85 &  3.18    \\ 
\hline
$2P$  & $B_{\Xi^-}$                  &  1.85 &  1.22    \\
      & $\Gamma^c_{\Xi^-}$           &  1.11 &  0.77    \\
      & $\sqrt{\langle r^2 \rangle}$ &  4.50 &  5.66    \\
\hline\hline 
\end{tabular}
\end{table}

Table~\ref{result} shows the results
for $1S$ and $2P$ bound states in $\Xi^- + ^{12}$C and 
$\Xi^- + ^{14}$N systems, where Coulomb interactions between 
$\Xi^-$ and $^{12}$C ($^{14}$N) are taken into account.
$B_{\Xi^-}$ and $\sqrt{\langle r^2 \rangle}$ are the 
binding energy and r.m.s. radius of $\Xi^-$, respectively.
Conversion widths $\Gamma^c_{\Xi^-}$ come from the imaginary parts 
included in $T=0$ $^1S_1$ and $^3P$ states.
The obtained $2P$ states become unbound, when the Coulomb 
interactions between $\Xi^-$ and $^{12}$C ($^{14}$N) are switched off. 
Namely these $2P$ states are so called Coulomb-assisted bound states.
They are specified by the fact that the values of
$\sqrt{\langle r^2 \rangle}$ are large due to their weak binding, 
but far smaller than those in $\Xi^-$ atomic states.
For instance, we have $B_{\Xi^-}$=0.175 MeV and 
$\sqrt{\langle r^2 \rangle}$=36 fm 
for the $\Xi^-$+$^{14}$N $3D$ state. 

Experimental information for $\Xi N$ interactions can be obtained from 
emulsion events of simultaneous emission of two $\Lambda$ hypernuclei 
(twin $\Lambda$ hypernuclei) from a $\Xi^-$ absorption point. 
The $\Xi^-$ produced by the $(K^-,K^+)$ reaction
is absorbed into a nucleus ($^{12}$C, $^{14}$N or $^{16}$O
in emulsion) from some atomic orbit, and by the following
$\Xi^-p \rightarrow \Lambda \Lambda$ process
two $\Lambda$ hypernuclei are produced.
Then, the energy difference between the initial $\Xi^-$ state 
and the final twin $\Lambda$ state gives rise to the binding
energy $B_{\Xi^-}$ between $\Xi^-$ and the nucleus.

Two events of twin $\Lambda$ hypernuclei (I)~\cite{Aoki93}
and (II)~\cite{Aoki95} were observed in the KEK E-176 experiment, 
and recently the new event (III)~\cite{Nakazawa} has been observed 
in the KEK E373 experiment. In the cases of (I) and (II),
each event has no unique interpretation for its reaction process. 
However, it is possible to find a consistent understanding 
for these two events as follows:
The events (I) and (II) were interpreted to be reactions of $\Xi^-$
captured by $^{12}$C.  
Assuming that the $\Xi^-$ is absorbed from the $2P$ orbit in each case,
we have consistently the following reactions 
\begin{eqnarray}
({\rm I})\!\!
&&\Xi^- + ^{12}{\rm\!C} \rightarrow ^9_\Lambda{\rm\!Be}+
^4_\Lambda{\rm\!H}\ (B_{\Xi^-}=0.82 \pm 0.17\ {\rm MeV}),
\nonumber\\
({\rm II})\!\!
&&\Xi^- + ^{12}{\rm\!C} \rightarrow ^9_\Lambda{\rm\!Be}^*+
^4_\Lambda{\rm\!H}\ (B_{\Xi^-}=0.82 \pm 0.14\ {\rm MeV}). \nonumber\\
\end{eqnarray}
Assuming that the $\Xi^-$ is captured from a $2P$ state, 
the calculated values of $B_{\Xi^-}(2P)$ in the $\Xi^- + ^{12}$C 
system (1.10 and 0.68 MeV for $\alpha=$0.0 and 0.1, respectively) 
turn out to be consistent with the values of $0.65 \sim 1.00$ MeV 
given by these data. In Ref.~\cite{Ehime}, this result was used 
to fit the strength of the $\Xi N$ interaction.
In these two events, however, a possibility cannot be 
ruled out that they are captured from $3D$ states.

The event (III) is uniquely identified as
\begin{eqnarray}
&&\Xi^- + ^{14}{\rm\!N} \rightarrow ^{10}_{\ \Lambda}{\rm\!Be}+ 
^5_\Lambda{\rm\!He}  \ ,
\end{eqnarray}
with $B_{\Xi^-}=4.38 \pm 0.25$ MeV, which is the first clear evidence
of a deeply bound state of the $\Xi^- -^{14}{\rm\!N}$ system. 
This value can be reproduced by taking $\alpha=0.14$,
assuming the $\Xi^-$ is captured from the $1S$ state.
However, it is far more probable that $^{10}_{\ \Lambda}$Be
is produced in some excited state.
In Ref.~\cite{Nakazawa}, the excitation energies are taken from
the theoretical calculations \cite{Hiyama12}~\cite{Millener12},
while the ground-state value of $B_{\Lambda}$ is taken from
the emulsion data. Their estimated values of $B_{\Xi^-}$ are
$1.8 \sim 2.0$ MeV and $1.1 \sim 1.3$ MeV, when $^{10}_{\ \Lambda}$Be
is in the first and second excited state, respectively.
Our calculated values of $B_{\Xi^-}(2P)$ 1.85 and 1.22 MeV for 
$\alpha=0.0$ and 0.1 are within the former and the latter regions, 
respectively. The similar value of $B_{\Xi^-}(2P)$ was predicted
in Ref.~\cite{Ehime}.
Thus, assuming that the $\Xi^-$ is in a $2P$ state, the experimental 
values of $B_{\Xi^-}$ can be explained reasonably by small tuning of 
our G-matrix interaction in both cases of possible two excitations 
of $^{10}_{\ \Lambda}$Be.
It should be noted that, assuming $\Xi^-$ captures from $2P$ states,
we could get a consistent interpretation for the three emulsion events
(I), (II) and (III) with use of the G-matrix interaction derived from ESC08c, 

It is well known that capture probabilities of $\Xi^-$ from $2P$ states
are far smaller than those from $3D$ states. In spite of this fact,
twin $\Lambda$ hypernuclei are produced dominantly after $2P$-$\Xi^-$ 
captures. As discussed in Ref.~\cite{Yam94}, the reason is because
sticking probabilities of two $\Lambda$'s produced after $2P$-$\Xi^-$
captures are substantially larger than those after $3D$-$\Xi^-$ captures.

 \onecolumngrid
\begin{center}
\begin{table}[ht]
\caption{Calculated values of $\Xi^-$ single particle
energies $E_{\Xi^-}$ and conversion widths $\Gamma^c_{\Xi^-}$
for $^{28}_{\Xi^-}$Mg ($^{27}$Al$+\Xi^-$) and
$^{89}_{\Xi^-}$Sr ($^{88}$Y$+\Xi^-$).
$\Delta E_L$ and $\Delta E_C$ are contributions from 
Lane terms and Coulomb interactions, respectively.
Entries are in MeV.
R.m.s. radii $\sqrt{\langle r^2 \rangle}$ are in fm.
Coulomb assisted bound states are marked by $(*)$.
}
\label{Gmat-X3}
\vskip 0.2cm
\begin{tabular}{|l|cccccc|}
\hline\noalign{\smallskip}
   &    & $E_{\Xi^-}$ & $\Delta E_L$ & $\Delta E_C$ 
  & $\Gamma^c_{\Xi^-}$ & $\sqrt{\langle r_\Xi^2\rangle}$  \\
\noalign{\smallskip}\hline\noalign{\smallskip}
$^{28}_{\Xi^-}$Mg 
              & $s$ & $-$7.35& $+$0.13 & $-$6.65 & 1.66 & 3.16 \\
              & $p$ & $-$3.86& $+$0.08 &   $(*)$ & 0.91 & 4.32 \\
              & $d$ & $-$0.92& $+$0.03 &   $(*)$ & 0.24 & 9.47 \\
\noalign{\smallskip}
\hline
$^{89}_{\Xi^-}$Sr 
              & $s$ & $-$15.5& $+$0.63 & $-$13.0 & 1.85 & 3.25 \\
              & $p$ & $-$11.9& $+$0.52 & $-$11.4 & 1.16 & 4.23 \\
              & $d$ & $-$8.61& $+$0.41 &   $(*)$ & 0.74 & 5.04 \\
              & $f$ & $-$5.37& $+$0.30 &   $(*)$ & 0.45 & 5.98 \\
\noalign{\smallskip}\hline\noalign{\smallskip}
\end{tabular}
\end{table}
\begin{figure}[ht]
\includegraphics*[width=15cm,height=8cm]{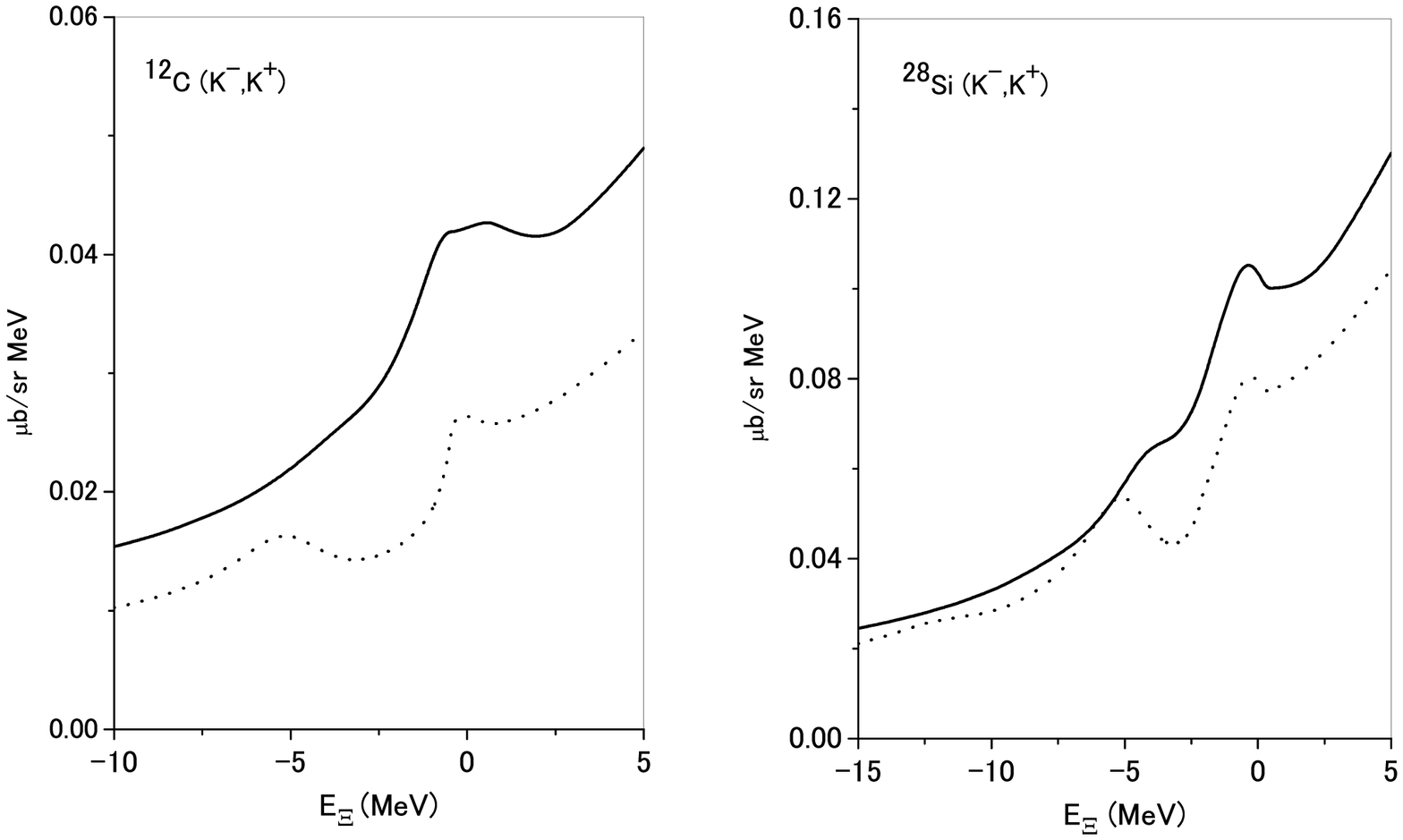}
\caption{$K^+$ spectra of $(K^-,K^+)$ reactions on $^{12}$C
(left panel) and $^{28}$Si (right panel) for ESC08c (solid)
and WS14 (dotted). }
\label{stfn}
\end{figure}
\end{center}
\twocolumngrid
Hereafter, calculations are performed using the G-matrix
interactions with $\alpha=0.1$.
Let us demonstrate the results for heavier systems
$^{28}_{\Xi^-}$Mg and $^{89}_{\Xi^-}$Sr,
being produced by $p(K^-,K^+)\Xi^-$ reactions on $^{28}$Si and
$^{89}$Y targets, respectively. In Table~\ref{Gmat-X3}, 
we show calculated values of $\Xi^-$ s.p. energies $E_{\Xi^-}$, 
conversion widths $\Gamma^c_{\Xi^-}$ and r,m.s radii 
$\sqrt{\langle r_\Xi^2\rangle}$ of solved $\Xi^-$ wave functions,
where $\Delta E_L$ and $\Delta E_C$ are contributions from Lane terms 
and Coulomb interactions, respectively.
It should be noted that the deep $s$ and $p$ states are owing to
large contributions from Coulomb attractions.

The BNL-E885 experiment~\cite{E885} suggests that a $\Xi^-$ 
s.p. potential in $^{11}_{\Xi^-}$Be is given by
the attractive Wood-Saxon potential with the depth 
$\sim \ -14$ MeV (called WS14).
In this case, the calculated value of $B_{\Xi^-}(2P)$ is
0.41 (0.79) MeV for the $\Xi^-+^{12}$C ($^{14}$N) system,
WS14 being slightly less attractive than the above $\Xi$-nucleus 
potentials suitable to the emulsion events of twin $\Lambda$ hypernuclei.

In order to investigate the possibility of observing 
$\Xi^-$ hypernuclear state, we calculate $K^+$ spectra of
$(K^-,K^+)$ reactions on some targets with use of
our G-matrix folding potentials. Calculations are performed with 
the Green's function method in DWIA~\cite{Tadokoro}. 
In Fig.~\ref{stfn}, we show the obtained $K^+$ spectra
for $^{12}$C and $^{28}$Si targets at forward-angle 
with an incident momentum 1.65 GeV$/c$.
We can see clearly the peaks of $p$- and $d$-bound states, 
respectively, in the cases of $^{12}$C and $^{28}$Si targets.
Here, the experimental resolution is assumed
to be 2 MeV. Solid and dotted curves are
for ESC08c and WS14, respectively.
Strong enhancement of the highest-$L$ state in the ESC08c
case is due to the $k_F$-dependent effects of
G-matrix interactions.
When the $\bar k_F$ values for the $p$ and $d$ states are 
taken as the same as those for the $s$ states, the obtained 
spectra for ESC08c become similar to those for WS14.
We conclude this section by making some remarks on the inclusion of the
three-body repulsive (TBR) and attractive (TBA) interactions for S=-2 systems.
In the case of the $\Lambda$-hypernuclei in paper II \cite{NRY12b} an 
important conclusion from the G-matrix analysis is that the experimental 
$B_\Lambda$ values and excited spectra can be reproduced in a natural way
by ESC08c. Although the multipomeron (MPP) repulsive contributions are
decisively important in the high density region, they should be almost canceled by the three-body attractions (TBA) in the normal density region.

In the case of the $\Xi$-hypernuclei it is shown here that the $\Xi N$
attraction in ESC08c is consistent with the $\Xi$-nucleus binding energies
given by the emulsion data of the twin $\Lambda$-hypernuclei.
As in the case of the $\Lambda$-hypernuclei, we can expect some role of
the MPP+TBA contribution. For a clear analysis, however, the experimental
data of $B_\Xi$ are too scarce. On the other hand, MPP contributions
are essential in the problem of $\Xi$-mixing in neutron star matter.
We defer the discussion and inclusion of the three-body interactions in the S=-2 
system, i.e. ESC08c+ model, to a future paper.

\section{Summary and Conclusion}
\label{sec:conc}
The ESC08c model potentials presented here are a major step in constructing
the baryon-baryon interactions for scattering and hypernuclei in the
context of broken SU(3)-symmetry using, apart from the 
gaussian repulsion from the Pomeron and inclusion of a systematic quark-core
effects for all baryon-baryon channels, generalized yukawian meson-exchange 
for the dynamics. The potentials are based on
(i) One-boson-exchanges, where the coupling constants at the
baryon-baryon-meson vertices are restricted by the broken SU(3) symmetry,
(ii) Two-pseudoscalar exchanges, (iii) Meson-Pair exchanges.
Each type of meson exchange (pseudoscalar, vector, axial-vector, scalar)
contains five free parameters: a singlet coupling constant, an octet
coupling constant, the $F/(F+D)$ ratio $\alpha$, a meson-mixing angle.
The potentials are regularized with gaussian cut-off parameters,
which provide a few additional free parameters.
As shown in paper I and II the $F/(F+D)$ parameters could be restricted, both 
for OBE and MPE, by the Quark-model predictions in the form of the 
$^3P_0$ quark-antiquark creation mechanism.

Although we performed truly simultaneous fits to the {\it N\!N} and {\it Y\!N} data, 
effectively
most of these parameters are determined in fitting the rich and    
accurate {\it N\!N} scattering data, while the remaining ones are fixed
by fitting also the (few) {\it Y\!N} scattering data. This still leaves
enough flexibility to accomodate the imposition of a few extra constraints.
As demonstrated here, the assumption of SU(3) symmetry for the couplings then
allows us to extend these models to the higher strangeness channels
(i.e., {\it YY} and all interactions involving cascades), without the
need to introduce additional free parameters. 
Like the NSC97 models, the ESC04 and ESC08 models are very powerful models 
of this kind, and the very first realistic ones.\\
The most striking prediction of ESC08c is the existence of the S=-2
deuteron $D^*$, below the $\Xi N$-threshold. 
The width is expected to be small since the decay must be isospin 
breaking and is electromagnetic and/or weak. The experimental search for 
baryon-baryon bound states by the Rome-Saclay-Vanderbilt collaboration
\cite{RSV82} in the mass range 21.-2.5 GeV/c$⁶2$ was negative.
It could be that the resolution in this experiment was insufficient
to detect a very narrow state near the $\Xi N$-threshold.
It is important to emphasize that the existence of the $D^*$-state is strongly
connected to the $\Xi$-nucleus attraction as indicated by experiments, 
see \cite{E885} and the recent emulsion-experiments results \cite{Nakazawa}.
In one of the ESC04-models, ESC04d, the bound S=-2 bound state occurred 
in the $\Xi N(^3S_1,I=0)$-channel, 
which is a member of an SU(3) octet $\{8_a\}$-irrep. The ESC08c result 
is much more natural, fitting nicely with the existence of a $\{10^*\}$ 
SU(3)-deuteron multiplet.

In order to illustrate the basic properties of these potentials,
we have presented results for scattering lengths, possible bound
states in $S$-waves, and total cross sections. 
Although the different versions ESC04 and ESC08 produce the $NN$ and $YN$ data 
well, there are considerable differences. In the NN-sector the quality of the fit to
the NN-data of the ESC08-models is superior to that for the ESC04-models.
Also, they lead to notable differences in the hypernuclear 
structures, especially in $S=-2$ systems. 
A typical example can be seen in their $\Xi N$ sectors: The derived
$\Xi$-nucleus potentials are different from each other even 
qualitatively. It is quite important that ESC04d and ESC08a,b,c solutions
predicts the existence of $\Xi$-hypernuclei
consistently with the indication given by the BNL-E885 experiment.
For a discussion $\Xi$-nucleus attraction in the case of the ESC04 and ESC08a,b we 
refer to \cite{Rij05b} and \cite{YMR10b} respectively.
The $\Xi$-nucleus attraction derived from ESC08c is owing to the 
situation that the $\Xi N$ interaction in the $^3S_1$ ($^{33}S_1$)
state is substantially attractive. This
feature is intimately related to tensor-potential giving a strong Lane term.
The mass dependence of $\Xi$ hypernuclei predicted by ESC08c is rather 
different from that by the OBE model such as NHC-D. The most striking
is that the peculier $\Xi N$ hypernuclear states are obtained by
ESC08c even in $s$- and light $p$-shell regions.

We finally mention that these ESC08 potentials also provide an
excellent starting point for calculations and predictions of multi-strange systems.
The extension of this work to the
$S=-3,4$-systems, i.e. comprising all $\{8\}\otimes\{8\}$ baryon-baryon
states, will be the topic of the last paper (IV) in this series.

\acknowledgments
We thank T. Motoba and E. Hiyama for many stimulating discussions. 

\newpage
\appendix 

\section{Baryon-baryon channels and SU(3)-irreps}
\label{app:C}
In Table~\ref{tab.irrep1} and Table~\ref{tab.irrep2}
we give the relation between the potentials
on the isospin basis and the potentials in the SU(3)-irreps.

\begin{table*}[ht]
\caption{ SU(3)-contents of the various potentials 
          on the isospin basis.}
\begin{ruledtabular}
\begin{tabular}{ccl} 
\multicolumn{3}{c}{Space-spin antisymmetric states $^{1}S_{0},
                    \ ^{3}P,\ ^{1}D_{2},...$} \\
\colrule
$ \Lambda \Lambda \rightarrow \Lambda\Lambda $  & $ I=0 $ & 
$V_{\Lambda\Lambda,\Lambda\Lambda}
 = \frac{1}{40}\left(27 V_{27}+8 V_{8_{s}}+5 V_{1}\right)$\\
$ \Lambda \Lambda \rightarrow \Xi N  $  & ,, & $V_{\Lambda\Lambda,\Xi N}
 = \frac{-1}{40}\left(18 V_{27}-8 V_{8_{s}}-10 V_{1}\right)$\\
$ \Lambda \Lambda \rightarrow \Sigma\Sigma $  & ,, & $V_{\Lambda\Lambda,\Sigma\Sigma}
 =\frac{\sqrt{3}}{40}\left(-3 V_{27}+8 V_{8_{s}}-5 V_{1}\right)$\\
$ \Xi N   \rightarrow \Xi N  $  & ,, & $V_{\Xi N ,\Xi N}
 = \frac{1}{40}\left( 12 V_{27}+8 V_{8_{s}}+20 V_{1}\right)$\\
$ \Xi N   \rightarrow \Sigma\Sigma $  & ,, & $V_{\Xi N ,\Sigma\Sigma}
 =\frac{\sqrt{3}}{40}\left( 2 V_{27}+8 V_{8_{s}}-10 V_{1}\right)$\\
$ \Sigma\Sigma \rightarrow \Sigma\Sigma  $  & ,, & $V_{\Sigma\Sigma,\Sigma\Sigma}
 = \frac{1}{40}\left( V_{27}+24 V_{8_{s}}+15 V_{1}\right)$\\
\colrule
$ \Xi N   \rightarrow \Xi N   $  & $I=1$ & $V_{\Xi N ,\Xi N }
 = \frac{1}{5}\left( 2 V_{27}+3 V_{8_{s}}\right)$\\
$ \Xi N   \rightarrow \Lambda\Sigma  $  & ,,    & $V_{\Xi N ,\Lambda\Sigma}
 = \frac{\sqrt{6}}{5}\left( V_{27} - V_{8_{s}}\right)$\\
$ \Sigma\Lambda \rightarrow \Sigma\Lambda  $  & ,,  & $V_{\Lambda\Sigma,\Lambda\Sigma}
 = \frac{1}{5}\left( 3 V_{27}+2 V_{8_{s}}\right)$\\
\colrule
$ \Sigma\Sigma  \rightarrow \Sigma\Sigma  $  & $I=2$ & $V_{\Sigma\Sigma,\Sigma\Sigma}
 = V_{27}$\\
\end{tabular}
\end{ruledtabular}
\label{tab.irrep1}
\end{table*}
 
\begin{table*}[ht]
\caption{ SU(3)-contents of the various potentials\\
 on the isospin         basis. }   
\begin{ruledtabular}
\begin{tabular}{ccl} 
\multicolumn{3}{c}{Space-spin symmetric states $^{3}S_{1},\ ^{1}P_{1},
                   \ ^{3}D,...$} \\
\colrule
$ \Xi N   \rightarrow \Xi N  $  & $ I=1$ & $V_{\Xi N,\Xi N}
 = \frac{1}{3}\left(V_{10}+ V_{10^{*}} + V_{8_{a}}\right)$\\
$ \Xi N   \rightarrow \Sigma\Lambda $  & $ ,,  $ & $V_{\Xi N ,\Sigma\Lambda}
 = \frac{\sqrt{6}}{6}\left(V_{10}- V_{10^{*}} \right)$\\
$ \Xi N   \rightarrow \Sigma\Sigma $  & $ ,, $ & $V_{\Xi N,\Sigma\Sigma}
 = \frac{\sqrt{2}}{6}\left(V_{10}+ V_{10^{*}} -2 V_{8_{a}}\right)$\\
$ \Sigma\Lambda \rightarrow \Sigma\Lambda $  & $ ,, $ & $V_{\Sigma\Lambda,\Sigma\Lambda}
 = \frac{1}{2}\left(V_{10}+ V_{10^{*}} \right)$\\
$ \Sigma\Lambda  \rightarrow \Sigma\Sigma $  & $ ,, $ & $V_{\Sigma\Lambda,\Sigma\Sigma}
 = \frac{\sqrt{3}}{6}\left(V_{10}- V_{10^{*}} \right)$\\
$ \Sigma\Sigma  \rightarrow \Sigma\Sigma $  & $ ,, $ & $V_{\Sigma\Sigma,\Sigma\Sigma}
 = \frac{1}{6}\left(V_{10}+ V_{10^{*}}+4 V_{8_{a}} \right)$\\
\colrule
$ \Xi N   \rightarrow \Xi N  $  & $ I=0$ & $V_{\Xi N,\Xi N}
 =  V_{8_{a}}$\\
\end{tabular}
\end{ruledtabular}
\label{tab.irrep2}
\end{table*}

\section{Meson-pair coupling constants}
\label{app:A}
In Table~\ref{prcop08c} we give the MPE-couplings
for model ESC08c.

\begin{table*}[hbt]
\caption{Pair coupling constants for model ESC08c, divided by
         $\protect\sqrt{4\pi}$. $I(M)$ refers to the isospin of the pair $M$ with 
         quantum-numbers $J^{PC}$.}
\begin{ruledtabular}
\begin{tabular}{ccccrrrrccrrrr}
 Pair & $J^{PC}$ & Type & $I(M)$ & 
 $N\!N\!M$ & $\Sigma\Sigma M$ & $\Sigma\Lambda M$ & $\Xi\Xi M$
 & $\hspace*{2ex}$
 & $I(M)$ & $\Lambda N\!M$ & $\Lambda\Xi M$ & $\Sigma N\!M$ & $\Sigma\Xi M$ \\
\colrule
$\pi\eta$& $0^{++}$& $g$ 
    & $ 1 $  & --1.2371  & --2.4742  &   0.0000  & --1.2371 
    && $1/2$ &   2.1427  & --2.1427  &   1.2371  &   1.2371  \\
  & & & $ 0$ & --2.1427  &   0.0000  &   0.0000  &   2.1427  
    &&    &   &   &   &  \\
$\pi\pi$&$1^{--}$& $g$ 
    & $1$    &   0.2703  &   0.5406  &   0.0000  &   0.2703 
    && $1/2$ & --0.4682  &   0.4682  & --0.2703  & --0.2703  \\
  &&& $ 0 $  &   0.4682  &   0.0000  &   0.0000  & --0.4682  
    &&    &   &   &   &  \\
$\pi\pi$&$1^{--}$& $f$ 
    & $1$    & --1.6592  & --1.3274  & --1.1495  &   0.3318 
    && $1/2$ &   1.7243  & --0.5748  & --0.3318  &   1.6592  \\
  &&& $ 0 $  & --0.5748  &   1.1495  & --1.1495  &   1.7243  
    &&    &   &   &   &  \\
$\pi\rho$&$1^{++}$& $g$ 
    & $1$    &   5.1287  &   4.1030  &   3.5533  & --1.0257 
    && $1/2$ & --5.3299  &   1.7766  &   1.0257  & --5.1287  \\
  &&& $ 0 $  &   1.7766  & --3.5533  &   3.5533  & --5.3299  
    &&    &   &   &   &  \\
$\pi\sigma$&$1^{++}$& $g$ 
    & $1$    & --0.2988  & --0.2391  & --0.2070  &   0.0598 
    && $1/2$ &   0.3106  & --0.1035  & --0.0598  &   0.2988  \\
  &&& $ 0 $  & --0.1035  &   0.2070  & --0.2070  &   0.3106  
    &&    &   &   &   &  \\
$\pi\omega$&$1^{+-}$& $g$ 
    & $1$    & --0.2059  & --0.1648  & --0.1427  &   0.0412 
    && $1/2$ &   0.2140  & --0.0713  & --0.0412  &   0.2059  \\
  &&& $ 0 $  & --0.0713  &   0.1427  & --0.1427  &   0.2140  
    &&    &   &   &   &  \\[2mm]
\end{tabular}
\end{ruledtabular}
\label{prcop08c}
\end{table*}

\section{BKS-phase parameters}              
\label{app:B}
In Tables~\ref{tab.bks.a}-\ref{tab.bks.c} 
we display the BKS-phase parameters for model ESC08c.

\begin{table*}[hbt]
\caption{ESC08c $^1S_0(\Lambda\Lambda \rightarrow \Lambda\Lambda)$ BKS-phase parameters
         in [degrees] as a function of the laboratory momentum $p_\Lambda$
         in [MeV]}
\begin{ruledtabular}
\begin{tabular}{r|rr|rr|rr|rrr}
$p_{\Lambda}$  & \multicolumn{1}{c}{$\delta(^1S_0)$}&\multicolumn{1}{c|}{$\rho(^1S_0)$} 
               & \multicolumn{1}{c}{$\delta(^3P_0)$}&\multicolumn{1}{c|}{$\rho(^3P_0)$}  
               & \multicolumn{1}{c}{$\delta(^3P_1)$}&\multicolumn{1}{c|}{$\rho(^3P_1)$}  
               & \multicolumn{1}{c}{$\delta(^3P_2)$}&\multicolumn{1}{c}{$\epsilon_2$}  
               & \multicolumn{1}{c|}{$\delta(^3F_2)$}\\                                      
\colrule   
 10  &   1.23 & ---  &--0.00 & ---   &  0.00 & ---  & 0.00  & 0.00 & 0.00 \\
 50  &   5.94 & ---  &--0.00 & ---   &  0.01 & ---  & 0.02  & 0.00 & 0.00 \\
 100 &  10.67 & ---  &--0.04 & ---   &  0.10 & ---  & 0.18  & 0.00 & 0.00 \\
 200 &  14.96 & ---  &--0.35 & ---   &  0.57 & ---  & 1.30  & 0.02 & 0.00 \\
 300 &  15.06 & ---  &--1.17 & ---   &  1.19 & ---  & 3.95  & 0.13 & 0.03 \\
 350 &  18.43 &13.90 &--1.66 &  0.71 &  1.40 & 0.29 & 6.39  & 0.30 & 0.07 \\
 400 &  11.31 &20.52 &--2.12 &  4.48 &  1.48 & 1.80 & 10.78 & 0.72 & 0.16 \\
 500 &   5.63 &22.04 &--3.56 &  9.92 &  1.07 & 4.59 & 14.55 & 2.81 & 0.36 \\
 600 & --9.11 &21.73 &--5.07 & 14.51 &--0.12 & 7.22 &--5.00 & 3.61 & 0.58 \\
 700 &--17.82 &20.65 &--5.92 & 19.01 &--1.84 & 9.51 &--8.75 & 4.04 & 0.83 \\
 800 &--25.64 &18.80 &--5.56 & 23.66 &--3.67 & 11.53&--10.75& 4.76 & 1.11 \\
 900 &--32.03 &15.64 &--3.43 & 28.44 &--4.47 & 13.85&--12.81& 5.81 & 1.65 \\
1000 &--36.88 &13.35 &  0.02 & 33.25 &--7.46 & 21.89&--14.18& 6.75 & 1.03 \\
\end{tabular}
\end{ruledtabular}
\label{tab.bks.a}  
\end{table*}

\begin{table*}     
\caption{ESC08c $^1S_0,^1P_1(\Xi N \rightarrow \Xi N, I=0$ BKS-phase parameters
         in [degrees] as a function of the laboratory momentum $p_\Lambda$
         in [MeV]}
\begin{ruledtabular}
\begin{tabular}{r|rr|rr}
$p_{\Xi}$  & \multicolumn{1}{c}{$\delta(^1S_0)$}&\multicolumn{1}{c|}{$\rho(^1S_0)$} 
             & \multicolumn{1}{c}{$\delta(^1P_1)$}&\multicolumn{1}{c}{$\rho(^1P_1)$}\\
\colrule   
 10  &   0.08 & 5.45 &   0.00 & 0.00 \\
 50  &   0.25 &11.85 &   0.05 & 0.00 \\
 100 & --0.30 &15.88 &   0.33 & 0.00 \\
 200 &--4.17  &19.78 &   1.23 & 0.00 \\
 300 &--10.09 &21.40 &   1.50 & 0.00 \\
 350 &--13.37 &21.79 &   1.22 & 0.00 \\
 400 &--16.72 &22.00 &   0.64 & 0.00 \\
 500 &--23.34 &22.00 & --1.31 & 0.00 \\
 600 &--29.54 &21.56 & --4.11 & 0.00 \\
 700 &--35.03 &20.72 & --7.50 & 0.00 \\
 800 &--39.55 &19.44 &--11.23 & 0.00 \\
 900 &--42.78 &17.51 &--15.13 & 0.00 \\
1000 &--43.87 &14.78 &--19.11 & 0.00 \\
\end{tabular}
\end{ruledtabular}
\label{tab.bks.b}  
\end{table*}

\begin{table*}[hbt]
\caption{ESC08c $^1S_0(\Xi N \rightarrow \Xi N, I=1)$ etc. BKS-phase parameters
         in [degrees] as a function of the laboratory momentum $p_\Xi$
         in [MeV]}
\begin{ruledtabular}
\begin{tabular}{r|rr|rr|rr|rrr}
$p_{\Lambda}$  & \multicolumn{1}{c}{$\delta(^1S_0)$}&\multicolumn{1}{c|}{$\rho(^1S_0)$} 
               & \multicolumn{1}{c}{$\delta(^3P_0)$}&\multicolumn{1}{c|}{$\rho(^3P_0)$}  
               & \multicolumn{1}{c}{$\delta(^3P_1)$}&\multicolumn{1}{c|}{$\rho(^3P_1)$}  
               & \multicolumn{1}{c}{$\delta(^3P_2)$}&\multicolumn{1}{c}{$\epsilon_2$}  
               & \multicolumn{1}{c|}{$\delta(^3F_2)$}\\                                      
\colrule   
 10  & --0.70 & ---  &--0.00 & ---   &  0.00 & ---  & 0.00  & 0.00 & --0.00 \\
 50  & --3.46 & ---  &--0.04 & ---   &  0.04 & ---  & 0.00  & 0.00 & --0.00 \\
 100 & --6.75 & ---  &--0.29 & ---   &  0.30 & ---  & 0.02  & 0.01 & --0.00 \\
 200 &--12.79 & ---  &--1.32 & ---   &  1.55 & ---  & 0.08  & 0.12 & --0.01 \\
 300 &--18.35 & ---  &--2.39 & ---   &  3.18 & ---  & 0.03  & 0.35 & --0.05 \\
 350 &--20.95 & ---  &--2.80 & ---   &  3.92 & ---  &--0.12 & 0.48 & --0.07 \\
 400 &--23.37 & ---  &--3.12 & ---   &  4.52 & ---  &--0.39 & 0.62 & --0.10 \\
 500 &--27.15 & ---  &--3.45 & ---   &  5.24 & ---  &--1.33 & 0.76 & --0.12 \\
 600 &--24.30 &17.88 &--3.31 & 0.61  &  5.55 & 1.35 &--2.82 & 1.12 & --0.13 \\
 700 &--37.01 &24.33 &--2.37 & 5.39  &  5.49 & 7.50 &--4.74 & 1.31 & --0.05 \\
 800 &--44.30 &25.12 &--0.89 & 11.40 &  4.39 & 11.77&--6.93 & 1.46 &   0.10 \\
 900 &--37.14 &25.06 &  0.28 & 18.12 &  2.47 & 14.93&--9.26 & 1.60 &   0.31 \\
1000 &--31.01 &24.57 &--0.21 & 24.24 &--0.00 & 17.31&--11.65& 1.73 &   0.58 \\
\end{tabular}
\end{ruledtabular}
\label{tab.bks.c}  
\end{table*}

\begin{table*}     
\caption{ESC08c $^1S_0,^3S_1-^3D_1(\Xi N \rightarrow \Xi N, I=0)$ 
         BKS-phase parameters in [degrees] as a function of the laboratory 
         momentum $p_\Xi$ in [MeV]}
\begin{ruledtabular}
\begin{tabular}{r|rr||rrr|rrr}
$p_{\Xi}$   & \multicolumn{1}{c}{$\delta(^1S_0)$}&\multicolumn{1}{c||}{$\rho(^1S_0)$} 
               & \multicolumn{1}{c}{$\delta(^3S_1)$}&\multicolumn{1}{c}{$\epsilon_1$} 
               & \multicolumn{1}{c|}{$\delta(^3D_1)$} 
               & \multicolumn{1}{c}{$\eta_{11}$}&\multicolumn{1}{c}{$\eta_{12}$} 
               & \multicolumn{1}{c}{$\eta_{22}$}\\
\colrule   
 10  &  0.08 &  5.45 &   6.19 & 0.00 &   0.00 &  1.00 & 0.00 & 1.00 \\
 50  &  0.25 & 11.85 &  27.52 & 0.20 &   0.00 &  1.00 & 0.00 & 1.00 \\
 100 &--0.30 & 15.88 &  42.78 & 0.87 & --0.00 &  1.00 &  0.03 & 1.00 \\
 200 &--4.17 & 19.78 &  51.04 & 2.47 & --0.08 &  0.99 &  0.11 & 0.99 \\
 300 &--10.09& 21.40 &  49.00 & 4.48 & --0.32 &  0.98 &  0.18 & 0.98 \\
 400 &--16.72& 22.00 &  43.88 & 7.06 & --0.87 &  0.97 &  0.24 & 0.97 \\
 500 &--23.34& 22.00 &  37.52 &10.12 & --1.77 &  0.95 &  0.28 & 0.95 \\
 600 &--29.54& 21.56 &  30.65 &13.41 & --2.81 &  0.92 &  0.29 & 0.92 \\
 700 &--35.03& 20.72 &  23.85 &16.71 & --3.47 &  0.90 &  0.27 & 0.90 \\
 800 &--39.55& 19.44 &  17.52 &19.87 & --3.32 &  0.89 &  0.23 & 0.89 \\
 900 &--42.78& 17.51 &  11.74 &22.77 & --2.24 &  0.89 &  0.18 & 0.89 \\
1000 &--43.87& 14.78 &   6.42 &25.31 & --0.37 &  0.91 &  0.14 & 0.91 \\
\end{tabular}
\end{ruledtabular}
\label{tab.bks.d}  
\end{table*}
\begin{table*}     
\caption{ESC08c $^1S_0,^3S_1-^3D_1(\Xi N \rightarrow \Xi N, I=1)$ 
         BKS-phase parameters in [degrees] as a function of the laboratory 
         momentum $p_\Xi$ in [MeV]}
\begin{ruledtabular}
\begin{tabular}{r|rr||rrr|rrr}
$p_{\Xi}$   & \multicolumn{1}{c}{$\delta(^1S_0)$}&\multicolumn{1}{c||}{$\rho(^1S_0)$} 
               & \multicolumn{1}{c}{$\delta(^3S_1)$}&\multicolumn{1}{c}{$\epsilon_1$} 
               & \multicolumn{1}{c|}{$\delta(^3D_1)$} 
               & \multicolumn{1}{c}{$\eta_{11}$}&\multicolumn{1}{c}{$\eta_{12}$} 
               & \multicolumn{1}{c}{$\eta_{22}$}\\
\colrule   
 10  &--0.70 &  0.00 & 173.90 & 0.00 & --0.00 &  1.00 &--0.00 & 1.00 \\
 50  &--3.48 &  0.00 & 151.56 & 0.03 & --0.00 &  1.00 &--0.00 & 1.00 \\
 100 &--6.80 &  0.00 & 131.47 & 0.11 & --0.01 &  1.00 &--0.00 & 1.00 \\
 200 &--12.91&  0.00 & 110.65 & 0.23 & --0.10 &  1.00 &--0.02 & 1.00 \\
 300 &--18.55&  0.00 & 101.81 & 0.28 & --0.27 &  1.00 &--0.05 & 1.00 \\
 400 &--23.69&  0.00 &  99.08 & 0.36 & --0.43 &  1.00 &--0.08 & 1.00 \\
 500 &--27.67&  0.00 & 101.75 & 0.75 & --0.44 &  0.99 &--0.13 & 0.99 \\
 600 &--25.39& 17.43 & 118.13 & 2.44 & --0.26 &  0.77 &--0.14 & 0.99 \\
 700 &--37.88& 23.74 & 116.81 & 3.32 &   0.06 &  0.45 &--0.16 & 0.98 \\
 800 &--43.46& 24.45 & 118.30 & 4.69 &   0.57 &  0.33 &--0.20 & 0.97 \\
 900 &--36.27& 24.31 & 121.57 & 6.81 &   1.45 &  0.28 &--0.23 & 0.95 \\
1000 &--30.08& 23.74 & 131.27 &10.53 &   2.15 &  0.34 &--0.26 & 0.93 \\
\end{tabular}
\end{ruledtabular}
\label{tab.bks.e}  
\end{table*}

\begin{table*}     
\caption{ESC08c $^1S_0,^3S_1-^3D_1(\Sigma\Lambda\rightarrow \Sigma\Lambda, I=1)$ 
         BKS-phase parameters in [degrees] as a function of the laboratory 
         momentum $p_\Xi$ in [MeV]}
\begin{ruledtabular}
\begin{tabular}{r|rr||rrr|rrr}
$p_{\Sigma}$   & \multicolumn{1}{c}{$\delta(^1S_0)$}&\multicolumn{1}{c||}{$\rho(^1S_0)$} 
               & \multicolumn{1}{c}{$\delta(^3S_1)$}&\multicolumn{1}{c}{$\epsilon_1$} 
               & \multicolumn{1}{c|}{$\delta(^3D_1)$} 
               & \multicolumn{1}{c}{$\eta_{11}$}&\multicolumn{1}{c}{$\eta_{12}$} 
               & \multicolumn{1}{c}{$\eta_{22}$}\\
\colrule   
 10  &  0.32 &  6.17 & --1.42 & 0.00 & --0.00 &  1.00 &--0.00 & 1.00 \\
 50  &  1.34 & 13.32 & --7.13 & 0.07 & --0.00 &  0.88 &--0.00 & 1.00 \\
 100 &  1.43 & 17.66 &--14.35 & 0.40 & --0.04 &  0.77 &--0.00 & 1.00 \\
 200 &--1.80 & 21.80 &--28.78 & 1.09 & --0.44 &  0.60 &--0.02 & 1.00 \\
 300 &--7.51 & 23.57 &--42.69 & 1.11 & --1.16 &  0.48 &--0.03 & 1.00 \\
 400 &--14.37& 24.31 &--55.47 & 0.85 & --1.90 &  0.39 &--0.03 & 1.00 \\
 500 &--21.64& 24.46 &--65.61 & 1.20 & --2.31 &  0.33 &--0.07 & 0.99 \\
 600 &--28.87& 24.18 &--67.63 & 2.90 & --1.25 &  0.31 &--0.17 & 0.97 \\
 700 &--35.84& 23.54 &--29.29 &13.38 & --3.45 &  0.19 &--0.16 & 0.83 \\
 800 &--42.45& 22.62 & --6.47 &13.61 & --9.76 &  0.19 &--0.08 & 0.87 \\
 900 &--41.35& 21.45 & --1.81 &11.92 &--13.71 &  0.28 &--0.07 & 0.88 \\
1000 &--35.53& 20.11 & --3.05 &10.41 &--17.07 &  0.36 &--0.09 & 0.89 \\
\end{tabular}
\end{ruledtabular}
\label{tab.bks.f}  
\end{table*}

\begin{table*}     
\caption{ESC08c $I=2,L=0, L=1$ $\Sigma^\pm\Sigma^\pm\rightarrow \Sigma^\pm\Sigma^\pm$ 
         BKS-phase parameters in [degrees] as a function of the laboratory 
         momentum $p_\Xi$ in [MeV]. In parentheses the phases without Coulomb are listed.}
\begin{ruledtabular}
\begin{tabular}{r|rrr||rrr}
$p_{\Sigma}$   & \multicolumn{1}{c}{$\delta(^1S_0)$}&\multicolumn{1}{c}{$\delta(^3P_0)$} 
               & \multicolumn{1}{c||}{$\delta(^3P_1)$}  
               & \multicolumn{1}{c}{$\delta(^3P_2)$}&\multicolumn{1}{c}{$\epsilon_2$} 
               & \multicolumn{1}{c}{$\delta(^3F_2)$} \\
\colrule   
 10  &--12.35 (0.94)    &--0.20 (  0.00) & --0.22 ( --0.00)&--0.21 (  0.00)&   0.00 &--0.00 (  0.00)\\
 50  &--13.72 (4.26)    &--3.08 (  0.19) & --3.41 ( --0.12)&--3.27 (  0.01)&   0.00 &--0.01 (  0.00)\\
 100 &--5.59  (6.46)    &--6.41 (  1.17) &--8.24  ( --0.71)&--7.44 (  0.11)&   0.04 &--0.59 (  0.00)\\
 200 &--3.03  (4.75)    &--1.01 (  4.46) &--8.24  ( --2.82)&--4.66 (  0.79)&   0.43 &--3.25 (  0.05)\\
 300 &--7.40  (--1.50)  &  1.54 (  5.86) &--9.17  ( --4.88)&--2.44 (  1.86)&   1.07 &--2.68 (  0.22)\\
 400 &--14.47 (--9.65)  &  0.36 (  3.95) &--10.07 ( --6.49)&--0.85 (  2.73)&   1.63 &--2.05 (  0.48)\\
 500 &--22.56 (--18.44) &--3.51 (--0.41) &--10.63 ( --7.53)&  0.08 (  3.18)&   1.94 &--1.49 (  0.76)\\
 600 &--30.91 (--27.29) &--8.89 (--6.14) &--10.59 ( --7.84)&  0.57 (  3.31)&   1.91 &--1.10 (  0.93)\\
 700 &--39.15 (--35.91) &--14.98(--12.50)&--9.76  ( --7.29)&  0.82 (  3.30)&   1.56 &--1.03 (  0.84)\\
 800 &--42.88 (--44.18) &--21.31(--19.04)& --8.14 ( --5.88)&  0.90 (  3.17)&   0.94 &--1.36 (  0.36)\\
 900 &--35.25 (--37.94) &--27.63(--25.53)& --5.96 ( --3.86)&  0.75 (  2.85)&   0.15 &--2.15 (--0.54)\\
1000 &--28.00 (--30.50) &--33.80(--31.84)& --3.57 ( --1.62)&  0.30 (  2.25)&   0.71 &--3.37 (--1.87)\\
\end{tabular}
\end{ruledtabular}
\label{tab.bks.g}  
\end{table*}

\end{document}